\documentclass[epj]{svjour}
\pdfoutput=1

\usepackage[T1]{fontenc} 
\usepackage{amsmath}
\usepackage{amssymb}
\usepackage{graphicx}
\usepackage{slashed}
\usepackage{feynmp}
\usepackage{color}
\newcommand{\sss}{\scriptscriptstyle}
\newcommand{\CA}{\ensuremath{\mathcal{C}_{\sss \mathcal{A}}}}
\newcommand{\Ccr}{\ensuremath{\mathcal{C}_{cr}}}
\newcommand{\CMag}{\ensuremath{\mathcal{C}_{\sss \mathcal{M} }}}
\newcommand{\CEle}{\ensuremath{\mathcal{C}_{el}}}

\newcommand{\OA}{\ensuremath{\mathcal{O}_{\sss \mathcal{A}}}}
\newcommand{\Ocr}{\ensuremath{\mathcal{O}_{cr}}}
\newcommand{\OMag}{\ensuremath{\mathcal{O}_{\sss \mathcal{M} }}}
\newcommand{\OEle}{\ensuremath{\mathcal{O}_{el}}}
\newcommand{\cw}{\ensuremath{c_{\sss W}}}
\newcommand{\sw}{\ensuremath{s_{\sss W}}}
\newcommand{\MW}{\ensuremath{m_{\sss W}}}
\newcommand{\MZ}{\ensuremath{m_{\sss Z}}}
\newcommand{\MX}{\ensuremath{m_{\sss \chi}}}
\DeclareMathAlphabet{\mathpzc}{OT1}{pzc}{m}{it}

\begin{document}
\title{\boldmath Light and Darkness: consistently coupling dark matter to photons via effective operators}
\author{Chiara Arina\inst{1} \and Andrew Cheek \inst{1} \and Ken Mimasu\inst{1,2} \and Luca Pagani\inst{3}}                     
\institute{Centre for Cosmology, Particle Physics and Phenomenology (CP3), Universit\'e catholique de Louvain, Chemin du Cyclotron 2, B-1348 Louvain-la-Neuve, Belgium  
\and 
Theoretical Particle Physics and Cosmology Group, Department of Physics,
King's~College~London, London WC2R 2LS, UK
\and 
Dipartimento di Fisica e Astronomia, Universit\`a di Bologna and INFN, Sezione di Bologna, via Irnerio 46, 40126 Bologna, Italy}
\date{Received: date / Revised version: date}
%
\abstract{
We consider the treatment of fermionic dark matter interacting with photons via dimension-5 and -6 effective operators, arguing that one should always use hypercharge gauge field form factors, instead of those of the photon. Beyond the simple observation that the electromagnetic form factor description breaks down at the electroweak scale, we show how the additional couplings to the $Z$ boson predicted by the hypercharge form factors modify the relic density calculation and indirect detection limits for dark matter masses of a few tens of GeV and above. Furthermore, constraints from the invisible $Z$ decay width can be competitive for masses below $10$ GeV.  We review the phenomenology of hypercharge form factors at the LHC as well as for direct and indirect detection experiments. We highlight where the electromagnetic and hypercharge descriptions lead to wildly different conclusions about the viable parameter space and the relative sensitivity of various probes, namely vector boson fusion versus mono-jet constraints from the LHC, and indirect versus direct searches, for larger dark matter masses.
We find that the dimension-5 operators are strongly constrained by direct detection bounds, while for dimension-6 operators LHC mono-jet searches are competitive or better than the other probes we consider.
} 
\maketitle
\section{Introduction}\label{sec:intro}
The cornerstone of the current standard model of Cosmology, $\Lambda$CDM, relies on dark matter~\cite{Aghanim:2018eyx}. Despite this, the nature of dark matter is very poorly known, only gravitational effects have been observed~\cite{Bertone:2004pz}, requiring that other interactions with Standard Model (SM) particles, are at most, very weak.
Of particular interest are the interactions of dark matter with the photon, given its status as the primary messenger of astrophysical and cosmological probes. UV models can generate these interactions at tree-level through extremely small couplings/mixings, and are known as milli-charged dark matter~\cite{Holdom:1985ag,Holdom:1986eq,Abel:2003ue,Batell:2005wa,Foot:2014uba}. If we assume dark matter to be electrically neutral, the general framework for describing these interactions is via electromagnetic form factors, which couple dark matter directly to the electromagnetic field strength tensor~\cite{Pospelov:2000bq,Sigurdson:2004zp,Ho:2012bg}. These are theoretically and experimentally well motivated, and arise in a plethora of models. A quintessential example is the $\gamma\gamma$ line signal, which is one of the main indirect detection signatures, and is ubiquitous in concrete models~\cite{Ibarra:2015fqa}, including supersymmetry, see {\it e.g.}~\cite{Cabral-Rosetti:2015cxa,Roszkowski:2017nbc,Bringmann:2018lay}, the Inert Doublet Model~\cite{Gustafsson:2007pc}, and extra-dimensions, see {\it e.g.}~\cite{Bertone:2010fn,Jackson:2013pjq}.

Heavy mediators that couple the SM to dark matter are a popular explanation for the relative weakness of its interactions. Allowing the use of an effective field theory (EFT) approach to assess the scenario in a fairly model independent way~\cite{Cao:2009uw,Beltran:2010ww,Fox:2011pm,Matsumoto:2014rxa,Belyaev:2018pqr}. The effective operators give a good description when the energy scale of the considered processes is well below the masses of the mediating particles, and their interactions respect the low-energy symmetries of the SM. The appropriate choice of such symmetries is a crucial aspect of a consistent EFT description, and ultimately depends on the relevant scales of the calculation at hand. In this paper we focus on the EFT for electromagnetic form factors, treating them as local, higher dimension operators mediated by heavy, new physics.

An important choice to be made is whether $U(1)_{\sss QED}$ or the full electroweak symmetry group, $SU(2)_{\sss W}\times U(1)_{\sss Y}$, is chosen as the low energy symmetry of the EFT. For operators involving a stable gauge singlet dark matter field, this amounts to either using $\mathcal{O}^{\mu\nu}F_{\mu\nu}$ or $\mathcal{O}^{\mu\nu}B_{\mu\nu}$, where $B_{\mu\nu}$ is the hypercharge gauge field strength and $\mathcal{O}^{\mu\nu}$ is a gauge singlet combination (usually a bilinear) of dark matter fields\footnote{The stability of dark matter ensures this, otherwise one can couple the new singlet to neutrinos as in Refs.~\cite{Magill:2018jla,Shoemaker:2018vii}.}. Hypercharge form factors are linear combinations of electromagnetic form factors and the corresponding $Z$ boson operators, weighted by appropriate factors of the cosine and sine of the Weinberg angle, $\cw$ and $\sw$:
\begin{align}
\label{eq:BtoFandZ}
\begin{split}
\mathcal{C}\mathcal{O}^{\mu\nu}B_{\mu\nu}=&\,\mathcal{C}^\gamma\mathcal{O}^{\mu\nu}F_{\mu\nu}+\mathcal{C}^{\sss Z}\mathcal{O}^{\mu\nu}Z_{\mu\nu},\\
    \mathcal{C}^\gamma=&\,\mathcal{C}\cw;\quad \mathcal{C}^{\sss Z}=-\mathcal{C}\sw,
\end{split}
\end{align}
where $\mathcal{C}$ denotes a generic Wilson coefficient. The two pictures apply in different ranges of energy. For computations at energies far below the electroweak scale, such as scattering processes relevant for dark matter direct detection, the $Z$ boson degree of freedom decouples and the two descriptions are identical. The reality, however, is that we are able to test these models over a wide range of scales thanks to, \emph{e.g.}, collider experiments. Furthermore, the dark matter mass itself is a free parameter and determines the relevant scale for thermal freeze-out and indirect detection constraints. Electromagnetic form factors for dark matter have been considered in direct and indirect detection as well as at colliders over several decades of dark matter mass, extending into the TeV range~\cite{Pospelov:2000bq,Sigurdson:2004zp,Barger:2010gv,Banks:2010eh,Fitzpatrick:2010br,Ho:2012bg,Weiner:2012cb,Weiner:2012gm,DelNobile:2014eta,Crivellin:2014gpa,Fichet:2016clq,Alves:2017uls,Kavanagh:2018xeh,Chu:2018qrm,Kang:2018oej,Florez:2019tqr,Chu:2020ysb}. However, it is clear that the hypercharge form factors should be the appropriate EFT here, given that relevant energies can far exceed the electroweak scale. 

In recent literature, a discrepancy has emerged in the treatment of the electromagnetic form factors beyond their validity, centered on the relevance of the $\chi \chi\rightarrow W^+W^-$ channel. On one hand, this channel has been exploited to yield impressive collider limits~\cite{Florez:2019tqr} in the vector boson fusion (VBF) channel. On the other hand, the weak boson channel has been assumed to be irrelevant for direct and indirect detection and subsequently ignored (see e.g. ~\cite{Kavanagh:2018xeh}). These treatments are in tension with each other and we argue that the correct treatment is somewhat of a synthesis. 

\begin{list}{\textbullet}{}
\item Simply ignoring the  $W^+W^-$ channel as in Ref.~\cite{Kavanagh:2018xeh} is not consistent because $\gamma \rightarrow W^+W^-$ is a vertex in the SM and will appear in annihilation. \\  
\item Electromagnetic form factors predict a large, unphysical growth of the $\chi \chi\rightarrow W^+W^-$  channel at energies above $\sim 80$ GeV. It can be avoided by adopting the hypercharge form factors, which restore the full SM gauge invariance to the effective description. This means that the collider limits in Ref.~\cite{Florez:2019tqr} result from the application of an EFT beyond its validity. \\
\item When including the hypercharge form factors, the $W^+W^-$ channel is tamed at high energies by $Z$-mediated diagrams, this results in a total annihilation cross-section similar to Ref.~\cite{Kavanagh:2018xeh}. However, below $100$ GeV the appearance of the $Z$-boson resonance substantially alters the phenomenology.  
\end{list}

The largest experimentally observable effects come in the form of resonant features around the $Z$-boson mass and additional constraints from its invisible decay width. Ultimately, the most stringent limits usually come from either direct detection or colliders, so modifications of indirect detection limits do not have tremendous consequences. Nevertheless, annihilation plays a central role in the relic density calculation, and Section~\ref{subsec:production} shows how, depending on which form factor you take, the relic line is substantially different.

This paper is organised as follows. In the next section we define both the electromagnetic and hypercharge effective interactions for Dirac and Majorana dark matter particles and discuss their validity range, focusing on the breakdown of the electromagnetic in favour of the hypercharge description. In section~\ref{sec:pheno_colliders} we revisit the collider constraints from VBF, quantifying the impact of switching from electromagnetic to hypercharge form factors. We also present the current and future sensitivity from mono-jet searches, which we now find are the best known way to look for such interactions. In section~\ref{sec:pheno_dm} we determine the most stringent constraints from dark matter direct and indirect searches and assess the prospect for future detection of the models. We finally summarise our findings and discuss the interplay between the various searches before concluding in section~\ref{sec:concl}.

\section{Dark matter hypercharge (and electromagnetic) effective field theory}\label{sec:gauge_inv}
The EFT framework relies on the presence of decoupled, new physics at an arbitrary high energy scale, $\Lambda$, that, in the low-energy limit, leaves behind the light SM particles plus a  dark matter candidate. In this paper we consider the effective interactions of a fermionic singlet dark matter particle with the hypercharge gauge boson field, which describe couplings between dark matter and both the photon and the $Z$ boson.
The hypercharge form factor operators up to dimension-6 are given by,
\begin{equation}
\mathcal{L}_{\mathrm{Majorana}}^{\chi}= \frac{\CA}{\Lambda^{2}} \frac{1}{2} \bar{\chi} \gamma^{\mu} \gamma^{5} \chi \cdot \partial^{\nu} B_{\mu \nu}
\label{eq:majorana_lag}
\end{equation}
for Majorana particles, indicated by $\chi$, and,
\begin{eqnarray}
\mathcal{L}_{\mathrm{Dirac}}^{\psi} & = &  2 \mathcal{L}_{\mathrm{Majorana}}^{\chi \rightarrow \psi} +
\left[
 \frac{\mathcal{C}_{\mathcal{M}}}{2 \Lambda} \bar{\psi} \sigma^{\mu \nu} \psi \cdot B_{\mu \nu} \right. \nonumber  \\
& & \left. +\frac{\mathcal{C}_{el}}{2 \Lambda} i \bar{\psi} \sigma^{\mu \nu} \gamma^{5} \psi \cdot B_{\mu \nu}
+\frac{\mathcal{C}_{cr}}{\Lambda^{2}} \bar{\psi} \gamma^{\mu} \psi \cdot \partial^{\nu} B_{\mu \nu}
\right]
\label{eq:dirac_lag}
\end{eqnarray}
for Dirac fermions, denoted by $\psi$~\footnote{We will loosely refer to dark matter as $\chi$ in our discussion whenever its nature need not be specified.}.  The $\mathcal{C}_j$ are the dimensionless Wilson coefficients for the dimension-6 anapole moment $(\CA)$, for the dimension-5 the electric and magnetic dipole moments $(\mathcal{C}_{el})$ and $(\mathcal{C}_{\mathcal{M}})$ and for the dimension-6 charge radius operator $(\mathcal{C}_{cr})$.  For Majorana particles, the only non-zero hypercharge interaction is the anapole moment, as demonstrated in~\cite{1958JETP....6.1184Z,Radescu:1985wf}. The relation of our Wilson coefficients to usual electromagnetic form factors, denoted by the `$\gamma$' superscript, can be found via Equation~\eqref{eq:BtoFandZ}.
We have implemented this EFT framework into \texttt{FeynRules}~\cite{Alloul:2013bka} and obtained the model files in the UFO format~\cite{Degrande:2011ua}, which will be used in the rest of our analysis~\footnote{The model files are publicly available at \texttt{https://feynrules.irmp.ucl.ac.be/wiki/EWFF4DM}.}.

\begin{figure*}
\centering
\includegraphics[width=0.2\textwidth]{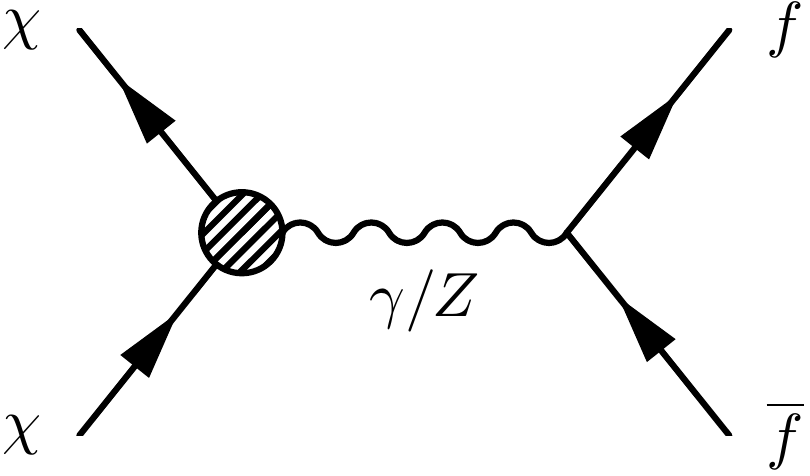}
\hspace{3em}
\includegraphics[width=0.22\textwidth]{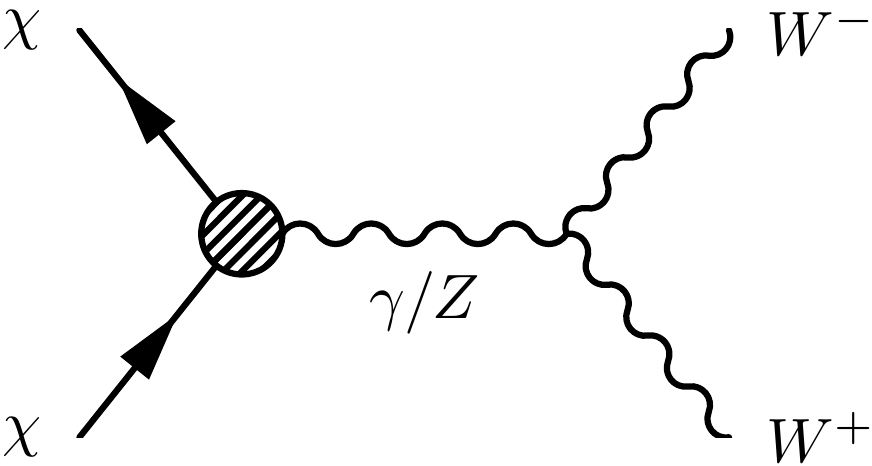}
\hspace{3em}
\includegraphics[width=0.2\textwidth]{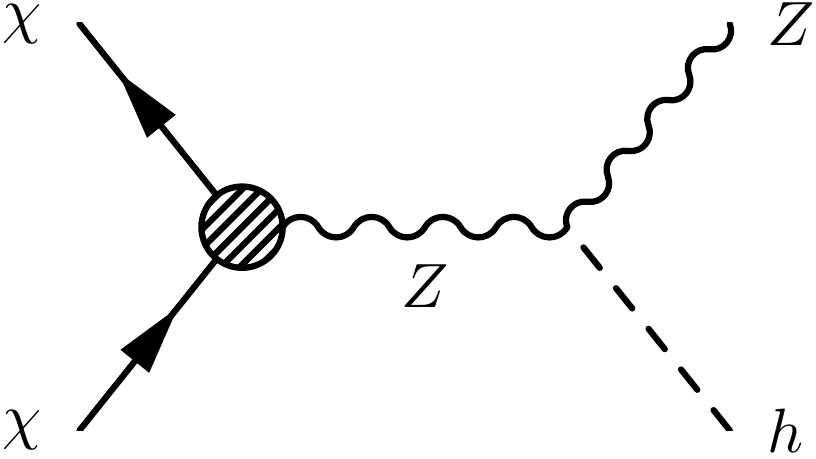}
\caption{Dark matter annihilation diagrams in the hypercharge EFT framework defined in Equations~\eqref{eq:majorana_lag} and~\eqref{eq:dirac_lag} at leading order, namely considering only one vertex insertion.}
\label{fig:annihilation_diags}
\end{figure*}

To highlight the types of interactions and scatterings that the dark matter form factors mediate, Figure~\ref{fig:annihilation_diags} depicts all possible Feynman diagrams for dark matter annihilation into two SM states, via a single insertion of the operators in Equations~\eqref{eq:majorana_lag} and~\eqref{eq:dirac_lag}, that is to say at leading order in the EFT expansion.
\subsection{Pushing electromagnetic form factors beyond their limits}
We have introduced the form factor operators and advocated the use of the hypercharge variants to safely test such dark matter models over a wide range of scales. In this section we will discuss two explicit examples in which the difference between the photon and hypercharge operators can have significant phenomenological consequences, rooted in the application of the photon-only operators beyond their validity.
\subsubsection{Dark matter scattering with $W$ bosons\label{sec:DMWW}}
The $\chi\chi\to W^+W^-$ scattering process is a simple example in which we see the difference between the two EFT descriptions and is particularly relevant for the phenomenology of EW-scale dark matter. Firstly, for dark matter masses above $\MW$, the $W^+W^-$ annihilation channel plays an important role in fixing the thermal relic abundance. Secondly, a recent phenomenological study~\cite{Florez:2019tqr} has found the VBF process to be a sensitive probe of `anapole' dark matter, which corresponds to the photon version of the $\CA$ operator in Equation~\eqref{eq:majorana_lag}. This process embeds the $W^+W^-\to\chi\chi$ amplitude, leading to a striking signature of two very forward jets recoiling against the missing energy reflecting the production of a pair of neutral, stable particles.

For concreteness, we focus on the scattering of a pair of Majorana dark matter candidates, interacting via an anapole moment. This form factor represents the interaction of a dark matter current with the current of the corresponding gauge  (photon or $B$) field. The middle diagram of Figure~\ref{fig:annihilation_diags} represents the contributions to this scattering process. It is clearest to first compute the contribution of the photon-only anapole moment, where only the photon is mediated, with Wilson coefficient $\CA^\gamma$. The amplitude for this process, with incoming momenta $p_1,p_2$ and outgoing momenta $p_3,p_4$, is
\begin{eqnarray}\label{eq:chichiWW_amp}
i M_{\sss A}^\gamma & = &  -\frac{\CA^\gamma}{\Lambda^2}\frac{i\,e}{k^2}\,
\bar{v}\left(p_2,\MX\right) \left(k^2\gamma^\mu\gamma^5-k^\mu\slashed{k}\gamma^5\right) \nonumber \\
& & u\left(p_1,\MX\right)
T^{\rho\sigma}_\mu\varepsilon_\rho(p_3)\,\varepsilon^{*}_{\sigma}(p_4),
\end{eqnarray}
where $k=p_1+p_2$, $u,v$ and $\varepsilon$ represent the spinors and polarisation vectors for $\chi$ and $W^\pm$ respectively, and the tensor structure of the $W^+W^-\gamma$ vertex has been abbreviated by $T^{\mu\rho\sigma}$. The corresponding squared matrix element, summed and averaged over final and initial state polarisations, in the high-energy limit ($\MW^2,\MX^2\ll s  < \Lambda^2$) is
\begin{equation}\label{eq:chichiWW_MA_highenergy}
\vert M_{\sss A}^\gamma\vert^2\sim \frac{2\pi\alpha_{\sss EW}}{9\MW^4}
\left(\frac{\CA^\gamma}{\Lambda^2}\right)^2s^4\,\sin^2\theta+\mathpzc{O}(s^3),
\end{equation}
where $s=k^2$ is the square of the centre of mass energy, $\theta$ denotes the scattering angle and $\alpha_{\sss EW}$ the EW fine structure constant. It implies a growth with energy of the underlying amplitude $|M_{\sss A}|\sim s^2$. Knowing that the contribution comes from a dimension-6 operator, one expects it to scale (at most) like $s$ in the amplitude. This is because $2\to2$ amplitudes are dimensionless and the only scales present in the high energy limit are $\Lambda$ and $s$, meaning that a dimension-6 operator should yield a $s/\Lambda^2$ behaviour. One can further see that the amplitude is not healthy as it does not actually admit a clean high-energy limit, diverging as $\MW\to0$. Using partial wave analysis as in Ref.~\cite{Kahlhoefer:2015bea}, the lowest partial wave of the helicity amplitude for the longitudinal $W$ boson configuration violates unitarity at a centre of mass energy
\begin{equation}
    \label{eq:CAn_pert_unit}
    \sqrt{s} \gtrsim 4.3\sqrt{\MZ\frac{\Lambda}{\sqrt{\CA^\gamma}}}.
\end{equation}
This implies that unitarity is violated below the cutoff ( $\sqrt{s}\lesssim \Lambda$ ), for
\begin{align}
    \label{eq:CAn_unit_viol}
    \Lambda \gtrsim \frac{1.7\,\text{TeV}}{\sqrt{\CA^\gamma}}.
\end{align}

Altogether, it is obvious that the treatment of this amplitude in the EFT is incomplete. This can be traced back to the fact that the photon-only anapole operator should strictly only be used in isolation at energy scales where the $W$ boson field is not a low-energy degree of freedom, \emph{i.e.}, below the EW scale. Instead, using the hypercharge anapole operator in its place yields a result with the expected high energy behaviour for a dimension-6 operator. The $Z$ boson mediated contribution, combined with the former as per Equation~\eqref{eq:BtoFandZ}, exactly cancels the leading high-energy behaviour of Equation~\eqref{eq:chichiWW_amp}, yielding the amplitude
\begin{eqnarray}
i\,M_{\sss A} & = &  \frac{\CA}{\Lambda^2}\frac{i\,e\,\MZ^2\,\cw}{k^2\left(k^2-\MZ^2\right)}\,\bar{v}\left(p_2,\MX\right)\left(k^2\gamma^\mu\gamma^5-k^\mu\slashed{k}\gamma^5\right) \nonumber \\
& & u\left(p_1,\MX\right)T_\mu^{\rho\sigma}\varepsilon_\rho(p_3)\,\varepsilon^{*}_{\sigma}(p_4),
\end{eqnarray}
whose matrix element squared has the high-energy limit
\begin{equation}\label{eq:chichiWW_MB_highenergy}
\vert M_{\sss A}\vert^2\sim \frac{2\pi\alpha_{\sss EW}}{\cw^2}
\left(\frac{\CA}{\Lambda^2}\right)^2s^2\,\sin^2\theta+\mathpzc{O}(s).
\end{equation}
Now the partial wave unitarity bounds take a more familiar form,
\begin{equation} \label{eq:CBan_pert_unit}
\sqrt{s}\gtrsim 18.9\,\frac{\Lambda}{\sqrt{\CA}},
\end{equation}
meaning only implausibly non-perturbative values of $\CA\gtrsim 360$ imply a violation of unitarity below the cutoff. The scattering process has therefore been partly `unitarised' by using the appropriate EFT description.

In summary, since the $\chi\chi\to W^+W^-$ scattering process has a relevant energy around or above the EW scale, the appropriate low energy symmetry of an EFT approach to new physics effects is $SU(2)_L\times U(1)_Y$. The photon-only anapole contribution to this process manifestly does not respect this symmetry, being only $U(1)_{\text{QED}}$ invariant. This violation of gauge invariance leads to two additional powers of `anomalous' energy growth, beyond the expectations dictated by dimensional analysis. This type of behaviour is common to all four form factor operator contributions to this scattering and is always cured by the description in terms of the corresponding hypercharge form factor. The charge radius operator gives identical predictions in this channel, with the same consequences for unitarity violation and ensuing bounds on $\Ccr^\gamma$ as Equations~\eqref{eq:CAn_pert_unit} and~\eqref{eq:CAn_unit_viol} that are relaxed for $\Ccr$ as in Equation~\eqref{eq:CBan_pert_unit}. Similar conclusions can be reached about the dimension-5 operators. Analogous effects due to non gauge-invariant descriptions of dark matter contact interactions with quarks and their unitarity-violating effects in mono-$W$ production have been pointed out in Ref.~\cite{Bell:2015sza}.

The implications of the differing treatments are discussed in the context of collider phenomenology in Section~\ref{sec:pheno_colliders} and the thermal relic abundance calculation in Section~\ref{subsec:production}. In the former, we will see that the above computation leads to the VBF production cross-section of $\chi\chi$ at the LHC being overestimated by several orders of magnitude while in the latter, the relic abundance predictions when $\MX \gtrsim \MW$ are drastically modified.
\subsubsection{Dark matter coupling to the Z\label{sec:DMcoupZ}}
The second aspect of the photon vs. hypercharge form factors is related to the fact that, as mentioned in the introduction, the hypercharge operators necessarily induce an additional dark matter coupling to the $Z$ boson. Since these can be viewed as a `completion' of the electromagnetic ones, for all the interesting phenomenological aspects of electromagnetic form factors, one generically expects a $Z$ boson form factor coupling of a similar magnitude. This entails a host of other experimental and theoretical probes of such dark matter models. The first major consequence is the presence of a $Z$-funnel region in the thermal relic density as a function of $\MX$, peaking at $\MX\sim45$ GeV, that alters the relationship between direct detection constraints and favoured regions for producing the correct relic density. The second, related consequence is that for $\MX<45$ GeV, the model can now be constrained by invisible $Z$ decays. Indirect LEP constraints on an additional invisible partial width, $\Gamma_{\text{inv.}}$, place a strong bound of $\Gamma_{\text{inv.}}< 2$ MeV which can potentially have a significant impact on the viable parameter space. The $Z$ boson partial decay widths into the dark matter candidate mediated by hypercharge form factors are:
\begin{eqnarray}
      \Gamma^{Z}_{\sss\mathcal{A}} & = & \frac{\CA^{2} \sw^{2} m_{Z}^{2}\left(m_{Z}^{2}-4 m_{\chi}^{2}\right)^{3 / 2}}{24 \pi \Lambda^{4}}\,, \nonumber \\
      \Gamma^{Z}_{cr} & = & \frac{\Ccr^{2} \sw^{2} m_{Z}^{2} \sqrt{m_{Z}^{2}-4 m_{\psi}^{2}}\left(m_{Z}^{2}+2 m_{\psi}^{2}\right)}{12 \pi \Lambda^{4}}\,, \nonumber \\
      \Gamma^{Z}_{el} & =  & \frac{\CEle^{2} \sw ^{2}\left(m_{Z}^{2}-4 m_{\psi}^{2}\right)^{3 / 2}}{24 \pi \Lambda^{2}}\,, \nonumber \\
      \Gamma^{Z}_{\sss\mathcal{M}} & = & \frac{\CMag^{2} \sw^{2} \sqrt{m_{Z}^{2}-4 m_{\psi}^{2}}\left(m_{Z}^{2}+8 m_{\psi}^{2}\right)}{24 \pi \Lambda^{2}}\,.
      \label{eq:Zinv}
\end{eqnarray}
This constraint from LEP corresponds to $\Lambda/C_{\mathcal{A}}^{1/2}\gtrsim 315$ GeV and $\Lambda/C_{cr}^{1/2}\gtrsim 370$ GeV for the anapole and charge radius respectively. For both dimension-5 interactions we have $\Lambda/C_{5} \gtrsim 1 $ TeV. These are shown along with the other constraints considered in this work in section~\ref{sec:concl}.

From the consideration of the scattering of dark matter with $W$ boson, one could have argued that the phenomenology of dark matter models was safely described by electromagnetic form factors for $\MX<\MW$. However, the correlation between the photon and $Z$ coupling predicted by the hypercharge form factor means that their relic density predictions will differ from those of the electromagnetic form factors and that additional $Z$-decay constraints will apply to the model for all masses below 45 GeV~\footnote{We note that the strict correlation between the $Z$ and photon form factors apply to EW-singlet dark matter. Dark matter candidates that are part of an $SU(2)$ multiplet can also have form factor interactions with the $W^{\pm}$ fields, which can de-correlate the two interactions. Naturally, if there is relatively light new physics around the EW scale but above $\MX$, the EFT description is not appropriate and these arguments do not exactly apply. It is nevertheless likely that realistic scenarios would predict the presence of both types of form factors.}.
To conclude this section, we have shown that hypercharge form factors are the favourable description for these type of dark matter effective interactions and that there does not appear to be any range of dark matter masses in which the electromagnetic form factor gives an adequate picture. They provide a consistent framework for calculating the dark matter production, annihilation and scattering processes relevant for theoretically an experimentally testing this scenario.

\section{Collider searches}

\begin{table}[t]
\caption{Summary of the kinematic selection criteria imposed in our VBF analysis.}
\label{tab:VBF_cuts}
\begin{center}
\begin{tabular}{|c|l||c|l|}
\hline
Variable & Cut & Variable & Cut \\
\hline
$\vert\eta(j)\vert$ & >\,3.0 &  $\vert\Delta\eta\vert$ & >\,7.0 \tabularnewline
$p_T(j)$ & >\,30.0 GeV & $E^{miss}_T$ & >\,175.0 GeV\tabularnewline
$N(j)$ & $\geq$\,2  &  $m_{jj}$ & >\,500.0 GeV\tabularnewline
\hline
\end{tabular}
\end{center}
\end{table}

\label{sec:pheno_colliders}
In this section we revisit the potential for the LHC to probe the parameter space of dark matter with electromagnetic form factors.
Reversing the dark matter annihilation diagrams of Figure~\ref{fig:annihilation_diags} suggests a number of potential production modes at hadron colliders including the traditional `mono-X' searches via the $q\bar{q}$ initial state and vector boson fusion mediated by the $W^+W^-$-initiated sub-amplitude. Several of these have been studied as a probe of the anapole dark matter coupling, $\CA^{\gamma}$~\cite{Gao:2013vfa,Alves:2017uls,Florez:2019tqr}. As discussed in Section~\ref{sec:DMWW}, although the VBF channel has been shown to be particularly sensitive to the electromagnetic anapole form factor, the consistency of the $W^+W^-\to\chi\chi$ amplitude requires a reformulation of the effective description in terms of the hypercharge form factor, particularly for dark matter masses above $\MW$.

We therefore begin with a reappraisal of this channel, quantifying the difference in its sensitivity to the photon and hypercharge form factors. The partial `unitarisation' of the $W^+W^-\to\chi\chi$ sub amplitude that occurs when going from the photon to the hypercharge form factor results in a drastic loss of sensitivity, to the point where mono-jet searches become the more stringent probes. We then go on to interpret the latest CMS mono-jet search~\cite{Sirunyan:2017jix}, presenting the strongest known limits from colliders on electroweak form factors for dark matter as well as projections for the high-luminosity LHC. Our analyses are performed with parton-level Monte Carlo simulations generated with \verb|Madgraph5_aMC@NLO|~\cite{mg5}, for proton-proton collisions at a centre of mass energy of $\sqrt{s}=13$ TeV, using our custom UFO model. \verb|MadAnalysis5|~\cite{ma5} was used to analyse some of the event samples.

\subsection{Electromagnetic form factors for dark matter in vector boson fusion}
\label{subsec:pheno_vbf}
\begin{figure}[t]
\begin{center}
    \includegraphics[width=0.4\columnwidth]{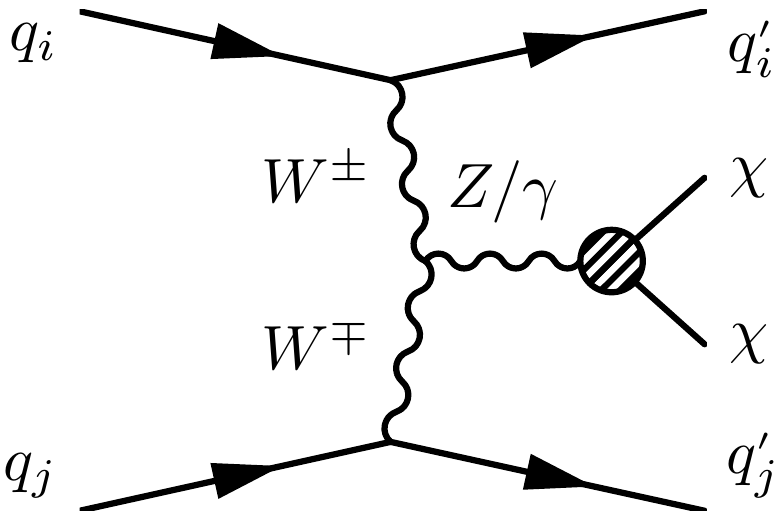}
    \hspace{1cm}
    \includegraphics[width=0.4\columnwidth]{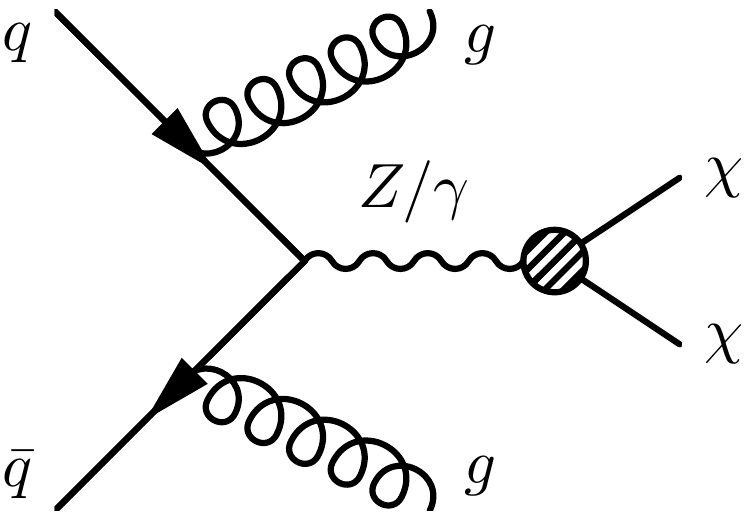}
\end{center}
\caption{\label{fig:VBF_diagrams}
\textbf{Left:} Example Feynman diagrams for pure-EW contributions to $p\,p\to\chi\chi\,j\,j$ mediated by EW form factors for dark matter. \textbf{Right:} Same as left for mixed QCD/EW contributions.
}
\end{figure}
The VBF production mode exhibits a striking signature of missing energy accompanied by a high invariant mass pair of forward jets, well-separated in rapidity. It is particularly well-known for being the most sensitive, direct way to search for invisible Higgs decay modes and has also been used to constrain dark matter production via the Higgs-portal interaction through both on- and off-shell probes~\cite{Craig:2014lda,Endo:2014cca,Han:2016gyy,Ruhdorfer:2019utl,Heisig:2019vcj}. More broadly, it offers a unique way to probe the interactions of a light dark sector with the EW gauge bosons via $W^+W^-\to\chi\chi$ scattering. In Section~\ref{sec:DMWW}, our study  of the different behaviour of this amplitude between the hypercharge and electromagnetic form factors suggests that the two different parametrisations will lead to very different phenomenology in this channel. The latter exhibits huge energy growth, beyond the expectations for a dimension-6 operator, signalling the breakdown of unitarity due to gauge symmetry violation at energies above the $W$-mass. In this section, we quantify the impact of this change in parametrisation on the prospects for constraining EW form factors for dark matter in VBF.

The starting point for our analysis is provided by~\cite{Florez:2019tqr}, in which the very promising prospects for VBF to constrain $\CA^{\gamma}$ were first identified. Their phenomenological analysis of signal and background distributions identified some efficient selection criteria to single out the phase space region in which the VBF signal dominates. These are summarised in Table~\ref{tab:VBF_cuts} and amount to the familiar requirements of exactly two jets with a large rapidity separation, $\vert\Delta\eta\vert$, as well as a large invariant mass $m_{jj}$ and a significant missing energy requirement. Our main goal is to quantify the difference between the limits obtained for the electromagnetic and hypercharge versions of each operator. We therefore reproduce a simple version of the kinematical analysis, taking into account the dominant source of SM background, namely $Z$ + jets with the $Z$ boson decaying into neutrinos.

The signal process, $p\,p\to \chi\chi jj$, has two contributions at tree-level of different coupling order, shown in Figure~\ref{fig:VBF_diagrams}. The first is the pure-EW contribution, arising at $\mathcal{O}(\alpha_{EW}^3)$, which includes the `true' VBF topology, and is the target of this analysis. 
The second, arising at $\mathcal{O}(\alpha_{S}^2\alpha_{EW})$ describes two QCD emissions from the underlying Drell-Yan-like production of the $\chi\chi$ final state.
Before applying the VBF selection, the total cross-section is much larger for the latter process. Once the tight selection requirements of Table~\ref{tab:VBF_cuts} are imposed, the signal rate is dominated by the VBF topology in the electromagnetic anapole case. However, this is no longer true for the hypercharge case.

\begin{figure*}[t]
    \centering
    \includegraphics[width=0.7\columnwidth]{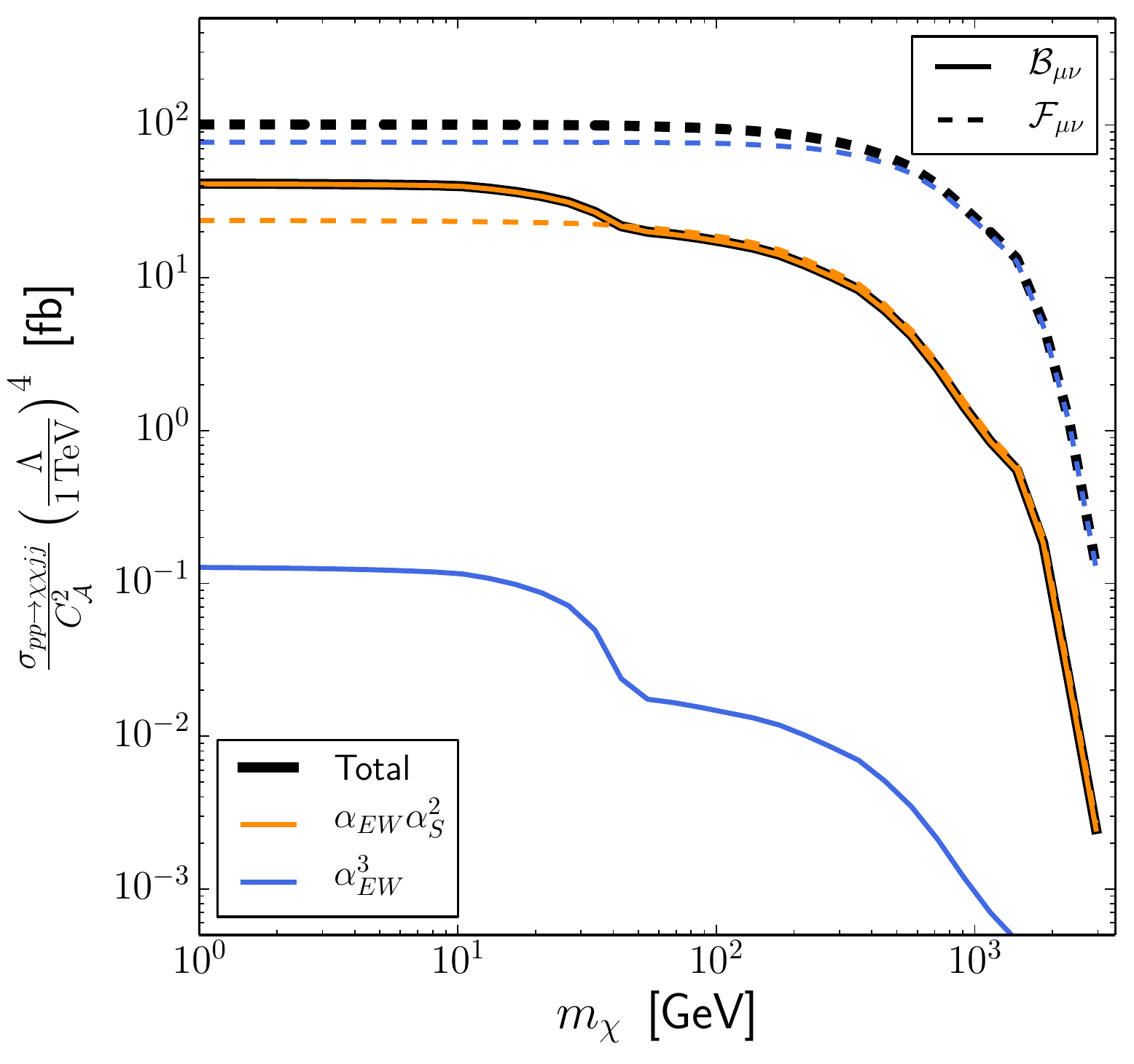}
    \includegraphics[width=0.7\columnwidth]{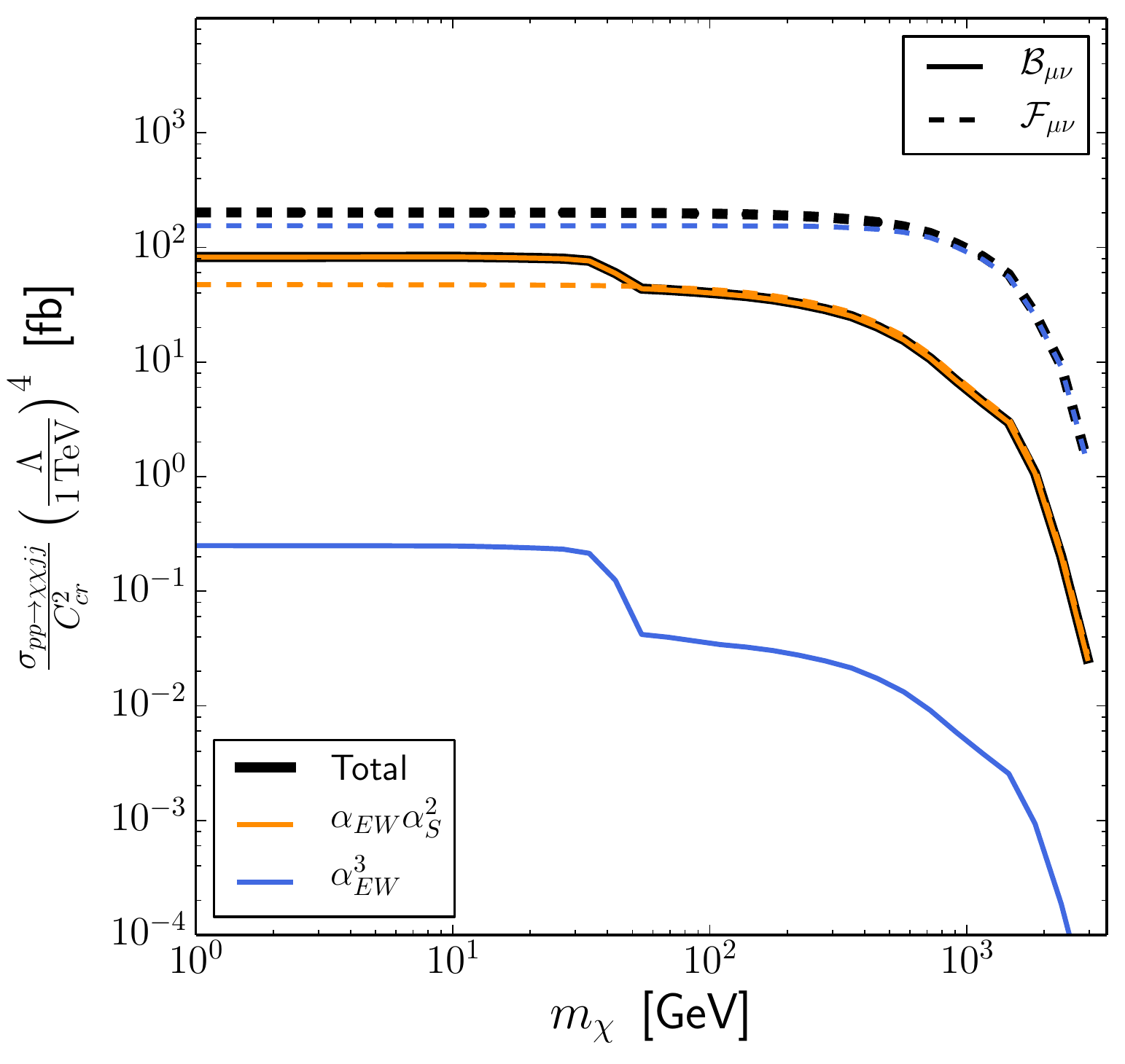}
    \includegraphics[width=0.7\columnwidth]{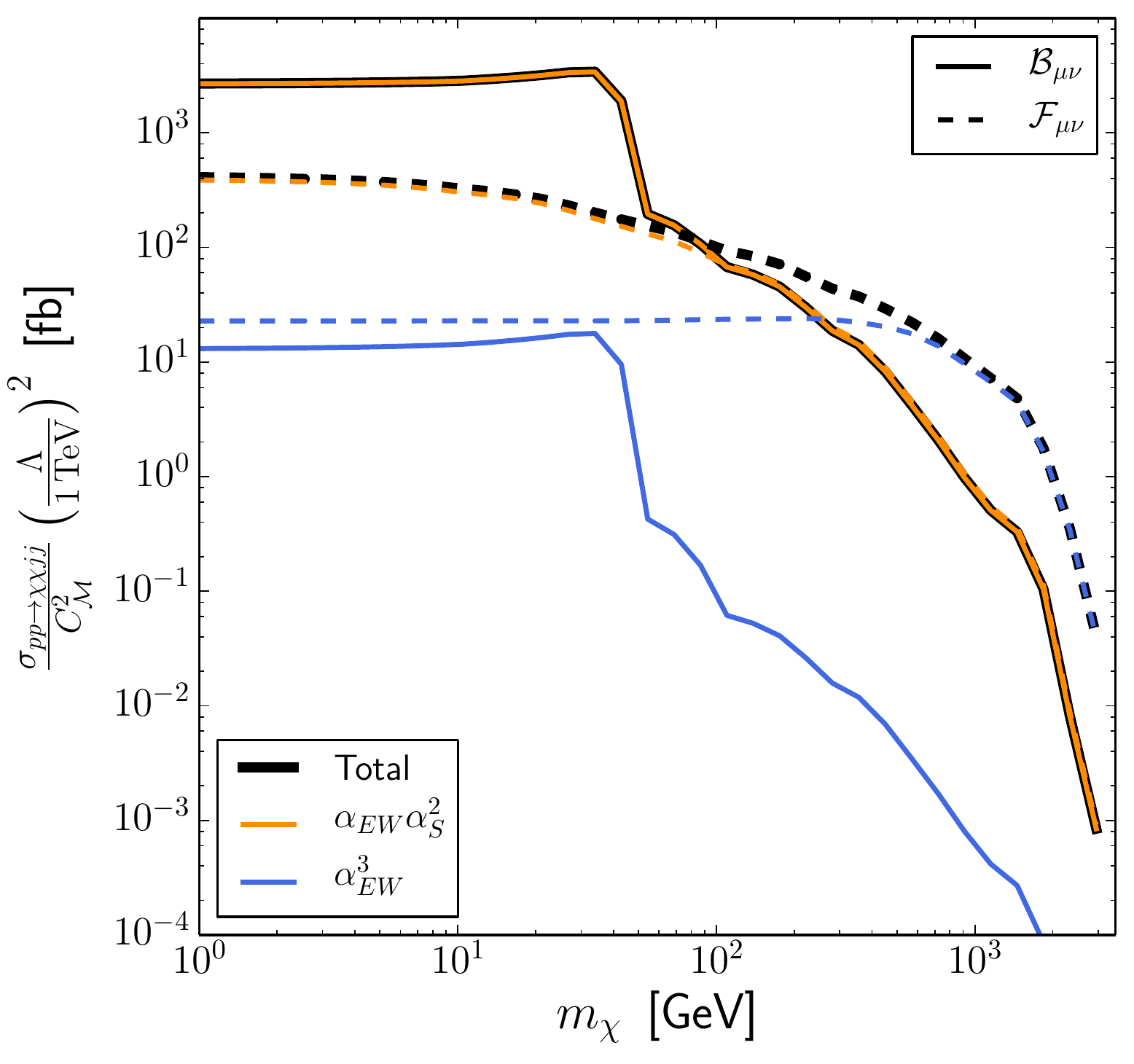}
    \includegraphics[width=0.7\columnwidth]{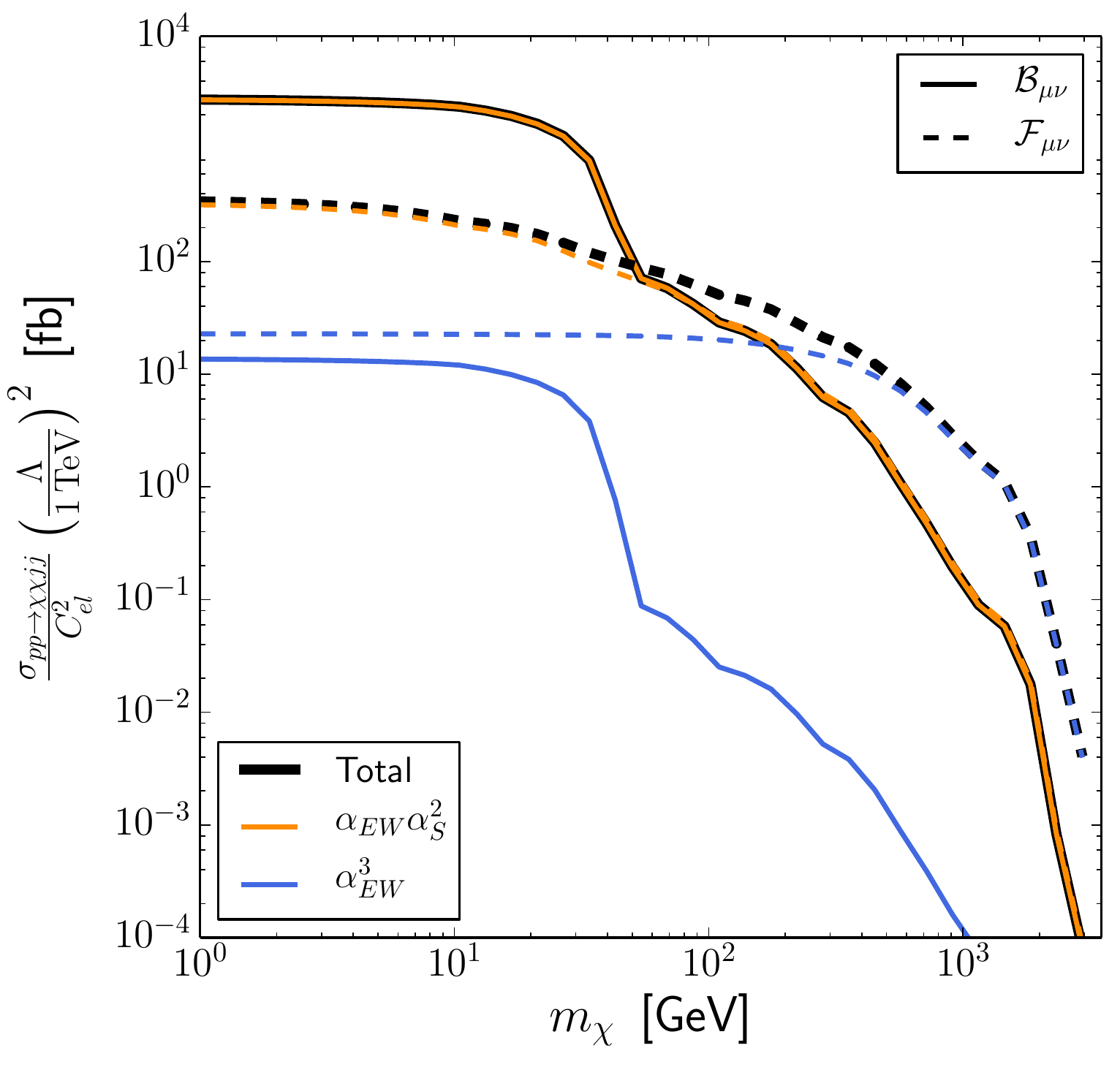}
    \caption{\textbf{Top left:} $pp\to\chi\chi j j$ cross-sections predicted by the hypercharge (solid line) and electromagnetic versions (dashed line) of the anapole moment, $\OA$, for the mixed QCD/EW contribution (orange), pure-EW contribution (blue) and total process (black). The rates are obtained for $C_i=1$ and $\Lambda=1$ TeV. \textbf{Top right and bottom left, right:} Same as top left for the charge radius operator, $\Ocr$, the magnetic dipole operator, $\OMag$ and the electric dipole operator $\OEle$ respectively. }
    \label{fig:ratio_dim5_6}
\end{figure*}

The two upper panels of Figure~\ref{fig:ratio_dim5_6} depict cross-sections of hypercharge and electromagnetic dimension-6 operators, for the two separate contributions and their sum as a function of dark matter mass. The cross-sections are estimated before the VBF cuts with only a di-jet invariant mass requirement of 100 GeV, to avoid on-shell vector bosons contributing two jets through their hadronic decays. We see that in the electromagnetic case the VBF contribution and the QCD-emission topology have roughly the same magnitude. Instead, in the hypercharge case the mixed QCD/EW contribution completely dominates the inclusive result. Furthermore, in all hypercharge cases, the cross-section exhibits a feature at 45 GeV, coming from the newly-present on-shell $Z\to\chi\chi$ contribution. The plots, especially in the hypercharge case, show that the mixed QCD/EW contribution to the signal can no longer justifiably be neglected in the determination of the sensitivity even though it comes from a different type of process than the one originally being targeted.  

The lower panels of Figure~\ref{fig:ratio_dim5_6} show the same cross-sections for the dimension-5 operators; here the purely VBF contribution is also suppressed for the electromagnetic form factors, and the two contributions only have comparable magnitudes in the limit of large dark matter masses.

After the stringent VBF selection, we find that for low values of dark matter masses, the signal rate is roughly equally divided between the pure-EW and mixed QCD/EW pieces. With increasing dark matter mass, the latter comes to completely dominate the signal cross-section.
\begin{figure*}[t]
    \centering
    \includegraphics[width=0.7\columnwidth]{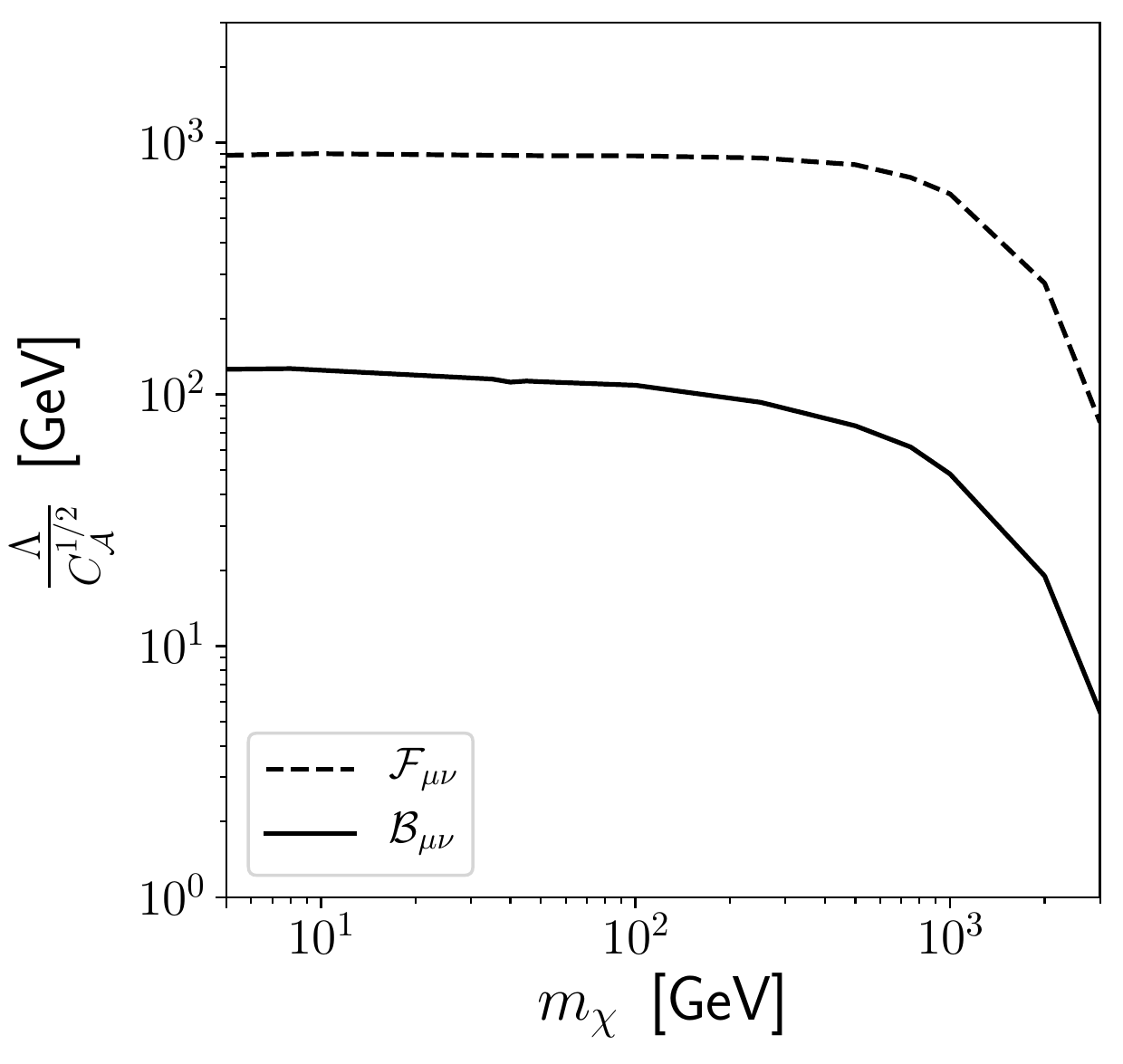} 
    \includegraphics[width=0.7\columnwidth]{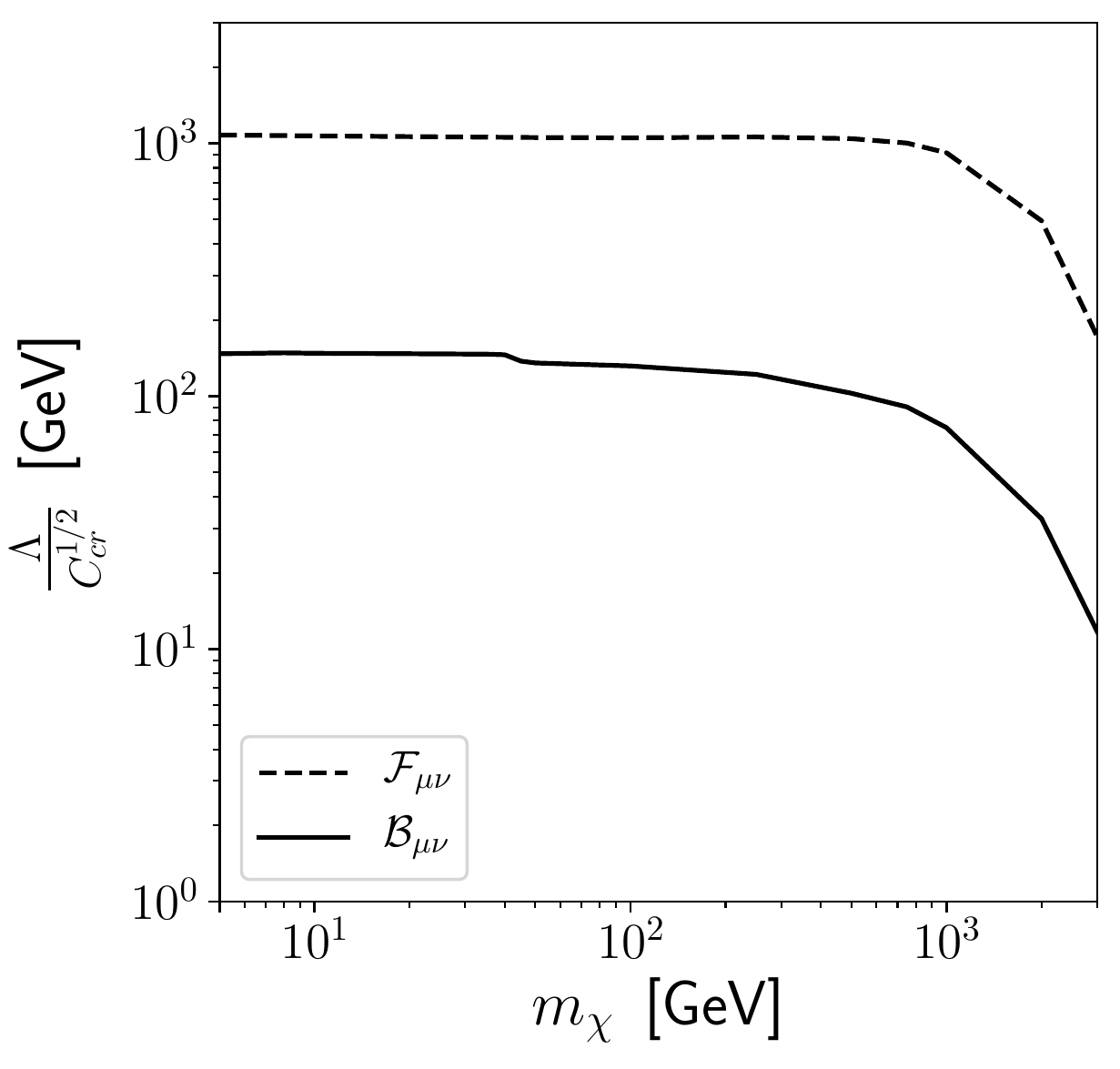}
    \includegraphics[width=0.7\columnwidth]{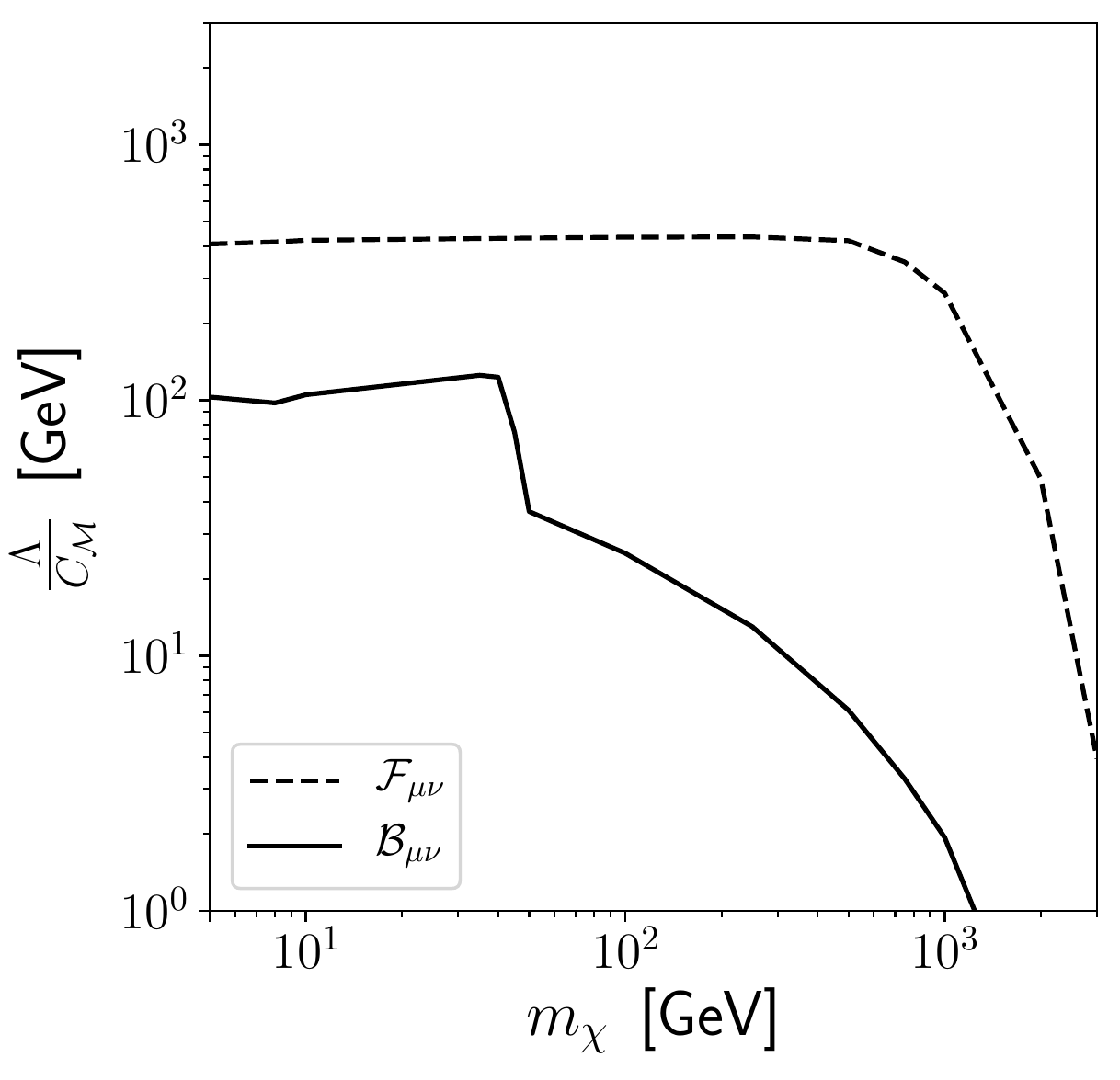} 
    \includegraphics[width=0.7\columnwidth]{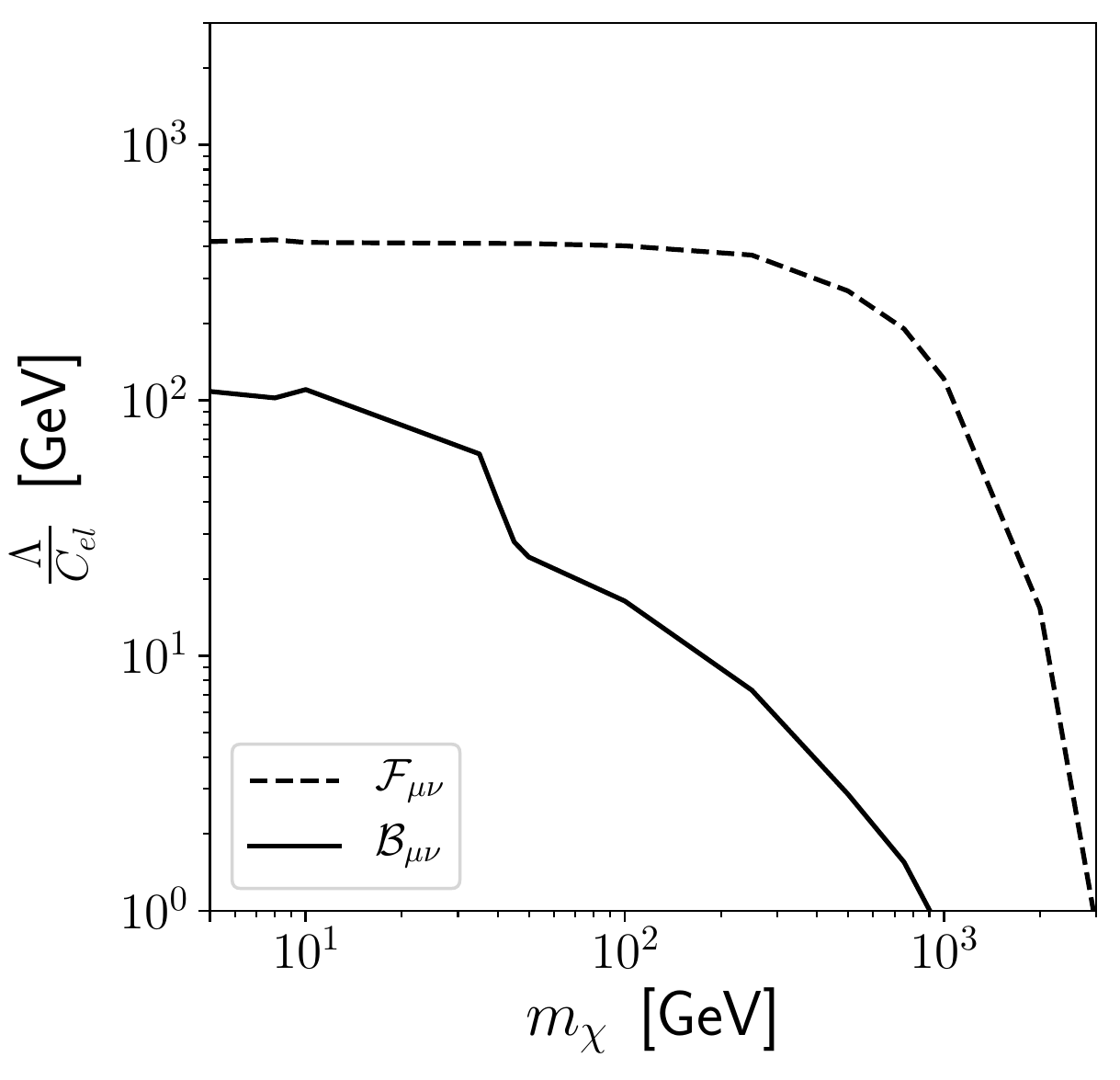}
    \caption{\textbf{Top left:} Vector boson fusion constraints for the electromagnetic (black dashed) and hypercharge (black solid) anapole moment. \textbf{Top right, bottom left and right:} Same as top left for the charge radius operator, and the magnetic and electric dipoles respectively.}
    \label{fig:VBF_limits}
\end{figure*}

In order to estimate the sensitivity, we perform a cut-and-count analysis assuming an integrated luminosity of 3000 fb$^{-1}$ collected at the LHC. We take the following definition of signal significance
\begin{equation*}
z=\frac{s}{\sqrt{s+b+(b/4)^2}},
\end{equation*}
where $s$ and $b$ denote the number of signal and background events in the signal region, respectively. The measure includes a 25\% relative systematic uncertainty on the background expectation, motivated in~\cite{Florez:2019tqr} as being typical for LHC VBF searches. The critical value, $z=2$ is used to determine our 95\% Confidence Level (C.L.) exclusion limit on the scale $\Lambda$ of a given operator, divided by the appropriate power of the Wilson coefficient.

The limits depicted in Figure~\ref{fig:VBF_limits}, quantify the drastic loss in sensitivity for the hypercharge form factors~\footnote{We note that one major difference between our analysis and that of~\cite{Florez:2019tqr} is the fact that we do not perform a binned-likelihood fit of the di-jet invariant mass distribution after the initial VBF selection cuts. While this is expected to somewhat improve the overall sensitivity, our main concern here is the difference between the electromagnetic and hypercharge form factors, as well as the fact that, ultimately, the loss in sensitivity negates the justification for considering this channel. }.
The TeV scale constraints for the dimension-6 operators are reduced by an order of magnitude. In the dimension-5 case, the loss is about a factor of 3 below $\MZ/2$ and again an order of magnitude above. We find that the drop in cross-section for the pure EW contribution is compounded by a loss in efficiency of the extreme VBF selection employed in this analysis, which was optimised for the electromagnetic, dimension-6 form factors, leading to a further worsening of prospects for this particular set of cuts. The obtained sensitivity to the hypercharge operators is therefore not likely to be optimal. Nevertheless, given the quartic(quadratic) dependence of the dimension-6(5) signal cross-section on the cutoff scale, $\Lambda$, it is extremely unlikely that an optimisation of signal to background would gain the orders of magnitude needed to recover comparable sensitivity. Finally, the relatively weak constraints also mean that the validity of the effective description is more likely to break down, considering the typical energy scales involved in LHC VBF processes. We therefore do not pursue this option, rather considering alternative options for the collider constraints on these scenarios.

An important observation is that, in contrast to the pure-EW contribution to $\chi\,\chi+2\,j$, the mixed QCD/EW cross-section is largely unaffected by the switch from electromagnetic to hypercharge form-factors. This is because the underlying new-physics process here is $q\bar{q}\to\chi\chi$, as opposed to $W^+W^-\to\chi\chi$ in the EW process (see Figure~\ref{fig:VBF_diagrams}).
In the dimension-6 case, the two operators in question can be removed by the photon/hypercharge gauge field equations of motion (or appropriate field redefinitions)
\begin{align}\label{eq:FandB_EOM}
    \partial^{\nu}F_{\mu\nu}=eJ_{\mu}^{\textrm{EM}},\quad
    \partial^{\rho} B_{\rho \mu}=eJ_{\mu}^{Y},
\end{align}
where $J_{\mu}^{\textrm{EM}}$ and $J_{\mu}^{Y}$ denote the electromagnetic and hypercharge currents, respectively. These consist mostly of fermion bilinears, meaning that both types of operators have a component that can be described by a linear combination of $q\bar{q}\chi\chi$ contact interactions. These four-fermion operators induce the mixed QCD/EW process, such that the hypercharge and electromagnetic operators are expected to have similar predictions up to $\mathcal{O}(1)$ factors of ratios of linear combinations of gauge charges and quark-antiquark parton luminosities.

The dimension-5 operators cannot be eliminated by equations of motion and can therefore only really be understood as momentum dependent interactions between the neutral gauge fields and a pair of dark matter particles. Once again, we do not expect a radical difference when switching to the hypercharge form-factor in this case apart from the observed additional on-shell $Z$ component of the cross-section in $q\bar{q}\to\chi\chi$ scattering. Here, it is the fact that hypercharge form factors partly unitarise the $W^+W^-\to\chi\chi$ scattering that this channel becomes so suppressed. In fact the VBF channel is not even particularly effective at constraining the dimension-5 electromagnetic form factors, let alone their hypercharge counterparts.

The clear dominance of the $q\bar{q}\to\chi\chi$ scattering for hypercharge form-factors lead us to the conclusion that this set of interactions is most likely to be better constrained by processes that explicitly target this amplitude, the most obvious of which is the well-known mono-$X$ channel.

\subsection{The latest on the mono-jet signature}
\label{subsec:pheno_monojet}
Without the explosive repercussions of gauge-violation, the most rudimentary collider searches are likely to be the most promising. Naturally, the mono-jet searches have been studied in the context of the effective operators before, see {\it e.g.} for the anapole interaction~\cite{Gao:2013vfa} and for more generic (dark matter) EFT studies~\cite{Bai:2010hh,Goodman:2010yf,Goodman:2010ku,Shoemaker:2011vi,Fox:2011pm,Fox:2011fx,Rajaraman:2011wf,Zhou:2013fla,Endo:2014mja,Mimasu:2014nea,Belyaev:2016pxe,Kahlhoefer:2017dnp,Belyaev:2018pqr}. Here we provide an update of these bounds using the results of the most recent CMS analysis~\cite{Sirunyan:2017jix} with a luminosity of $35.9$ fb$^{-1}$ as well as a projection into the high-luminosity LHC phase. As discussed in the previous section, unlike with VBF processes, the growth of off-shell $\gamma/Z$ plays no role in either the electromagnetic or EW form factors, leading to similar limits for both operator types. However, we hope by now that the reader in convinced that the $B$-field interactions are the only physically meaningful ones, and we therefore present just these results.

The selection requirements of the CMS analysis on the single jet are, $p^{\rm{T}}_{\rm{jet}} > 250$ GeV and $|\eta| < 2.5 $. The missing transverse energy distribution, $p_T^{\textrm{miss}}$ (equal to the jet $p_T$ at leading-order), is then used to constrain the production of invisible particles produced in association with a single jet or a boosted, hadronic vector boson. In order to calculate the current and future limits we make use of a binned $\chi^2$ statistic,
\begin{align}
    \chi^2 \equiv
( \vec{n}_{\text{exp}} + \kappa \,\vec{n}_{\text{sig}} - \vec{n}_{\text{obs}})
\cdot \mathbf{V}^{-1}\cdot
( \vec{n}_{\text{exp}} + \kappa \,\vec{n}_{\text{sig}} - \vec{n}_{\text{obs}} ),
\end{align}
comparing the data ($\vec{n}_{\text{obs}}$) with the expected SM background distribution reported by the analysis ($\vec{n}_{\text{exp}}$), incorporating the predicted shape of the dark matter signal contribution
 ($\vec{n}_{\text{sig}}$). The new physics interaction is denoted by $\kappa=(c/\Lambda)^2$ or $\kappa=(c/\Lambda^2)^2$ for dimension-5 and dimension-6 interactions respectively and the covariance matrix for the data, $\mathbf{V}$, contains the reported statistical and systematic uncertainties for the $p_T^{miss}$ distribution and their correlations. The shape of the signal distribution depends only on the dark matter mass, $m_\chi$, as the coupling strength, $\kappa$, can be factorised from the process. Since our observables depend linearly on the parameter of interest, the $\Delta\chi^2$ can be written as
\begin{equation}
    \Delta\chi^2(\kappa)= (\kappa -\kappa_{\rm{min}})\cdot \mathbf{F}\cdot(\kappa -\kappa_{\rm{min}}),
\end{equation}
where $\kappa_{\rm{min}}$ is the value of $\kappa$ that minimizes the $\chi^2$ and $\mathbf{F}$ is the Fisher information matrix, that quantifies the shape of the likelihood in $\kappa$ arounds its maximum for a given $m_\chi$. It depends on the normalised signal for each dark matter mass, and the covariance matrix, $\mathbf{V}$. We use this form to derive upper limits on $\kappa$ using the critical value of $\Delta\chi^2=3.84$.
\begin{figure*}[t]
    \centering
    \includegraphics[width=0.7\columnwidth]{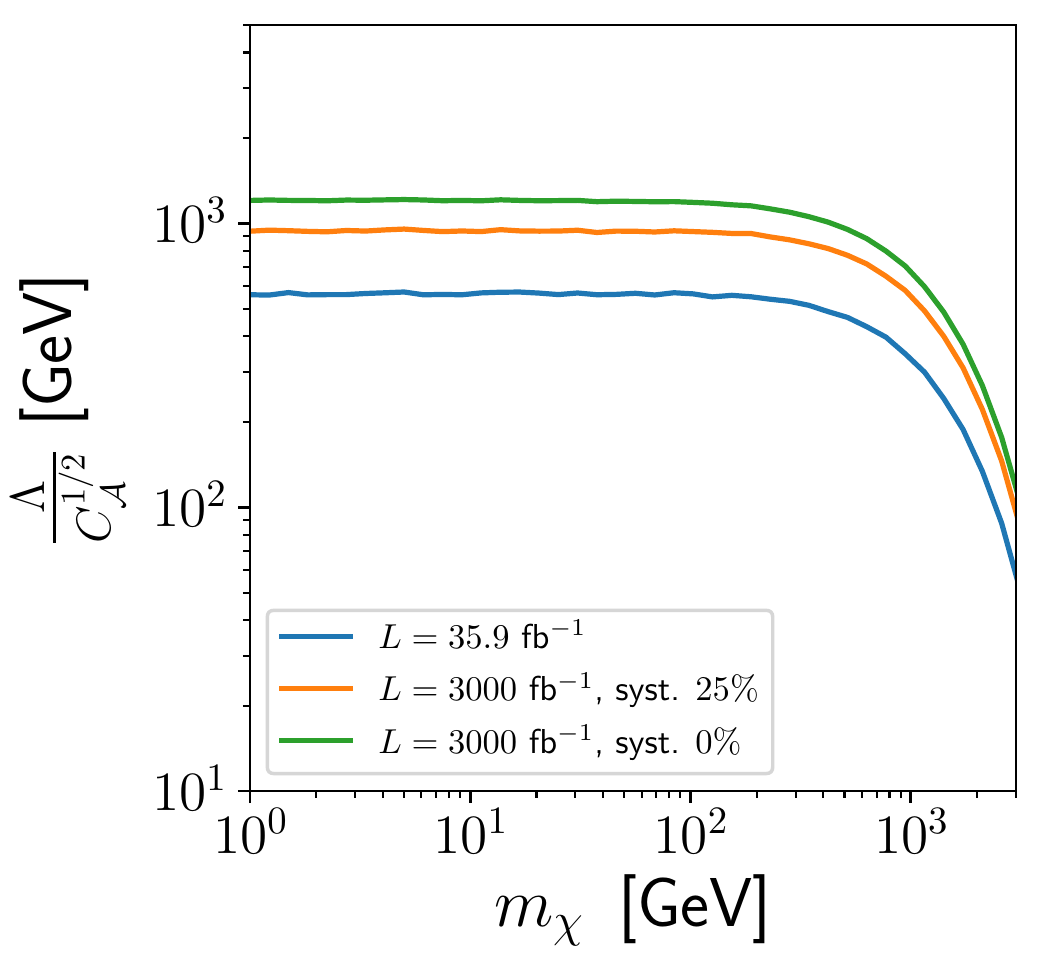}
    \includegraphics[width=0.7\columnwidth]{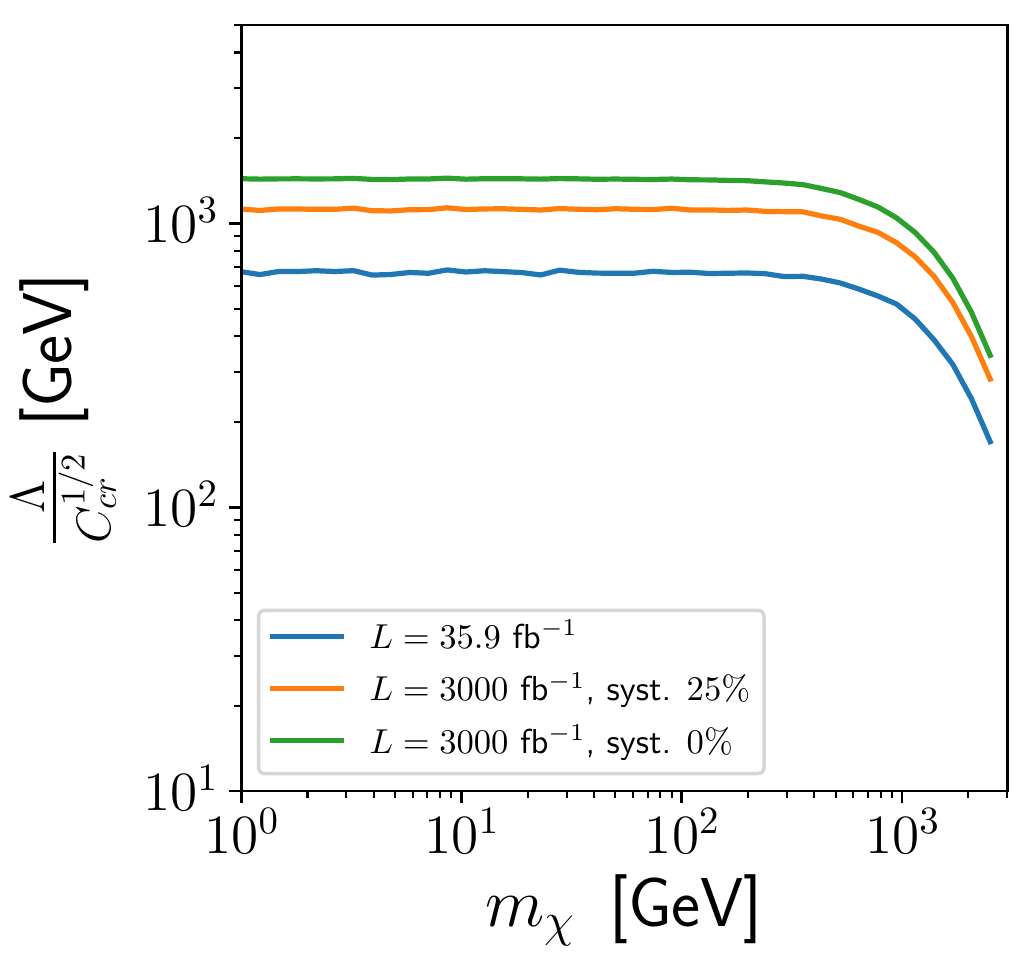}
    \includegraphics[width=0.7\columnwidth]{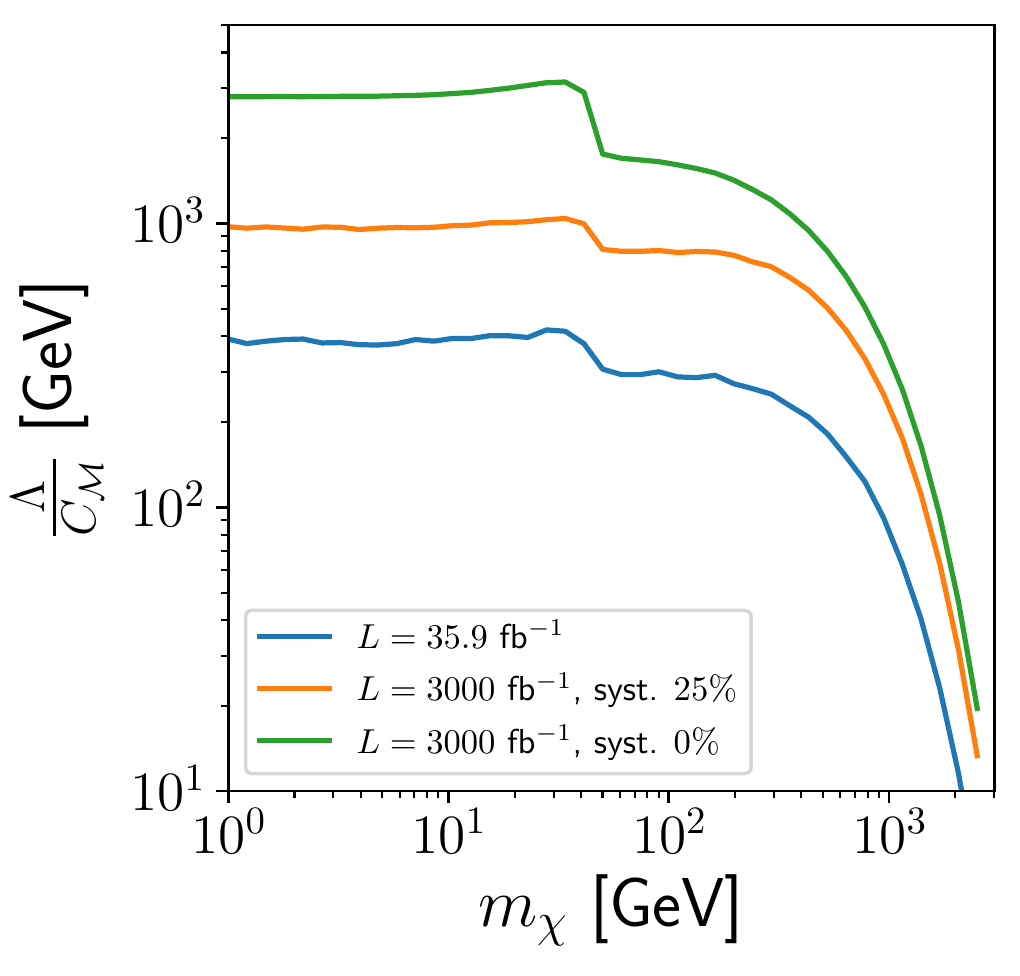}
    \includegraphics[width=0.7\columnwidth]{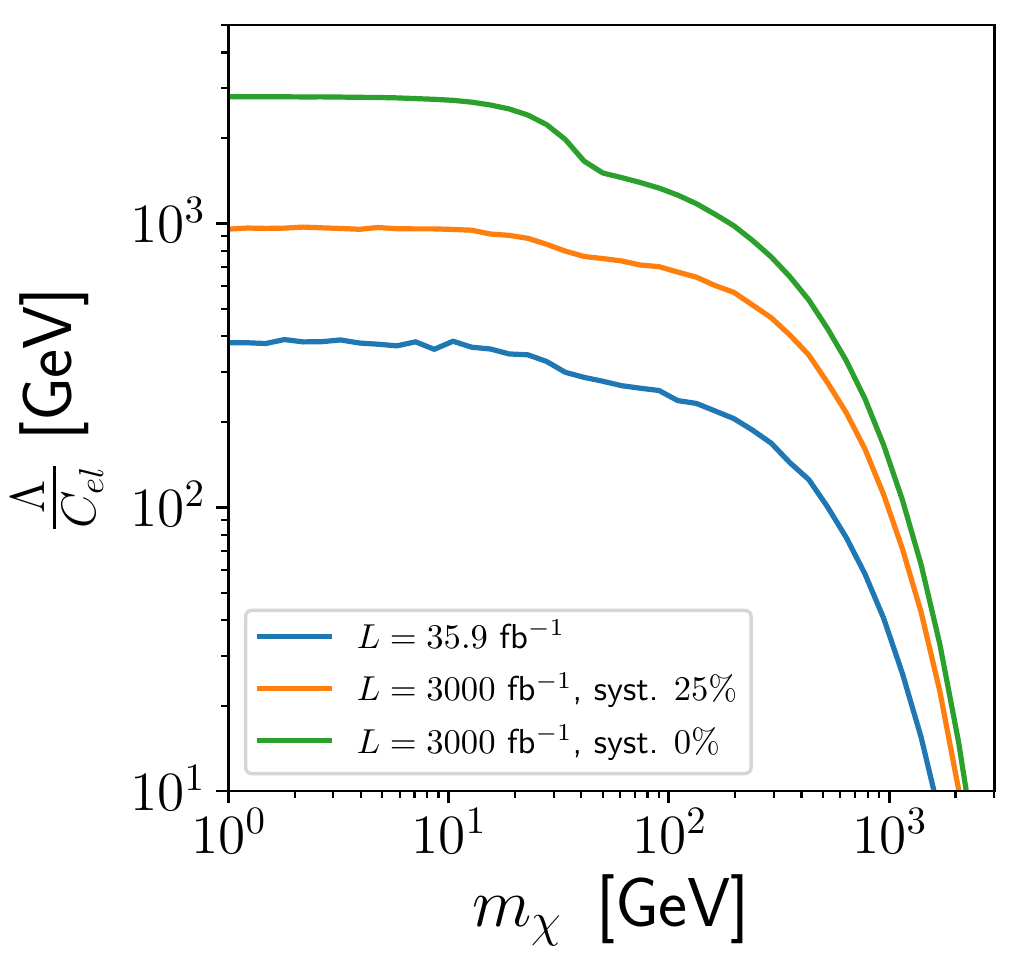}
    \caption{\textbf{Top left:} Current mono-jet LHC constraint (blue) and projected high luminosity LHC reach with $25\%$ (orange) and $0\%$ (green) systematic uncertainty, for the anapole moment. \textbf{Top right, bottom left and right:} Same as top left for the charge radius operator, and magnetic and electric dipoles respectively.}
    \label{fig:monoJ_lims}
\end{figure*}

\begin{figure*}[t]
    \centering
    \includegraphics[width=0.7\columnwidth]{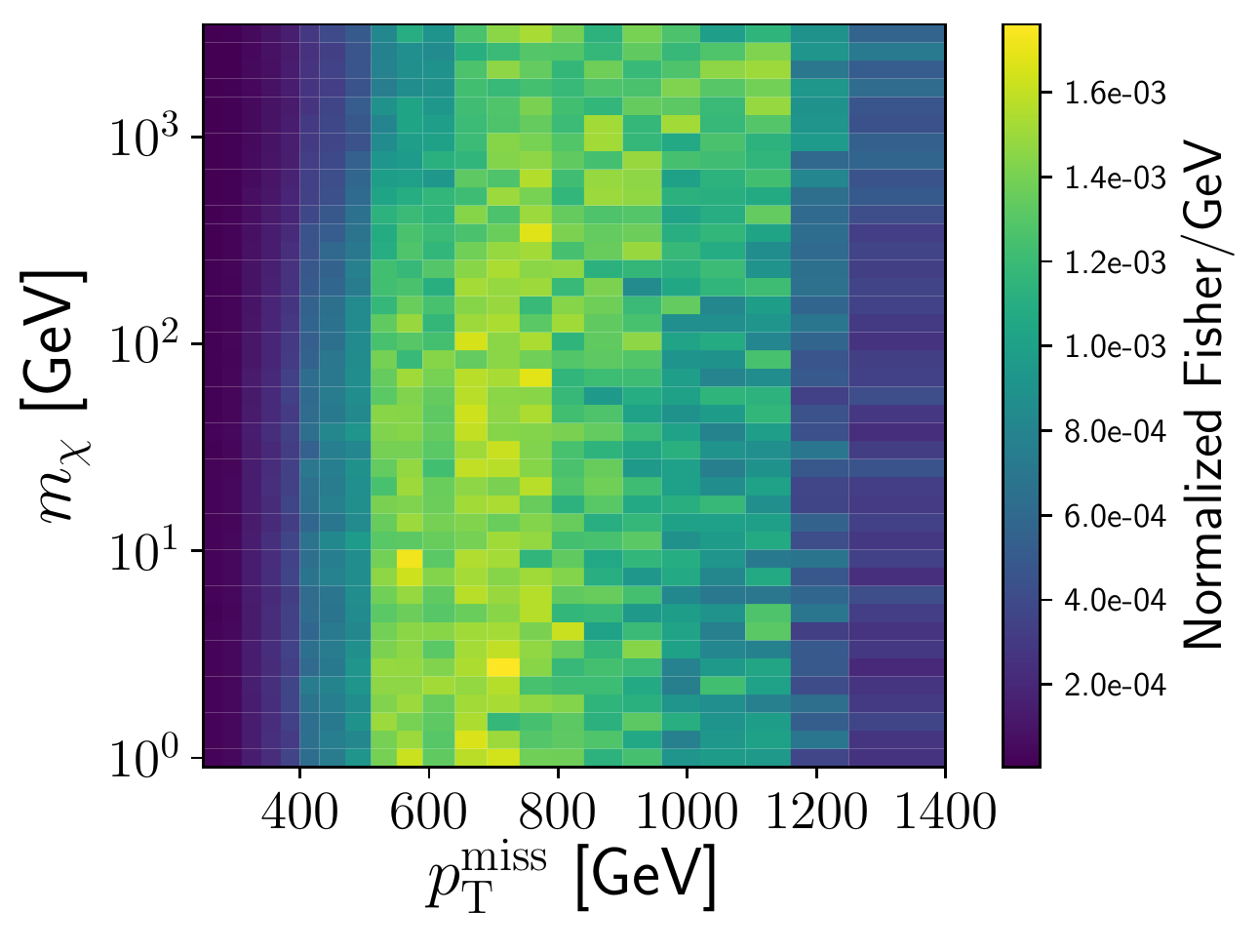}
    \includegraphics[width=0.7\columnwidth]{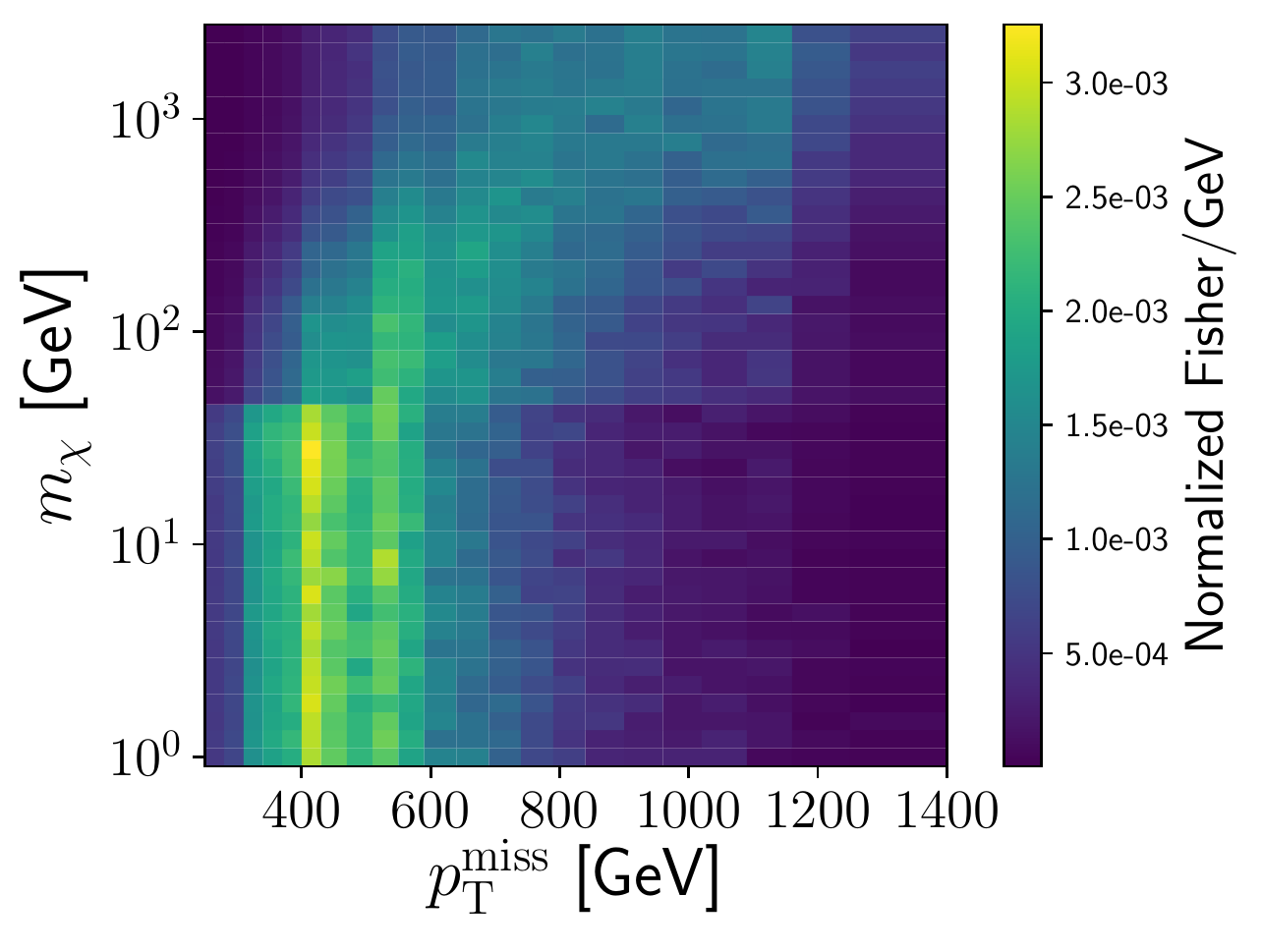}
    \includegraphics[width=0.75\columnwidth]{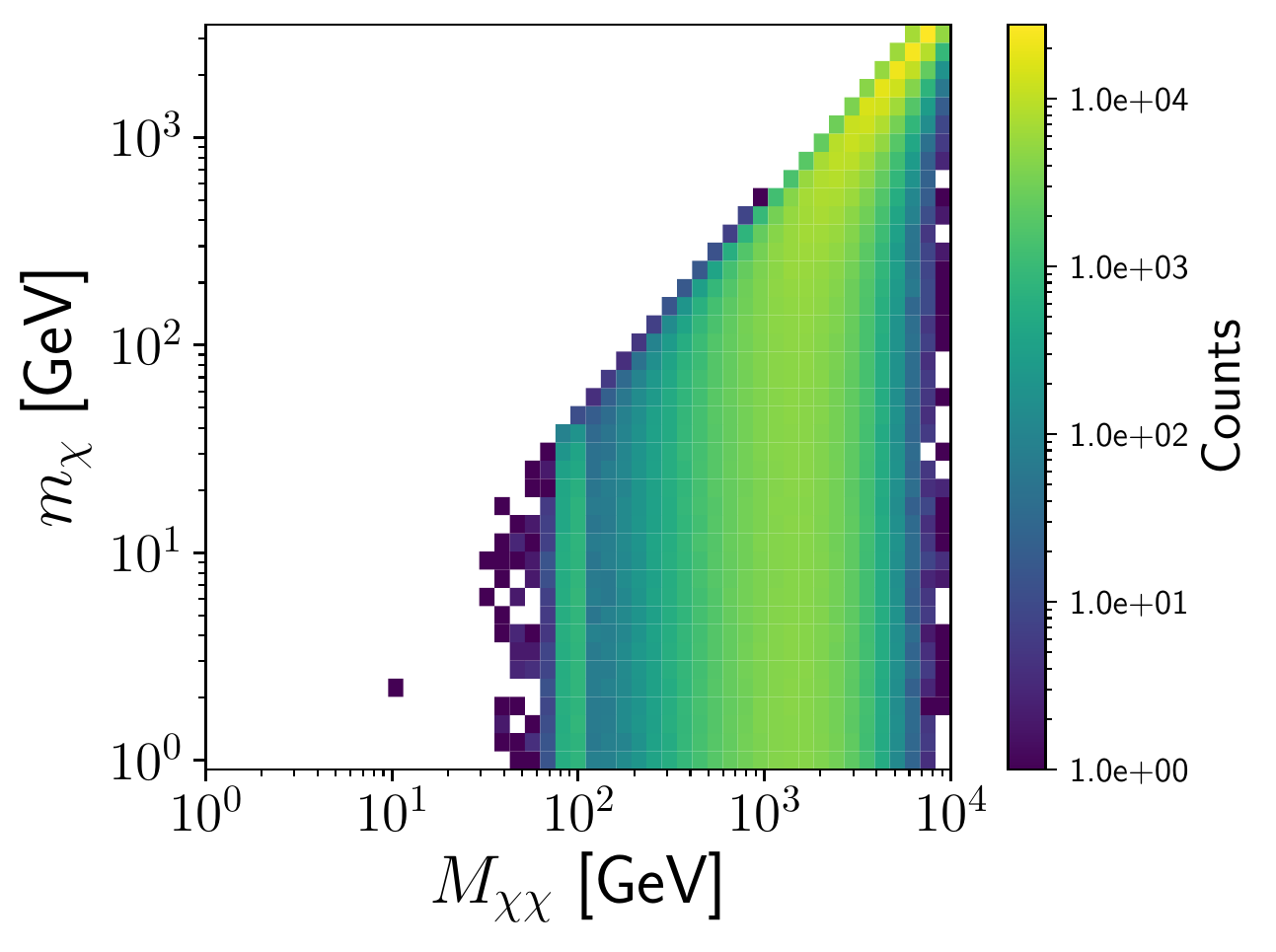}
    \includegraphics[width=0.75\columnwidth]{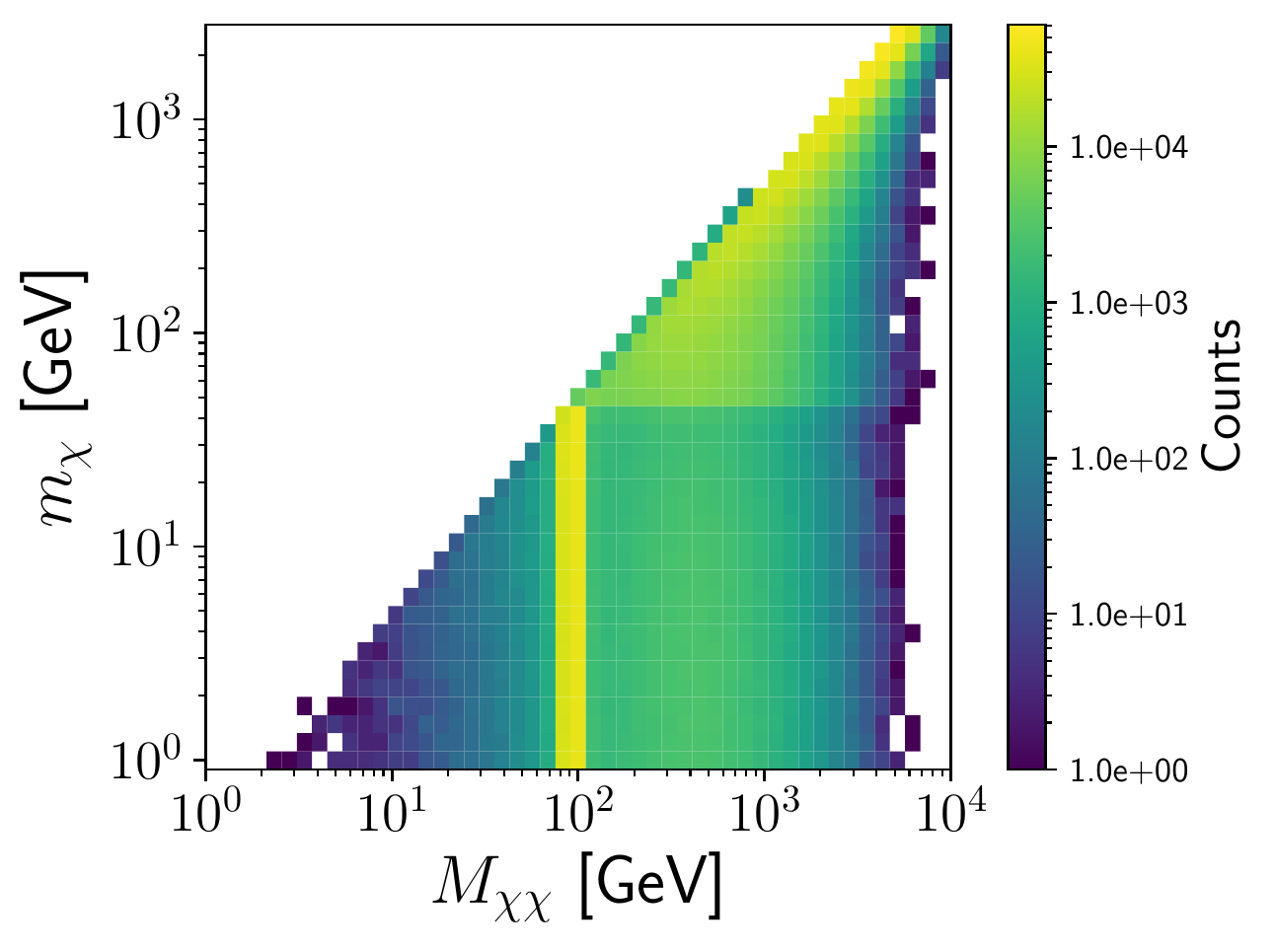}
    \caption{\textbf{Top row:} 2D heatmap showing how the Fisher information per GeV is distributed over the $p_T^{\text{miss}}$ for the anapole and the magnetic dipole interaction on the left and right respectively. The Fisher information is normalized to 1 for each value of $m_{\chi}$. \textbf{Bottom row:} The invariant mass bins of the mono-jet search, for the anapole and magnetic dipole studies.}
    \label{fig:2Dfish}
\end{figure*}

The results presented in Figure~\ref{fig:monoJ_lims} provide the most up-to-date limits on the dimension-5 and -6 dark matter form-factor operators, presented in units of GeV. They are within an order of magnitude of each other, with charge radius interaction achieving the strongest limit. The constraints start to degrade around $m_{\chi}\sim1$ TeV for dimension-6 interactions and $m_{\chi}\sim 300$ GeV for dimension-5. This occurs due to the fact that the dimension-6 operators grow more with energy and therefore populate more easily the high-$p_T$ bins in the distribution, where systematic uncertainties in the background expectations are less important. Also shown are projections for 3 ab$^{-1}$ of LHC data, assuming either reduced (25\%) or no (0\%) systematic uncertainties. The former is considered to be a reasonable expectation factoring in improvements in theoretical and experimental methods over the next 15 years. The high energy bins are also less sensitive to changes in systematic errors in the background. This is why, for dimension-5 operators, the projected sensitivities are much more responsive to changes in assumed systematic uncertainties. This picture is illustrated by the top row in Figure~\ref{fig:2Dfish}, which shows how the Fisher information, and therefore the constraining power, is distributed over the bins as a function of $m_\chi$. This visualisation is only possible without including the correlations between the bin uncertainties, which has an overall small effect on our limits. One can see a clear bias for lower $p_T$ in the dimension-5 case, particularly for $m_\chi$ below the $\MZ/2$.

One of the important aspects to consider when treating collider data with an EFT approach is whether the energy at which we are deriving limits, is sufficiently below the new physics scale, $\Lambda$. The subtleties associated with this have been examined~\cite{Bai:2010hh,Fox:2011pm,Fox:2012ee,Busoni:2013lha,Buchmueller:2013dya,Bauer:2016pug} and have led to the preferential adoption of simplified models~\cite{Abdallah:2015ter,Primulando:2015lfa,DeSimone:2016fbz}. Of course, since the EFT approach does not, \emph{a priori}, predict the values of the Wilson coefficients, $\mathcal{C}$, and constraints apply to the combination $\mathcal{C}/\Lambda^n$, one is technically safe from such worries up to a point. It is only when matching these operators with a particular UV model that care must be taken. A naive version of the argument would take the highest bin value used in the analysis and requiring, $\Lambda > p_{\rm{T}}^{\text{max}}$. Limits of the form
$$C/\Lambda^n < \text{limit},$$
can then be recast as a lower bound on the Wilson coefficient that admits a valid interpretation,
\begin{equation}
    \label{eq:validity}
    C_{j}>\left(\frac{\rm{max\,bin}}{\rm{limit}}\right) \hspace{3em} \textrm{and} \hspace{3em}  C_{j}>\left(\frac{\rm{max\,bin}}{\rm{limit}}\right)^{2},
\end{equation}
for dimension-5 and -6 respectively. Taking the maximum bin of $1.4$ TeV, gives a minimum $C\sim\mathcal{O}(1)$. However, looking at Figure~\ref{fig:2Dfish} we see that much of the information is found in lower energy bins, which may give more room for lower values of $C$ and $\Lambda$ to be consistent with these results. 

That said, the $p_T^{\text{miss}}$  is not the only independent energy scale of the process that can be used to assess the validity. Although unobservable, the $\chi\chi$ invariant mass is arguably a more accurate representation of the energy being probed, given that can be identified with the momentum flow through the effective vertex. One would naively associate this quantity with $2m_\chi$, since particle pair production tends to preferentially occur close to the kinematic threshold. However, for `low' DM masses below a few hundred GeV, the high energy of the LHC collisions coupled with the valence quark PDF in the initial state and the energy dependence of the interactions bias this quantity to much higher than expected values. The bottom row of Figure~\ref{fig:2Dfish} shows the distribution of invariant mass, $M_{\chi\chi}$, for different values of $m_{\chi}$. We find that for dimension-6 operators, many of our signal events populate invariant masses around 1 TeV, for dark matter masses up to 500 GeV, at which point the expected $m_{\chi\chi}\sim2m_\chi$ relation is restored. For dimension-5, the dominant value for $M_{\chi\chi}$ is $\sim m_Z$ when $m_{\chi}\lesssim 45$ GeV, above which the behaviour is similar to dimension-6. This is consistent with our previous estimates from the Fisher information in the case of low dark matter masses, but calls for care when interpreting the limits when $m_\chi$ reaches $\mathcal{O}(\text{TeV})$ and beyond. As an example, using Eq.~\ref{eq:validity} with aforementioned estimates of $m_{\chi\chi}$ yields minimum Wilson Coefficients of around $C_5 \gtrsim 0.03$ for the dimension 5 operators in the low mass case, when $m_{\chi} < 45$ GeV. For the dimension-6 operators, the higher typical scale leads to $C_6 \gtrsim 1$ for $m_\chi < 1$ TeV, above which the limits are significantly weakened anyway.
Overall, the collider sensitivity lends itself to interpretations in relatively strongly coupled scenarios, and would not be sufficient to admit loop-induced Wilson coefficients, as is often the case for EW form factors induced by weakly-coupled UV completions.

\section{Phenomenology of dark matter with hypercharge form factors}
\label{sec:pheno_dm}

In this section we present the dark matter phenomenology of the hypercharge EFT framework. We show the interplay among the most constraining direct and indirect dark matter searches, and discuss the prospects for detection by assessing the sensitivity of future probes to the model parameter space.

\subsection{Dark matter production}
\label{subsec:production}
An important aspect of any dark matter analysis is to assess its creation in the early universe. Here we consider that this is achieved via the standard thermal freeze-out mechanism, where initially the interactions between the dark matter and the SM particles are effective enough to keep it in thermal equilibrium with the SM plasma. Eventually, when the expansion of the universe dilutes dark matter enough, annihilations become ineffective and the dark matter freezes out with a relic density $\Omega_{\chi}$, which depends on the cross-section like,
\begin{equation}
    \Omega_{\chi} h^{2}\propto\frac{1}{\left\langle\sigma_{\mathrm{ann}} v\right\rangle}
    \label{eq:relic_approx},
\end{equation}
where $h$ is the reduced Hubble parameter and $\langle \sigma_{\mathrm{ann}} \,v\rangle$ is the thermally averaged annihilation cross-section evaluated at the freeze-out temperature.

\begin{figure}[t]
    \centering
    \includegraphics[width=\columnwidth]{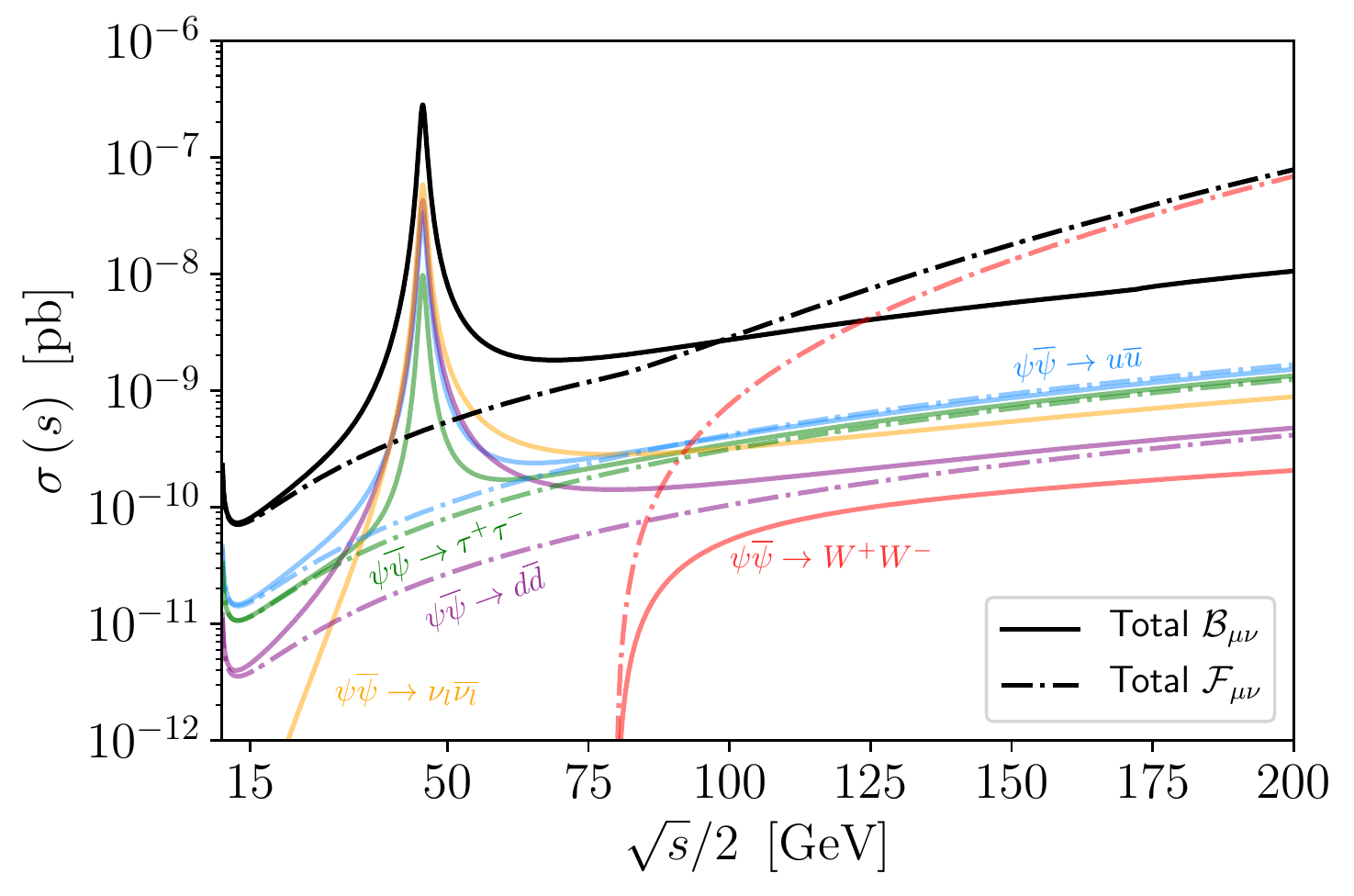}
    \caption{Annihilation cross-section of the electromagnetic (dashed) and hypercharge (solid) charge radius interaction, as a function of the centre-of-mass energy. The dark matter mass is set to $m_{\chi}=10$ GeV, the Wilson coefficient is $\mathcal{C}_{{cr}}=1.0$ and $\Lambda=10$ TeV. The different contributing SM channels are shown in color as labelled.}
    \label{fig:sigma_cr}
\end{figure}

Dark matter annihilations into SM particles are different for the hypercharge and electromagnetic form factors, referring back to Figure~\ref{fig:annihilation_diags}: in the former case $\chi \chi \rightarrow \textrm{SM} \,\textrm{SM}$ processes are also mediated by the $Z$ boson in s-channel, whose contribution can potentially be resonant, while this is not the case for the electromagnetic form factors. In Figure~\ref{fig:sigma_cr}, we show the annihilation cross-section into various SM final states as a function of the centre-of-mass energy ($\sqrt{s}/2$), for the charge radius operator (qualitatively the picture is similar for all the other interactions). The first striking difference between the gauge invariant and violating case is the $Z$-funnel region which leads to the following conclusions:
(i) the use of the photon interaction only is not accurate already at energies as low as 15 GeV, (ii) the $Z$ resonance dominates $\langle \sigma_{\mathrm{ann}} \,v\rangle$ roughly in the range between 15 and 60 GeV and includes the $\chi \chi \rightarrow \nu_l \,\nu_l$ process with a similar relevance to annihilation into quarks, while this channel is not present at all in the case of the electromagnetic operators.
The second deviation between the two dark matter models is found above the $Z$-funnel, where the hypercharge $\langle \sigma_{\mathrm{ann}} \,v\rangle$ returns a proper energy dependence ($\propto s$), but the electromagnetic operator exhibits an annihilation cross-section which grows with a gradient much greater than $s$ in energy, as as discussed in section~\ref{sec:DMWW}. At high energies, the hypercharge form factor annihilations are dominated by fermionic final states, which is in contrast to the electromagnetic case, where $W^+W^-$ annihilation dominates.
Ultimately these differences result in a different viable parameter space when searching for the correct relic abundance, to which we now turn.

We compute the effective coupling that provides the correct relic density for a given dark matter mass using the \texttt{MadDM} 3.0~\cite{Ambrogi:2018jqj} tool, matching the value of $\Omega_{\chi} h^{2}$ to the value measured by the Planck satellite~\cite{Ade:2015xua}.
\begin{figure*}[t]
    \centering
    \includegraphics[width=0.7\columnwidth]{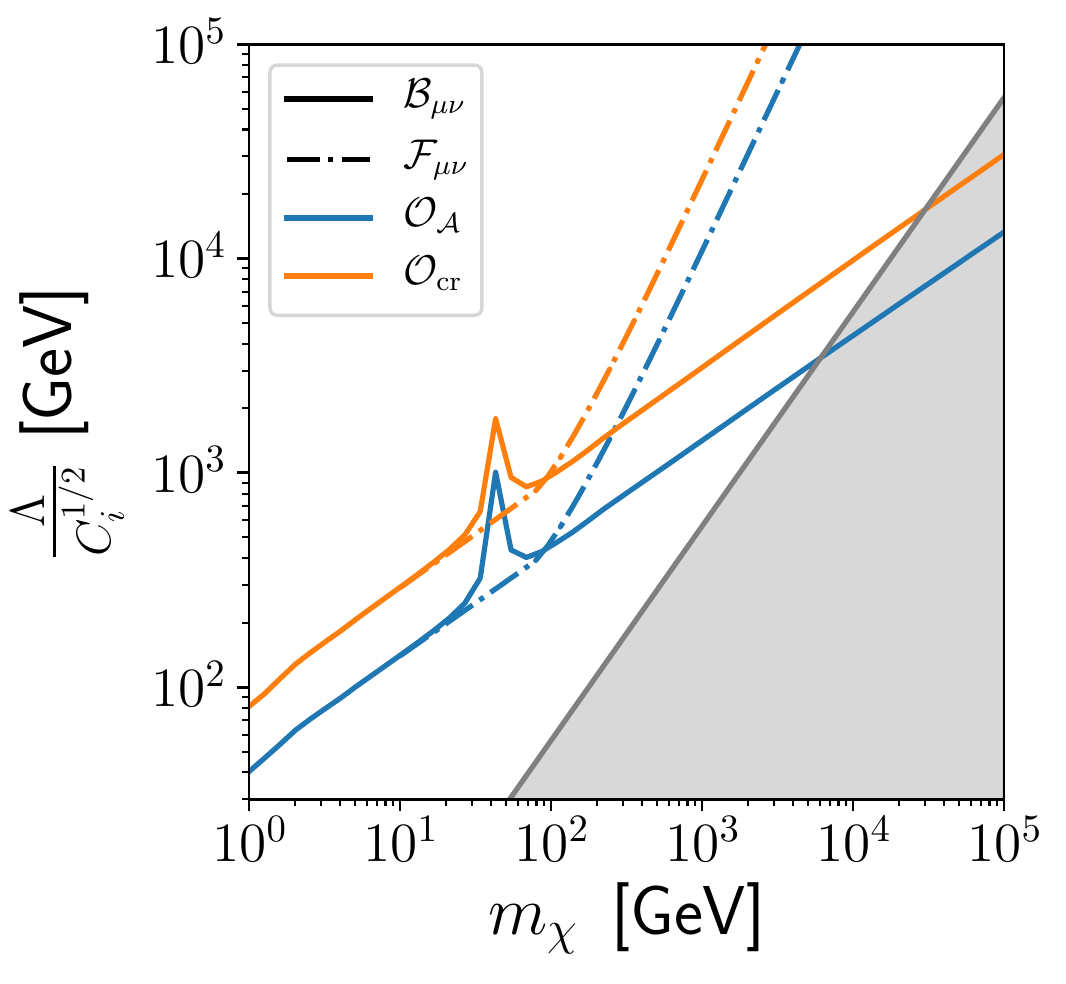}
    \includegraphics[width=0.7\columnwidth]{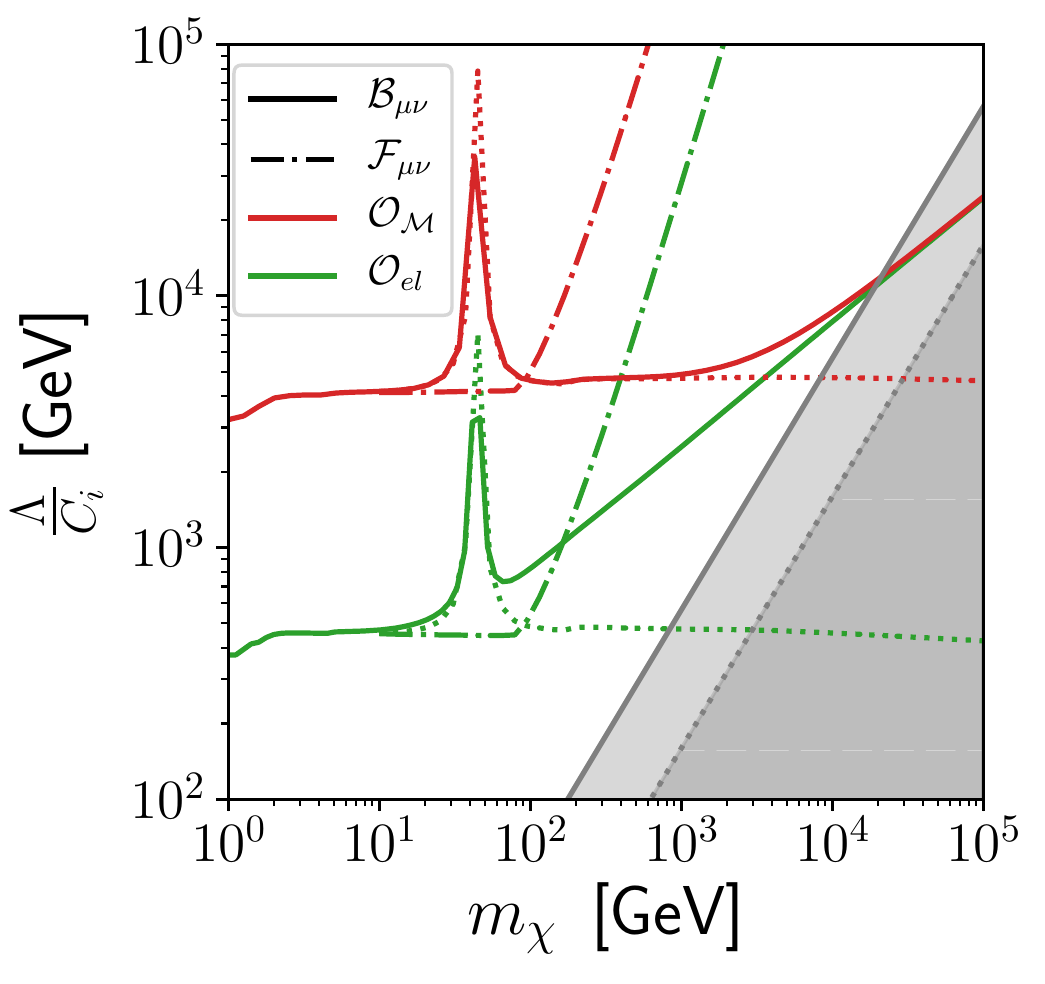}
    \caption{$\Lambda/\sqrt{\mathcal{C}_j}$ values producing the correct relic abundance (colored lines) as a function of the dark matter mass. \textbf{Left:} Results for dimension-6 operators. \textbf{Right} Same as left for dimension-5 operators. The color code is as following: \textbf{blue} lines for anapole moment, $\mathcal{O}_{\mathcal{A}}$, \textbf{orange} lines for charge radius interaction, $\mathcal{O}_{cr}$, \textbf{red} lines for magnetic moment, $\mathcal{O}_{\mathcal{M}}$, and \textbf{green} lines for electric dipole moment, $\mathcal{O}_{el}$. Hypercharge operators are denoted by solid lines, while photon only interactions are given by dot-dashed lines. The grey region denotes the breakdown of perturbativity for the EFT at dimension-6 for both panels. \textbf{Dotted} lines are for hypercharge interactions that do not include double insertion contributions, {\it i.e.} the pure dimension-5. The dotted grey region is the area where perturbativity breaks down for dimension-5 operators.}
    \label{fig:relic_lines}
\end{figure*}
The left of figure~\ref{fig:relic_lines} shows the results for the dimension-6 interactions. The results corroborate Figure~\ref{fig:sigma_cr} insofar as the electromagnetic and hypercharge interactions start to diverge around $m_{\chi}\sim 30$ GeV for the whole $Z$-funnel and that above $m_{\chi}\sim 90$ GeV the electromagnetic anapole has a steeper gradient. We plot the dark matter mass up to $100$ TeV as this is the generic thermal relic bound~\cite{PhysRevLett.64.615} calculated using partial wave analysis for a generic $s$-wave cross-section. For instance all anapole interactions are $p$-wave hence partial wave unitarity might break down at higher dark matter masses with respect to the case of $s$-wave unitarity, see {\it e.g.}~\cite{Zavala:2014dla}.

We also consider naive perturbativity limits on the parameter space by taking,
\begin{equation}
  \frac{\mathcal{C}_5}{\Lambda}\sqrt{s}\leq 4\pi, \hspace{3em} \textrm{and} \hspace{3em} \frac{\mathcal{C}_6}{\Lambda^2}s\leq 4\pi,
\end{equation}
for dimension-5 and dimension-6 vertices respectively. For the annihilation process we set $\sqrt{s}\sim 2m_\chi$ to get the relations,
\begin{equation}
  \frac{m_{\chi}}{2\pi}\leq \frac{\Lambda}{\mathcal{C}_5}, \hspace{3em} \textrm{and} \hspace{3em} \frac{m_{\chi}}{\sqrt{\pi}}\leq \frac{\Lambda}{\sqrt{\mathcal{C}_6}}.
\end{equation}
These constraints (shown in the Figures by the grey area) are a loose statement on whether it makes sense to treat the effective couplings perturbatively. Beyond these regions, one may worry that loop contributions could be comparable to the tree-level ones that we have computed. They are independent of the scale of new physics $\Lambda$ since one can always compensate any restriction on by varying the Wilson coefficients $\mathcal{C}_j$. If one is willing to make more specific assumptions, {\it i.e.} how this effective term relates to a UV complete model, different constraints can be drawn.

The left panel in figure~\ref{fig:relic_lines} shows that for the dark matter relic density to be set by the hypercharge moments, the perturbative description of the scattering starts to break down around $\sim 20$ TeV and $\sim 6$ TeV for the charge radius and anapole interactions respectively.

\begin{figure}
\centering
\includegraphics[width=0.4\columnwidth]{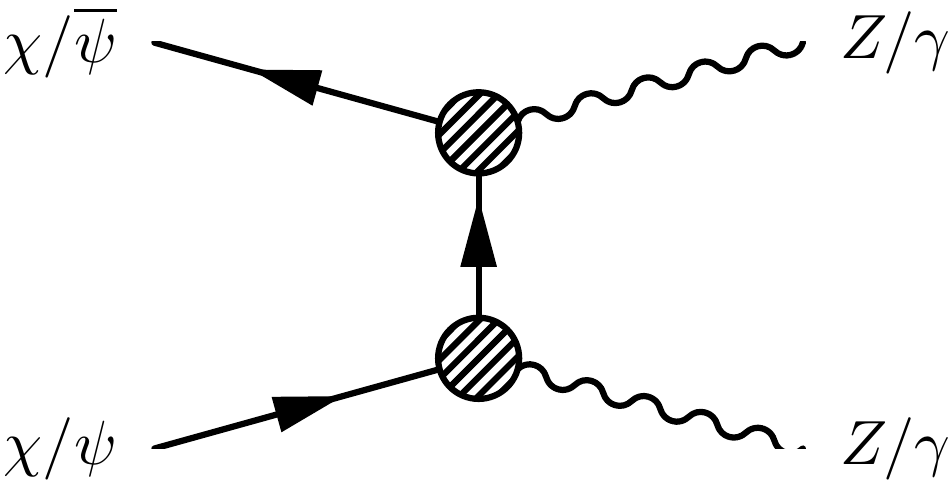}
\caption{Dimension-6 or -8 annihilation channels which occur through insertion of more than one effective vertex. Possible final states are $\gamma\gamma$, $\gamma Z$ and $ZZ$.}
\label{fig:higher_order}
\end{figure}

For both dimension-6 operators, the $s$-channel annihilation channels shown in Figure~\ref{fig:annihilation_diags}, are the only ones available. It is tempting to also consider $t$- and $u$-channel processes with two effective vertices, as shown in Figure~\ref{fig:higher_order}. These annihilation processes with di-photon and $\gamma Z$ final states, which are of primary relevance for dark matter line searches~\cite{Ackermann:2015lka,Rinchiuso:2019rrh}, as well as the $ZZ$ final state, which contributes for instance to continuum gamma ray searches. From an EFT perspective, however, care should be taken when considering processes at higher orders in the Wilson coefficients. This is because higher order contributions to a process with multiple operator insertions can be of the same order in $1/\Lambda$ as direct contributions from (typically neglected) higher dimension operators. It can even be the case that a multiple insertion diagram can be fully described in terms a higher dimensional operator contribution. Indeed, multiple insertions of an EFT operator in, \emph{e.g.}, loop calculations require the theory to be renormalised to higher orders in $1/\Lambda$. These effects are truly of higher dimension and should generally be considered on the same footing as higher dimensional operators.

For the dimension-5 operators in~\eqref{eq:dirac_lag}, however, it turns out that there is no dimension-6 interaction that mediates $\chi\chi\rightarrow\gamma\gamma, \gamma Z, ZZ$ annihilation. This means that, up to $1/\Lambda^2$ in the EFT, these processes are described only by the square of the dimension-5 couplings. New operator contributions only arise starting at dimension-7, in the form of so-called Rayleigh operators, $\chi\chi F^{\mu\nu}F_{\mu\nu}$~\cite{Weiner:2012cb,Weiner:2012gm,Latimer:2017lwm}.
It is therefore justified to include `double insertions' as part of the model, {\it i.e.} to consider the magnetic and electric dipoles up to dimension-6, when describing the new physics contributions to dark matter annihilation onto neutral gauge bosons. In the rest of the analysis we will consider the phenomenology of the pure dimension-5 operators and their double insertions separately, to understand the parameter regions in which each type of contribution is relevant.
In the right panel of Figure~\ref{fig:relic_lines} we present the correct relic abundance lines for both the magnetic and electric dipole interactions (red and green respectively). We show the curves for both the electromagnetic (black dot-dashed) and hypercharge field strength tensors (black dotted), but this time including the double insertion processes (black solid). The phenomenology is rather similar to the case of the dimension-6 operators, except that for the hypercharge dimension-5 operators the value of $\Lambda/\mathcal{C}_j$ at large dark matter masses is flat. This is due to the fact that dimension-5 operators do not grow with energy, while dimension-6 operators grow $\propto s$. Therefore the dimension-6 double insertion process start to dominate the hypercharge EFT model at large dark matter masses, above $10^3$ GeV or $100$ GeV for the magnetic and electric vertex respectively. The additional diagrams grow with energy and the relic lines exhibit a slope, similar to the anapole and charge radius cases.
Given their respective energy growths, the single and double-insertion scenarios have different naive perturbativity bounds shown by the two grey regions. As expected, the dimension-6 bounds take out a little bit more of the viable parameter space. In Figure~\ref{fig:relic_lines} (right panel), we see that for the magnetic dipole, including the double insertion processes slightly restrains the viable parameter space for thermal relic: the dark matter mass upper bound due to perturbative unitary of $30$ TeV for dimension-5 operator only becomes 20 TeV for dimension-6 vertices. Conversely, the parameter space of the electric dipole is enlarged, moving the upper bound from perturbative unitarity on the dark matter mass from 3 TeV up to 20 TeV.

An important consequence of the unphysical growth in the cross-section for the photon only interactions is that the steep gradient in Figure~\ref{fig:relic_lines} make it appear as if the correct relic can be obtained to arbitrarily high values of $m_{\chi}$. Of course, this is not the case as the $W^+W^-$ scattering cross-section would violate unitarity at much lower masses.

As stated at the beginning of the section, the computation of the dark matter relic abundance is based on the thermal freeze-out assumption with a standard cosmological history. In the Figure~\ref{fig:relic_lines}, the region below the black curve denotes under-abundant dark matter, while above it the dark matter is over-abundant. In this latter case its  annihilation cross-section is small and dark matter decouples too early from the thermal bath with a large abundance, the later you decouple the more dark matter undergoes matter suppression. In the next section we will however consider that the dark matter candidate under consideration makes up all the dark matter regardless of the abundance that results from thermal freeze-out. In particular this assumption concerns the local dark matter densities that enter the fluxes computations for indirect and direct detection experiments. For under-abundant dark matter this scenario could be realised {\it e.g.} by additional non-thermal contributions to dark matter production, for example the late gravitino decay in supersymmetric models is a popular mechanism to augment the neutralino relic density and bring it to the Planck measured value~\cite{Allahverdi:2012wb}. Over-abundant dark matter could be accommodated by a non-standard cosmological history which for instance modifies the expansion rate of Universe, see {\it e.g.}~\cite{Cirelli:2008pk}, or UV completion of phenomenological models could provide more efficient annihilations at early times, see {\it e.g.}~\cite{Cerdeno:2019vpd} and references therein. Finally, alternative scenarios exist that can produce the correct relic density, such as the freeze-in mechanism~\cite{McDonald:2001vt,Hall:2009bx} and forbidden dark matter~\cite{DAgnolo:2015ujb}, which might highlight different regions of the EFT model parameter space. Notice however that those scenarios typically select dark matter candidates which are fairly light, close or below the GeV mass scale. This region is in great tension with the measurements of the $Z$ boson invisible decay width for the hypercharge EFT model, however a throughout analysis of the phenomenology from {\it e.g.} the freeze-in mechanism is beyond the scope of this work.
We believe it is instructive to show in the analysis the parameter space denoting the correct thermal freeze-out relic abundance, as interesting benchmark from a model building point of view.

\subsection{Direct searches}
\label{subsec:direct}

Direct dark matter searches seek to measure the recoil of target nuclei hit by a dark matter particle passing by in underground detectors.
The momentum transfer in the elastic collision is limited by kinematics and the galactic escape velocity in the detector frame, $v_{\textrm{esc}}$,

\begin{equation}
    q_{\textrm{max}}^2=4 \mu_{\chi\,T}^2 v_{\mathrm{esc}}^{2},
\end{equation}
where $\mu_{\chi\,T}$ is the reduced mass of the incoming dark matter and the target nucleus. The largest $q^2$ value is achieved in the large $m_{\chi}$ limit, which with $v_{\textrm{esc}}\sim700$ km s$^{-1}$~\cite{Piffl:2013mla}, is $q\leq 500$ MeV. This maximum value, is well above much of the signal, which is in the [1-10] MeV range. Therefore results for the electromagnetic moments will be directly applicable to the hypercharge EFT model, as the $B$-field strength tensor is simply related to both the photon and $Z$ boson as in \eqref{eq:BtoFandZ}. In the following, we briefly review how the non relativistic operators relevant for direct detection are obtained from the EW EFT operators in equations~\eqref{eq:majorana_lag} and~\eqref{eq:dirac_lag}.

When calculating the direct detection contribution for the dimension-5 operators, one has to consider the scattering amplitude with the full propagator. The interaction vertices for the SM quarks with the photon and the $Z$ are
\begin{equation}
    \mathcal{L}_{\textrm{int}}\supset e Q_{q} \bar{q} A_{\mu} q + \frac{g}{4 \cw}\bar{q} \gamma^{\mu}\left(V_{q}-A_{q} \gamma^{5}\right) q Z_{\mu},
\end{equation}
where $g$ is the EW coupling, $Q_q$ is the electromagnetic charge and the vector ($V_q$) and axial ($A_q$) couplings of each quark to the $Z$, they are parameterised by~\cite{Arcadi:2014lta}
\begin{equation}
V_{q}=2\left(-2 Q_{q} \sw^{2}+T_{q}^{3}\right) \quad {\rm and} \quad A_{q}=2 T_{q}^{3}\,,
\end{equation}
where $T_{q}^3$ is the weak isospin number.
Therefore, for the pure vector coupling to the $Z$ boson, the billinear structure is the same as for the photon, and hence produces the same operator responses from nucleons $\mathcal{O}^{\textrm{DD}}$ as outlined in~\cite{Fan:2010gt,Fitzpatrick:2012ix,DelNobile:2018dfg,Kavanagh:2018xeh}. An important difference is for the coefficients for such operators, which for the $Z$ contributions, have a suppression factor,
\begin{equation}
    \frac{C^{\textrm{N}}_{B_{\mu}\gamma^{\mu}}}{{C^{\textrm{N}}_{A_{\mu}\gamma^{\mu}}}}=\frac{1}{4m_W^2}\frac{V_N}{Q_N},
\end{equation}
where the couplings in the Lagrangian in \eqref{eq:majorana_lag} and \eqref{eq:dirac_lag} are assumed to be the same. The super and sub scripts, $N$, refer to the nucleon level interaction, namely the parameters $V_N$ and $Q_N$ are summed values from the quark level coefficients $V_q$ and $Q_q$ respectively. The exact values of which can be found in Ref.~\cite{Bishara:2017pfq}. The axial vector couplings give rise to novel responses not present in the photon case~\footnote{In the non-relativistic formalism of~\cite{Fitzpatrick:2012ix}, these responses are the operators $\mathcal{O}_9$ and $\mathcal{O}_{14}$.}, however, due to the suppression coming from the mediator, the contributions are sub-dominant.

For the dimension-6 operators, when considering the electromagnetic interactions, calculations are simplified by making use of the equations of motion (see Equation~\ref{eq:FandB_EOM}),
\begin{equation}
    \partial^{\nu}F_{\mu\nu}=eJ_{\mu}^{\textrm{EM}}\approx e\sum_{q=u, d, s} Q_{q} \bar{q} \gamma_{\mu} q\,.
    \label{eq:F_EOM}
\end{equation}
The approximation comes because we are interested in calculating low energy scattering with nucleons. For the $B$-field, there is a similar expression,
\begin{equation}
    \partial^{\rho} B_{\rho \mu}=\frac{1}{2}g^{\prime} H^{\dagger} i \stackrel{\leftrightarrow}{D}_{\mu} H+g^{\prime} (J_{\mu}^{\textrm{EM}}-J_{\mu}^{3}),
    \label{eq:B_EOM}
\end{equation}
where $J_{\mu}^{3}$ is the current $J_{\mu}^{3}=\sum_{i} \bar{f}_{i}^{L} \gamma^{\mu} T^{3} f_{i}^{L}$, $H$ is the Higgs field, $T^3$ is the weak isospin value of the fermion $f$ and $g^{\prime}$ is the hypercharge coupling. After EW-symmetry breaking  Equation~\eqref{eq:B_EOM} becomes,
\begin{equation}
    \partial^{\rho} B_{\rho \mu}=\frac{egv^2}{4}Z_{\mu}+g^{\prime} \left(J_{\mu}^{\textrm{EM}}-J_{\mu}^{3}\right),
    \label{eq:B_EOM_EWSB}
\end{equation}
where $v$ is the Higgs vacuum expectation value and $g$ is the weak charge.
Using the $Z$ equation of motion,
\begin{equation}
    Z_{\nu}=\frac{e}{s_W}\Pi^{\mu\nu}J^Z_{\mu}=\frac{e}{s_W}\Pi^{\mu\nu}\left(J^3_{\mu}- s_W^2J_{\mu}^{\textrm{EM}}\right),
\end{equation}
where $\Pi^{\mu\nu}$ is the Z propagator, and expanding in the large $m_{Z}$ limit, at the lowest order the result from the photon field is recovered
\begin{equation}
    \partial^{\rho} B_{\rho \mu}\approx ec_WJ_{\mu}^{\textrm{EM}}+\mathcal{O}\left(\frac{q^2}{m_{Z}^2}\right)\,,
\end{equation}
which is the relevant term at the energy scales of direct detection. Since the effects of the $Z$ mediated interactions are much weaker than the photon, the results here are the same as in other works, where, in the non-relativistic effective theory basis, the dark matter-nucleon interactions go like
\begin{eqnarray}
    \bar{\chi} i \sigma^{\mu \nu} \gamma^{5} \chi B_{\mu \nu} &\longrightarrow& Q_{N} e \frac{4}{q^{2}} m_{\chi} m_{N}^{2} \mathcal{O}_{11}^{DD}\,,\nonumber\\
     \bar{\chi} \sigma^{\mu \nu} \chi B_{\mu \nu} &\longrightarrow& 2 e m_{\chi} m_{N}\left[\frac{Q_{N}}{4 m_{\chi}} \mathcal{O}_{1}^{DD} +Q_{N} m_{N} \frac{\mathcal{O}_{5}^{DD}}{q^{2}}\, \right. \nonumber \\
     & & \left.+\frac{g_{N}}{2 m_{N}}\left(\mathcal{O}_{4}^{DD}-\frac{m_{N}^{2} \mathcal{O}_{6}^{DD}}{q^{2}}\right)\right]\,,\nonumber\\
     \bar{\chi} \gamma^{\mu} \chi \partial^{\nu} B_{\mu \nu} &\longrightarrow& 4 m_{\chi} m_{N} e Q_{N} \mathcal{O}_{1}^{DD}\,,\nonumber\\
    \bar{\chi} \gamma^{\mu} \gamma^{5} \chi \partial^{\nu} B_{\mu \nu} &\longrightarrow& 4 m_{\chi} m_{N} e \left(2 Q_{N} \mathcal{O}_{8}^{DD}-g_{N} \mathcal{O}_{9}^{DD}\right),
\end{eqnarray}
where the factors $Q_N$, $m_N$ and $g_N$ are the charge, mass and the magnetic moments of the nucleons respectively.
\begin{figure*}[t]
    \centering
    \includegraphics[width=0.7\columnwidth]{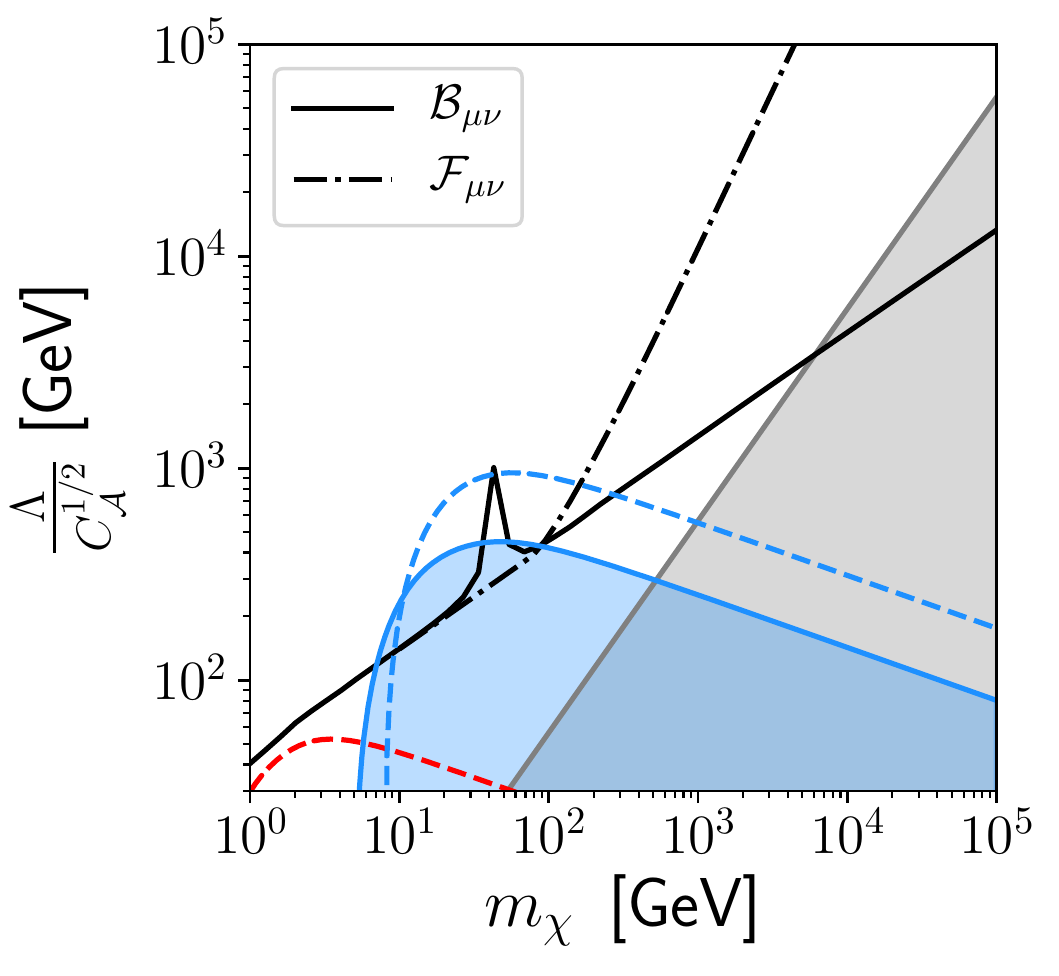}
    \includegraphics[width=0.7\columnwidth]{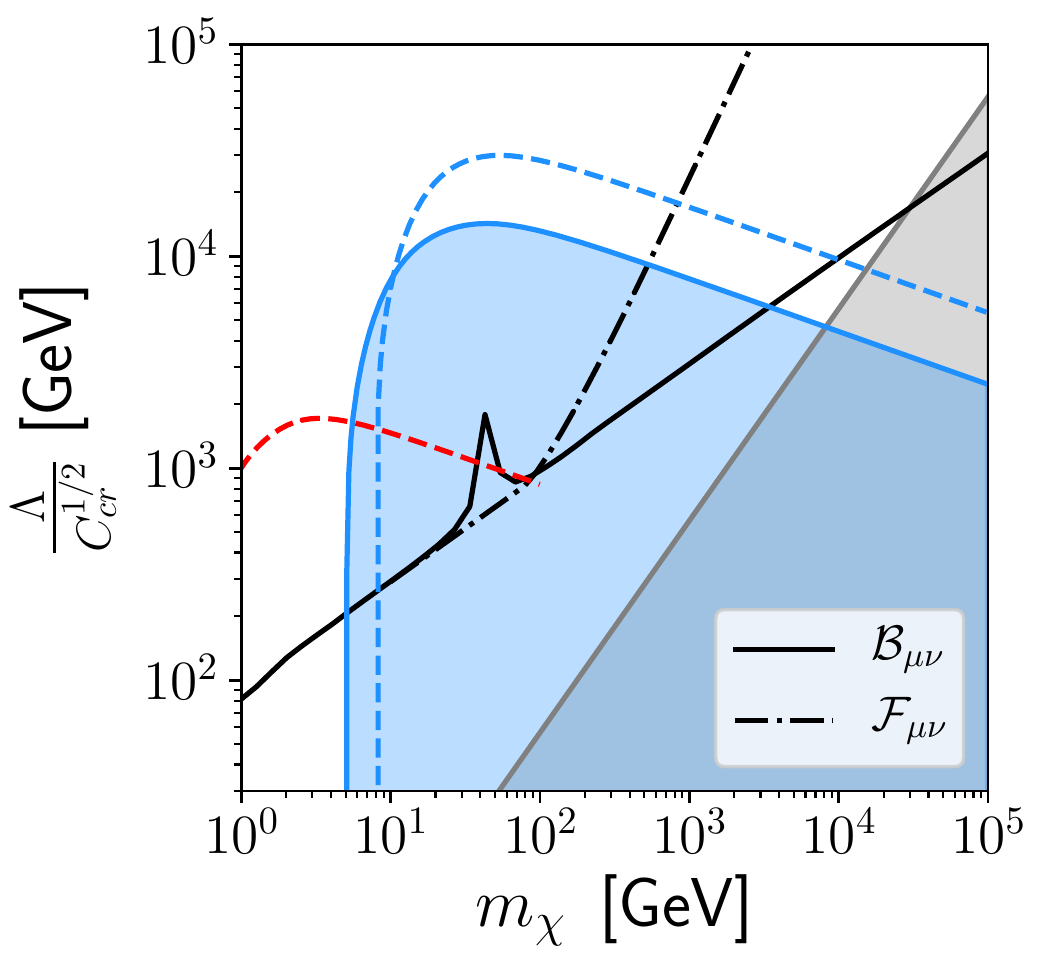}
    \includegraphics[width=0.7\columnwidth]{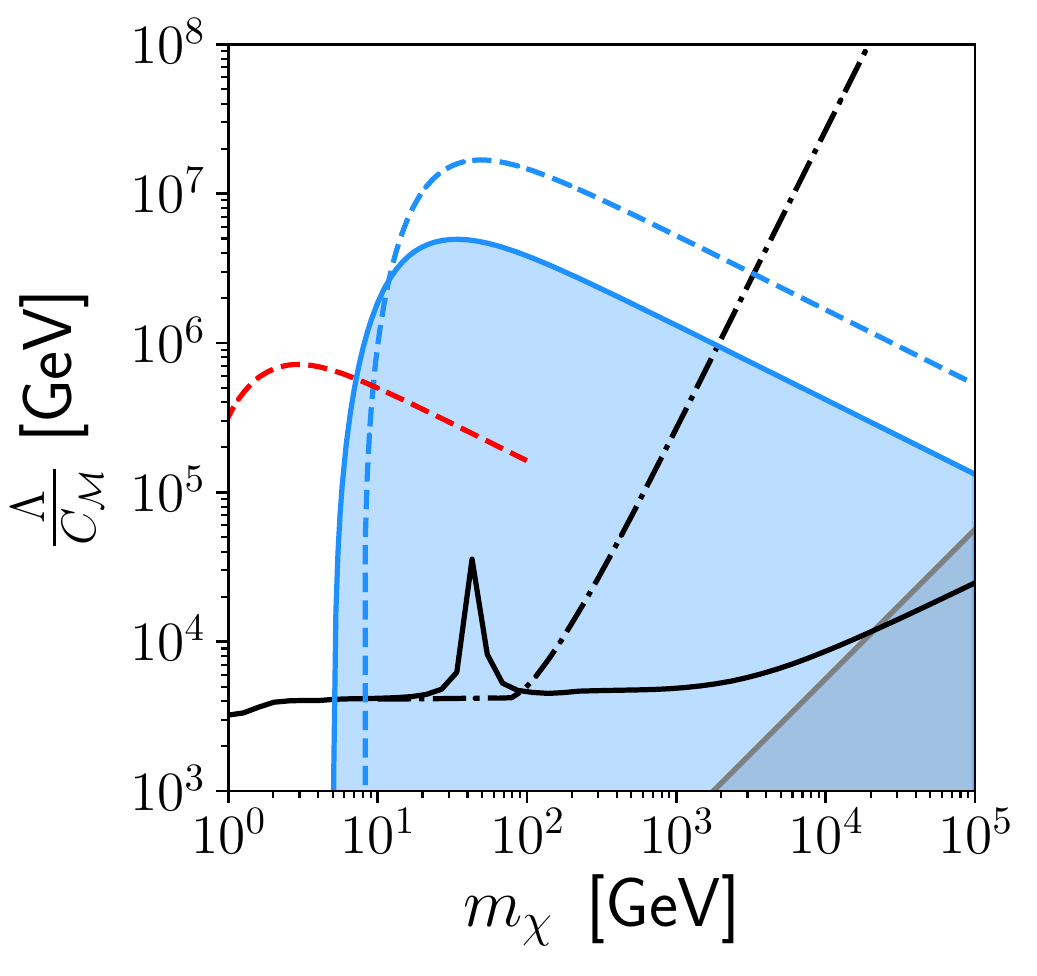}
    \includegraphics[width=0.7\columnwidth]{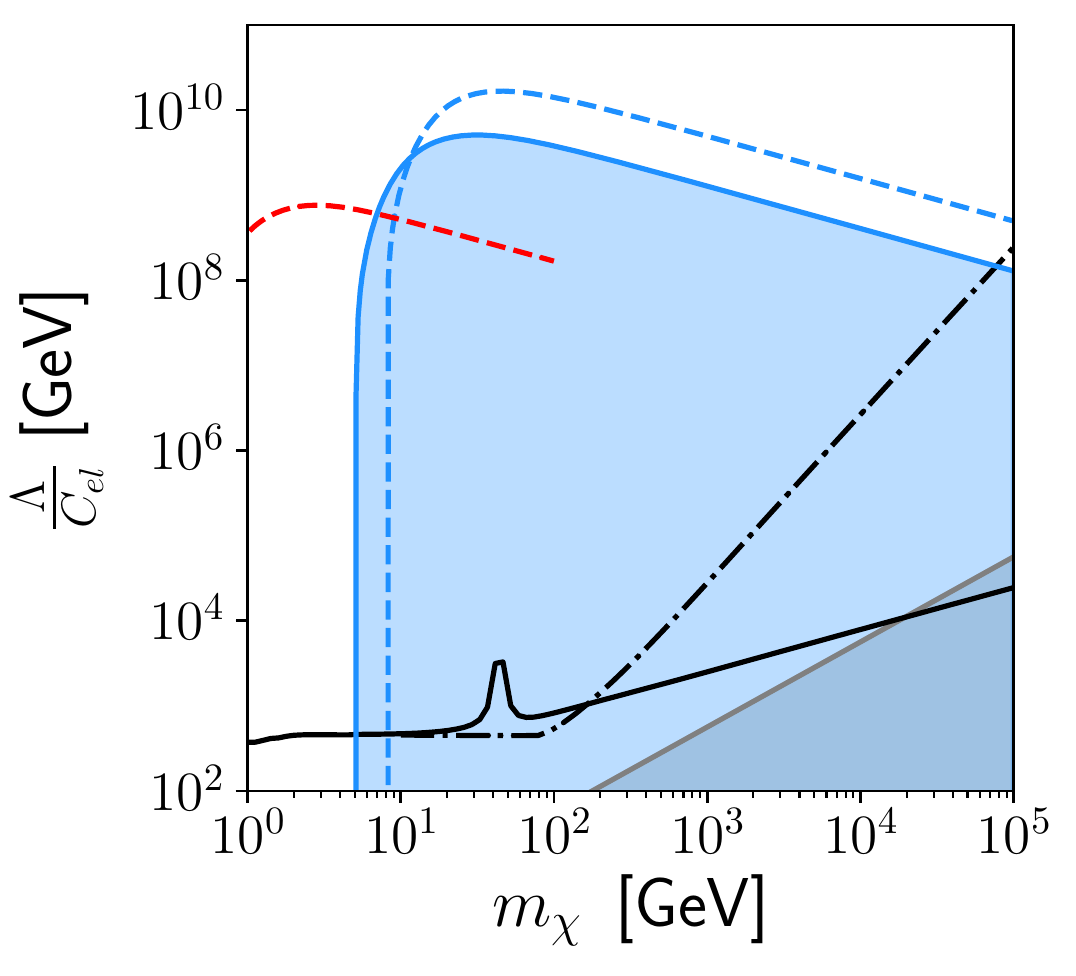}
    \caption{Direct detection limits at 95\% confidence level (CL) on $\Lambda/\mathcal{C}_j$ (or  $\Lambda/\mathcal{C}^{1/2}_j$) as a function of the dark matter mass $m_\chi$, coming from the current XENON1T exclusion limit \textbf{(blue)} as well as projected sensitivities of the future LZ experiment \textbf{(blue dashed)} and SuperCDMS \textbf{(red dashed)}. \textbf{Top left:} Constraints on the hypercharge anapole moment. \textbf{Top right, bottom left and bottom right:} Same as top left for the charge radius moment, the magnetic  and electric dipole moments respectively. The scale of the y-axis has been substantially altered from Figure~\ref{fig:relic_lines} due to the strength of the direct detection limits.}
    \label{fig:prod_and_DD}
\end{figure*}

For those unfamiliar with the basis defined above, $\mathcal{O}_1^{DD}$ is the canonical spin-independent interaction which receives a coherent contribution from the nucleons in the target nucleus by way of $A^2$, where $A$ is the atomic number. The same occurs for $\mathcal{O}_{11}^{DD}$ which however is also momentum suppressed. From this, we can anticipate that the anapole response will be the weakest.

Using the \texttt{RAPIDD} tool~\cite{Cerdeno:2018bty}, we recast the current exclusion limit by XENON1T~\cite{Aprile:2018dbl} as well as future LZ~\cite{Akerib:2015cja,Mount:2017qzi} projected sensitivity in terms of the hypercharge EFT model. In order to match with results coming from colliders and indirect searches, we present the 95\% confidence level (CL) exclusions and projections, as opposed to the direct detection community standard of 90\% CL. XENON1T has the strongest exclusion limit at intermediate and large dark matter masses while LZ will likely be the most sensitive detector built in a near future in the same mass range. For the XENON1T results, we make use of the prescription given in appendix A of~ \cite{Kavanagh:2018xeh}. We apply the same procedure to derive the projected limit for the  LZ experiment and for an exposure of 1000 days, as LZ will be a dual phase time projection chamber consisting of 5.6 tons of xenon similar to XENON1T. Our result is consistent with the one obtained in appendix D of~\cite{DEramo:2016gos}. To assess the sensitivity of direct detection at low dark matter masses, we simulate the SuperCDMS experiment following~\cite{Agnese:2016cpb,Kahlhoefer:2017ddj,Workgroup:2017lvb,Bozorgnia:2019mjk}. We use specifically  the high-voltage design of the experiment which will be able to access very low threshold energies thus enabling greater sensitivity to light dark matter.

We have not included any current bounds in the parameter space below $m_{\chi}\sim 6$ GeV, which would likely come from CRESST-III~\cite{Abdelhameed:2019hmk,Abdelhameed:2019mac} or DarkSide-50~\cite{Agnes:2018ves}. The situation here is more complicated and could even be most constrained through electron recoils. Refs.~\cite{Catena:2019gfa,Catena:2020tbv} has computed electron recoil bounds for the anapole, magnetic and electric moments and shown that at $m_{\chi}\sim 1$ GeV, Xenon1T~\cite{Aprile:2018dbl} results are most sensitive. The interplay between electronic and nuclear recoils as well as the multitude of ongoing experiments is something we leave for future work.

We report in Figure~\ref{fig:prod_and_DD} the constraints at 95\% (CL) and future reach from direct detection for the hypercharge EFT model and find that the basic picture does not change with respect to the case of the photon only interaction~\cite{Kavanagh:2018xeh} barring a few remarks. The blue shaded region shows that the current constraints from XENON1T are able to exclude large regions of viable parameter space. Indeed the black solid lines, denoting the couplings that predict the correct dark matter abundance via freeze out, are completely ruled out for the magnetic and electric dipole moments, as in previous works~\cite{DelNobile:2012tx}. Our findings here show that even considering the extra dimension-6 contributions do not allow you to evade constraints.

\subsection{Indirect searches}
\label{subsec:indirect}
\renewcommand{\arraystretch}{2.}
\begin{table*}[t!]
\centering
\caption{Analytic expressions for the cross-section ($\sigma$), and velocity-weighted, non-relativistic annihilation cross-sections ($\sigma v$) into SM final states ($XX$), for dimension-5 and dimension-6 operators of the hypercharge EFT for Dirac dark matter. The cross-section is provided in the high-energy limit with massless SM particles, $\beta^2 = (1-4 m_\psi^2/s)$ and $N_c$ is the number of colours. In the low velocity limit $\beta\approx v $, so cross-sections that are $s$-wave will be $\mathcal{O}(1/\beta)$ at the lowest order. The $\psi \bar{\psi} \to ZZ$ annihilation cross-section is given by $(\sw^{4}/\cw^{4})\sigma_{\gamma\gamma}$.  The Majorana anapole can be obtained by dividing the Dirac anapole moment by two, following Equations~\eqref{eq:majorana_lag} and~\eqref{eq:dirac_lag}.}
\begin{tabular}{|c|c||c|c|}
\hline
Operator & $\psi \bar{\psi} \rightarrow XX$ &  $\sigma_{XX} (s\gg m_f,\, m_Z ) $& $(\sigma v)_{XX}$  \\
\hline
$\OMag$  & $ f \bar{f}$&  $\frac{\left(3-2 \beta^{2}\right) e^{2} N_{c} C_{\mathcal{M}}^{2} Y_{f}^{2}}{48 \pi \beta \Lambda^{2} \cw^{2}}$ & $\frac{e^{2} N_{c} C_{\mathcal{M}}^{2} Y_{f}^{2}}{16 \pi \Lambda^{2} \cw^{2}}$ \\
& $ W^{+} W^{-} , Zh$ &$\frac{\left(3-2 \beta^{2}\right) e^{2} C_{\mathcal{M}}^{2}}{384 \pi \beta \cw^{2} \Lambda^{2}}$ & $\frac{e^{2}  C_{\mathcal{M}}^{2} }{128 \pi \cw^{2} \Lambda^{2} }$ \\
&  $ \gamma \gamma$ & $\frac{C_{\mathcal{M}}^{4}\cw ^{4} s\left(\left(9-7 \beta^{2}\right) \beta^{2}-6 \beta\left(\beta^{2}-1\right)^{2} \tanh ^{-1}(\beta)\right)}{96 \pi\beta^{3 } \Lambda^{4}}$  & $\frac{\cw^{4} C_{\mathcal{M}}^{4}  m_{\psi}^{2}}{8 \pi \Lambda^{4} }$\\
&  $ \gamma Z$ & $\frac{\sw^{2}}{\cw^{2}}\left(2\sigma_{\gamma\gamma}+3\left(\beta^2-1\right)\log{\left(\frac{3+\beta}{1-\beta}\right)}\right)$ & $\frac{\sw^{2}}{\cw^{2}}2(\sigma v)_{\gamma\gamma}$\\
  \hline
$\OEle$ & $ f \bar{f}$&$\frac{ e^{2} \beta N_{c} C_{el}^{2} Y_{f}^{2}}{48 \pi \cw^{2} \Lambda^{2}}$ & $\frac{e^{2} N_{c} C_{\mathrm{el}}^{2} Y_{f}^{2}}{48 \pi \cw^{2} \Lambda^{2}}\cdot v^2$\\
 & $W^{+} W^{-}, Zh$ & $\frac{e^{2} \beta C_{el}^{2}}{384 \pi  \cw^{2}\Lambda^{2}}$ & $\frac{e^{2} C_{\mathrm{el}}^{2} }{384 \pi \cw^{2} \Lambda^{2} } v^2$ \\
&  $ \gamma \gamma$ & $\frac{C_{el}^{4}\cw ^{4} s\left(\left(9-7 \beta^{2}\right) \beta^{2}-6 \beta\left(\beta^{2}-1\right)^{2} \tanh ^{-1}(\beta)\right)}{96 \pi\beta^{3 } \Lambda^{4}}$  & $\frac{\cw^{4} C_{el}^{4}  m_{\psi}^{2}}{8 \pi \Lambda^{4} }$\\
&  $ \gamma Z$ & $\frac{\sw^{2}}{\cw^{2}}\left(2\sigma_{\gamma\gamma}+3\left(\beta^2-1\right)\log{\left(\frac{3+\beta}{1-\beta}\right)}\right)$ & $\frac{\sw^{2}}{\cw^{2}}2(\sigma v)_{\gamma\gamma}$\\
\hline
$\Ocr$ & $ f \bar{f}$ & $\frac{\left(3-\beta^{2}\right) e^{2} s N_{c} C_{cr}^{2} Y_{f}^{2}}{48 \pi \beta \cw^{2} \Lambda^{4}}$ & $\frac{e^{2} N_{c} C_{cr}^{2} Y_{f}^{2}m_{\psi}^{2}}{4 \pi \cw^{2} \Lambda^{4}}$\\
 & $W^{+} W^{-}, Zh$&$\frac{\left(3-\beta^{2}\right) e^{2} s C_{cr}^{2}}{384 \pi \beta \cw^{2} \Lambda^{4}}$& $\frac{e^{2}  C_{cr}^{2}m_{\psi}^{2}}{32 \pi \cw^{2} \Lambda^{4} }$\\
  \hline
$\OA$ & $ f \bar{f}$ & $\frac{e^{2} s \beta C_{\mathcal{A}}^{2} N_{c}Y_{f}^{2}}{24 \pi \Lambda^{4} \cw^{2}}$ & $\frac{e^{2}  C_{\mathcal{A}}^{2} N_{c} Y_{f}^{2}  m_{\chi}^{2}}{6 \pi \Lambda^{4} \cw^{2}}\cdot v^2$\\
& $W^{+} W^{-}, Zh$&$\frac{e^{2} s \beta C_{\mathcal{A}}^{2}}{192 \cw^{2} \pi \Lambda^{4}}$& $\frac{e^{2}  C_{\mathcal{A}}^{2}m_{\chi}^{2}}{48 \pi \Lambda^{4} \cw^{2}}\cdot v^{2}$\\
\hline
\end{tabular}
\label{tab:analytic}
\end{table*}
\renewcommand{\arraystretch}{1.}

Indirect searches rely on the annihilation of dark matter into SM particles, which subsequently decay, shower and hadronise to lead to a continuum spectrum of gamma rays, cosmic rays (positrons and antiprotons) and neutrinos, see {\it e.g.}~\cite{Gaskins:2016cha} for a review and references therein. Alternatively, dark matter can annihilate into the diphoton or $\gamma Z$ final state, producing the smoking-gun signature of a sharp gamma-ray line feature at the dark matter mass~\cite{Bouquet:1989sr,Bergstrom:1989jr,Rudaz:1989ij}. In our analysis, we derive exclusion bounds for our models using the continuum annihilation spectra, including the $ZZ$ final state, and the $\gamma\gamma,\,  \gamma Z$ line final states. These latter three annihilation channels all come from the double insertion diagrams depicted in Figure~\ref{fig:higher_order}, with the last two being unique to the hypercharge form factors.

Ultimately, all annihilations share a typical energy scale that is set by the dark matter mass, as the late time environments that provide the strongest constraints ({\it i.e.} dwarf spheroidal galaxies, dSPhs, and the Galactic Centre) are much cooler than at the time of freeze-out, being characterised by relative velocities $v/c$ ranging roughly from $10^{-5}$ to $10^{-3}$. Therefore, whether or not the annihilation cross-section is $s$- or $p$-wave is hugely important for determining how indirect constraints map onto the hypercharge EFT model parameter space. We have computed analytically the (velocity averaged) annihilation cross-sections for all operators in~\eqref{eq:majorana_lag} and~\eqref{eq:dirac_lag} and report in Table~\ref{tab:analytic} their expression in the limit of massless SM particles. These expressions are in accord with~\cite{Kavanagh:2018xeh}. Clearly, the anapole moment is $p$-wave hence we will not consider it further in this section. The electric dipole moment annihilation into SM fermions and gauge bosons is suppressed by $p$-wave while the related dimension-6 operator leading to $\gamma\gamma$, $\gamma Z$ and $ZZ$ are not, hence only these latter channels will be accounted for in the analysis. The magnetic dipole and the charge radius interaction, on the other hand, are $s$-wave, meaning that their annihilation strength is unaltered throughout the thermal history of the universe.
\begin{figure*}[t!]
    \centering
    \includegraphics[width=0.7\columnwidth,]{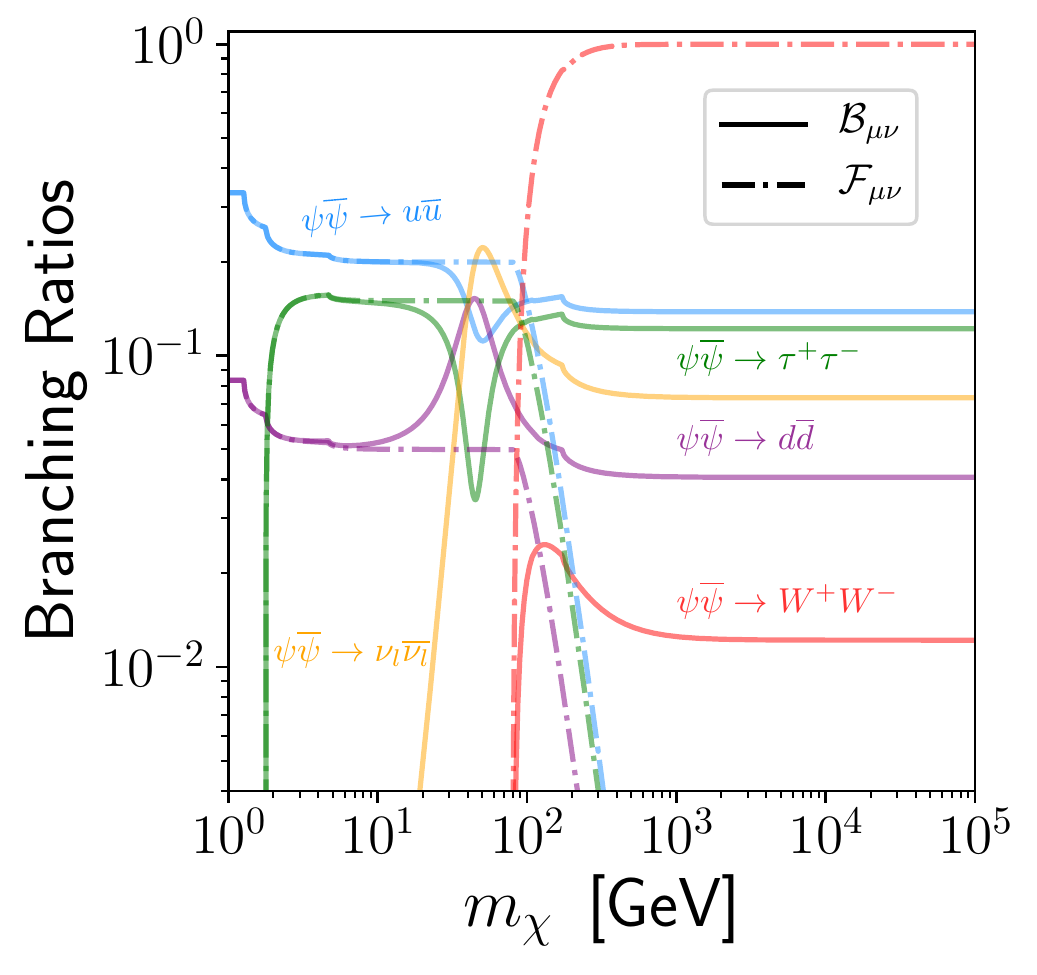}  
    \includegraphics[width=0.7\columnwidth,trim={0.1cm 0.2cm 0.2cm 0.1cm},clip]{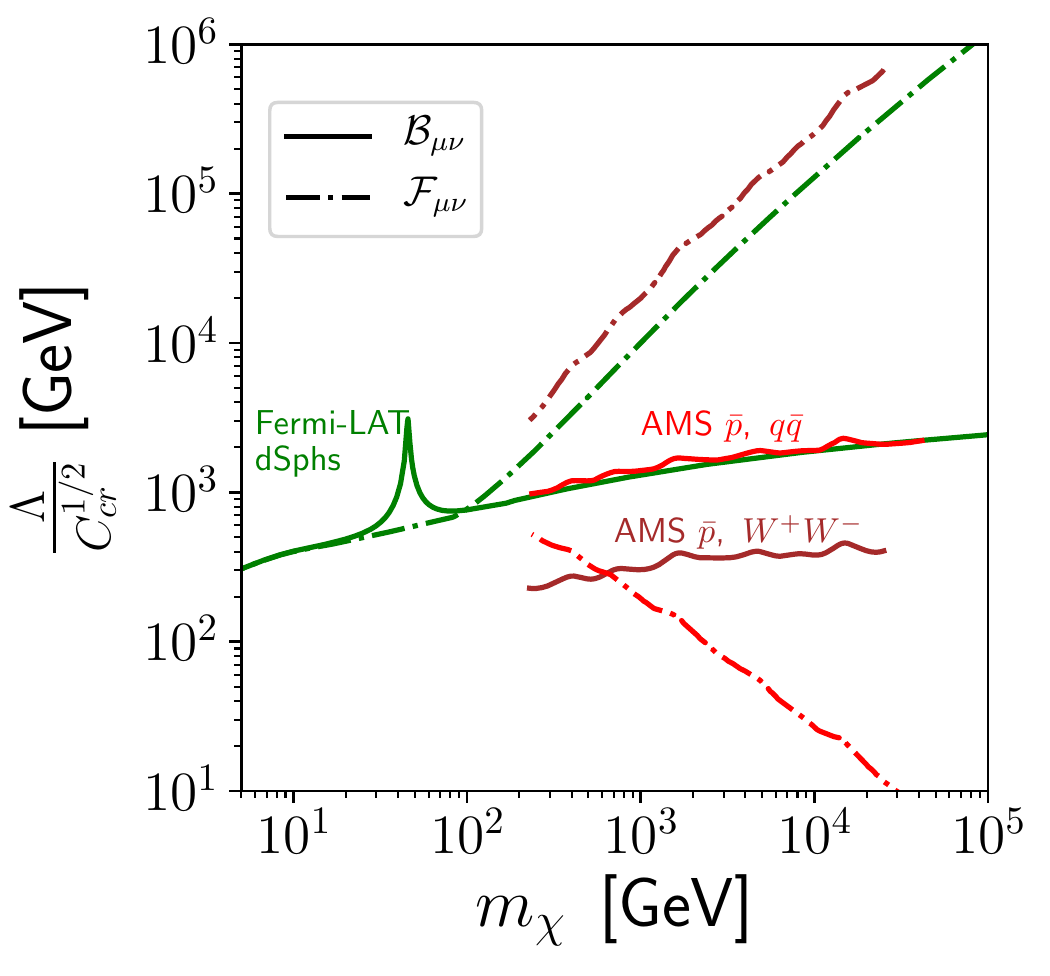}
    \caption{\textbf{Left:}  Branching ratios of the velocity averaged annihilation cross-section into all possible SM final states as a function of the dark matter mass, for the charge radius effective interaction. The line and colour scheme is the same as Figure~\ref{fig:sigma_cr}. \textbf{Right:} Exclusion limits (excluded the region below the curve) at 95\% CL coming from Fermi-LAT dwarf spheroidal galaxies~\cite{Ambrogi:2018jqj} (Fermi-LAT dSphs) and AMS 02 cosmic ray measurements~\cite{Cuoco:2017iax} (AMS $\bar{p}$, for the $q\bar{q}$ and $W^+W^-$ final states) in the plane $\Lambda/\mathcal{C}^{1/2}_{cr}$ versus the dark matter mass. The results are shown for both the cases of hypercharge (solid) and photon only (dashed) EFT dark matter model (dashed) as labelled.}
    \label{fig:sigma_branching}
\end{figure*}

From Table~\ref{tab:analytic}, all operators manifest a hierarchy between the $f\bar{f}$ and $W^+W^- (Zh)$ final states in the high energy limit,  which differs only by factors of their respective hypercharges, as already shown in Figure \ref{fig:sigma_cr}. This is supported by Figure~\ref{fig:sigma_branching} on the left, which shows these branching ratios for the charge radius interaction in particular, however all operators (neglecting the double insertions for the moment) have qualitatively the same trend.
We have included the dashed lines for the electromagnetic moment to reiterate the importance of taking the gauge invariant interaction. By now, the reader will be familiar with the large growth for the $W^+W^-$ channel, which erroneously provides also in the case of indirect detection an important constraint. The effect of interpreting indirect detection limits in the context of hypercharge instead of photon only EFT models is shown for the charge radius interaction on the right of Figure~\ref{fig:sigma_branching}:  Fermi-LAT dSph  bounds are the most constraining for hypercharge operators, while in the case of photon only operators cosmic-ray antiprotons~\cite{Cuoco:2017iax} from AMS 02 originating from $W^+W^-$ dominate. It should be all but unsurprising at this point that by taking the gauge violating effective interactions, indirect constraints get larger and larger as the mass of the dark matter increases, even surpassing the strong sensitivity of the direct detection constraints around 500 GeV.
Previously in the literature~\cite{Kavanagh:2018xeh} it has been claimed that the $W^-W^+$ final state is sub-dominant, which is clearly not the case for the electromagnetic interaction. Ultimately, their results considering fermionic final states only are much more in line with the one obtained with a correct treatment of the hypercharge operators, as these final states clearly dominate the annihilation cross-section, while bosonic final states  have $\rm BR \simeq 10^{-2}$.
What is of more interest is the appearance of both the $Z$-funnel region in Figures~\ref{fig:sigma_cr} and~\ref{fig:sigma_branching} and the presence of more annihilation channels in the hypercharge EFT case. For indirect detection, the monochromatic neutrino channel is of interest because it is a clean astrophysical messenger and its branching ratio is sizeable ($\sim 0.1$) at high energies and dominates in the $Z$-funnel region.

\begin{figure*}[t!]
    \centering
    \includegraphics[width=0.7\textwidth,trim={0.1cm 0.3cm 0.2cm 0.1cm},clip]{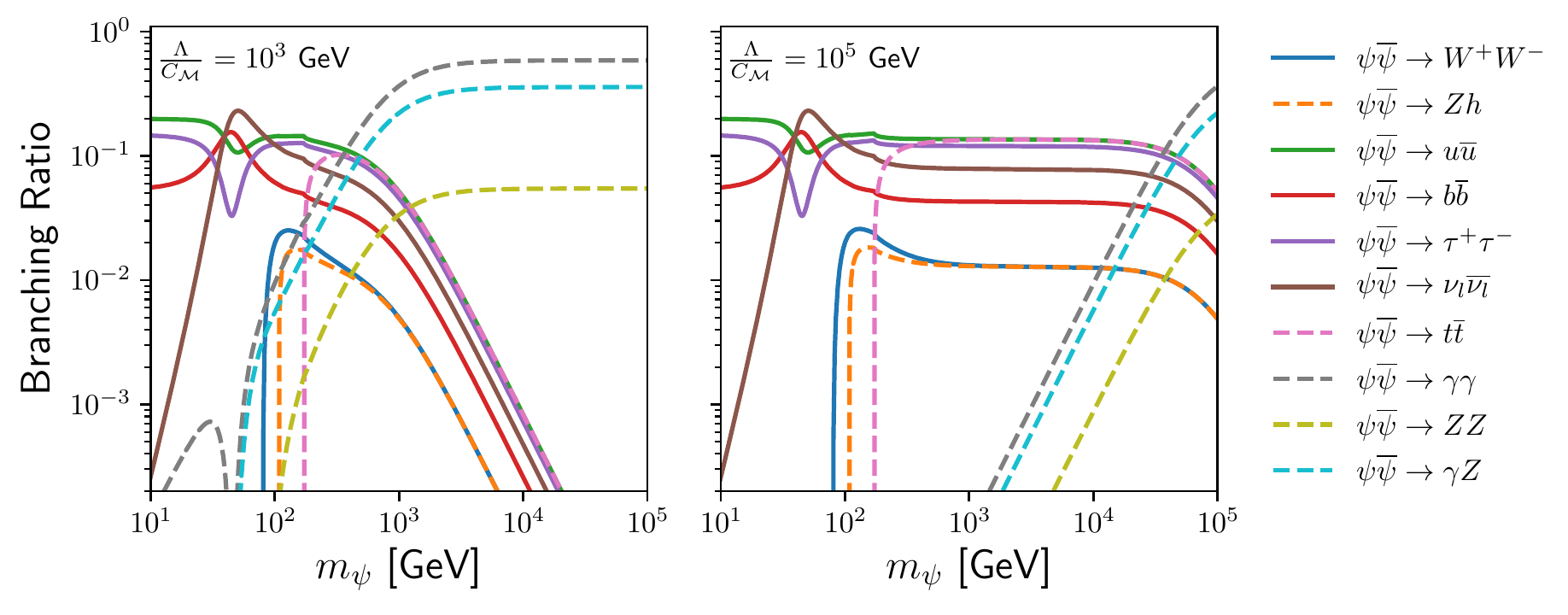}
    \caption{\textbf{Left:}  Branching ratios of the velocity averaged annihilation cross-section into all possible SM final states as a function of the dark matter mass, for the magnetic effective interaction with $\Lambda/\mathcal{C_M} = 10^3$ as labelled, including the dimension-6 contributions. The line and colour scheme is the same as Figure~\ref{fig:sigma_cr}. \textbf{Right:} Same as left for $\Lambda/\mathcal{C_M} = 10^5$.}
    \label{fig:sigma_branching_di}
\end{figure*}
The inclusion of double insertions for dimension-5 operators, as argued in section~\ref{subsec:production}, is both legitimate and correct. Doing this predictably complicates the picture as, for instance, the branching ratio of a specific annihilation channel no longer depends only on dark matter mass. Now some annihilation processes are proportional to different powers of effective coupling, {\it i.e.}
\begin{equation}
    \frac{\langle \sigma v \rangle_{\gamma\gamma}}{\langle \sigma v \rangle_{f\bar{f}}}\propto \frac{\Lambda^2}{\mathcal{C}_j^2}\,,
    \label{eq:branch_ratio_double}
\end{equation}
meaning that bigger couplings lead to the diphoton channel dominating for lower values of dark matter mass. In Figure~\ref{fig:sigma_branching_di} we show for two specific values of $\Lambda/\mathcal{C}_i$. The relative branching ratios of $\gamma\gamma$, $\gamma Z$ and $ZZ$ are the same in both cases, as expected from the fact that this is dictated by EW symmetry breaking. However, the onset of the dimension-6 operator is delayed to larger dark matter masses for larger $\Lambda/\mathcal{C}_i$, as expected from~\eqref{eq:branch_ratio_double}. Notice that the $\gamma\gamma$ final state has the largest branching ratio, followed closely by $\gamma Z$, while the $ZZ$ final state is one order of magnitude lower.

Before going into the details on how indirect detection constraints delimit the model parameter space we briefly describe what are the exclusion limits we consider in the analysis and how they have been recasted for the effective moments.\\[1ex]

\noindent \textit{Charged lepton and quark final states}:\\[0.3ex]
\noindent We consider the 95\% CL gamma-ray continuum bounds from Fermi-LAT dSphs and antiproton bounds from AMS 02 (AMS $\bar{p}$).

The dSph Fermi-LAT constraint has been obtained using~\texttt{MadDM}, which performs a statistical analysis for determining the limit given our specific model. For future constraints coming from land and space based telescopes on gamma-ray measurements from dSphs we have considered the Cherenkov Telescope Array (CTA)~\cite{Acharya:2017ttl} and the Large Synoptic Survey Telescope (LSST) + Fermi-LAT white paper~\cite{Drlica-Wagner:2019xan}. For CTA, we obtain a projection by using the sensitivity for the $\tau^+\tau^-$ annihilation channel shown in~\cite{Acharya:2017ttl}, which is the most constraining for the hypercharge EFT models. The LSST+Fermi-LAT (LSST+LAT) projection assumes that it is accurate to scale the $b\bar{b}$ projection to other fermionic channels. A more thorough analysis would likely achieve better projections, but this is beyond the scope of this work. Whenever relevant, we include into the gamma-ray continuum limits the contribution of the $ZZ$ and $\gamma Z$ final states (the latter contribution is scaled by 1/2 to take into account the fact that only one $Z$ boson is emitted).

Antiproton bounds have been obtained from~\cite{Cuoco:2017iax}, by rescaling each final state with BR of our models. Notice that the astrophysical uncertainties are huge, and even when profiled out their inclusion makes an envelope that can shift the bound up or down by a factor of roughly four. Nonetheless, these are competitive bounds for dark matter masses larger than 250 GeV.

Lastly we also consider the Planck limits~\cite{Slatyer:2015jla} on dark matter annihilating into $e^+e^-$, which are competitive with dSphs bounds at very low dark matter masses. These bounds are based on the fact that annihilating dark matter injects electromagnetically interacting particles during the dark ages, which can potentially modify the residual ionization fraction, broaden the last  scattering  surface and  modify the anisotropies of the cosmic microwave background.\\[1ex]
\noindent \textit{Neutrino exclusion limits}:\\[0.3ex]
\noindent A comprehensive study on the constraints from dark matter to SM neutrinos from the Galactic Centre or from the diffuse isotropic background has been performed in~\cite{Arguelles:2019ouk}, from which we take constraints from Antares~\cite{Adrian-Martinez:2015wey}. The IceCube constraints are a combination of~\cite{Aartsen:2016pfc} together with~\cite{ElAisati:2017ppn}, which are derived for neutrino lines. Future projections from the neutrino telescopes KM3NeT~\cite{Adrian-Martinez:2016fdl} and Hyper-Kamiokande~\cite{Abe:2018uyc} are also shown, again derived in the most optimistic scenario of neutrino lines. The experimental constraints have usually been obtained for a NFW dark matter density profile hypothesis~\cite{Navarro:1996gj,Navarro:2008kc}. Each limit has been rescaled accordingly with the branching ratio into neutrino lines of the model. All bounds are shown at 90\% CL. The $ZZ$ and $\gamma Z$ final states produce a continuum neutrino spectrum, hence a rescaling of the above mentioned limits and projections is not correct. The implementation of this final state would imply a full recasting of experimental limits using their likelihoods, however this goes beyond the scope of this work and our results remain unchanged, as the $ZZ$ contribution into neutrinos is a subdominant component anyway.\\[1ex]

\noindent \textit{Gamma-ray lines}:\\[0.3ex]
\noindent We have considered two lines searches at 95\% CL from the Galactic Centre by Fermi-LAT~\cite{Ackermann:2015lka} (Fermi $\gamma\gamma, \gamma Z$) and HESS~\cite{Rinchiuso:2019rrh} (HESS $\gamma\gamma, \gamma Z$), both obtained with the Einasto dark matter density profile assumption~\cite{Einasto:2009zd}. The experimental exclusion bounds have been rescaled again by the corresponding branching ratio of the processes $\psi \bar{\psi} \rightarrow \gamma\gamma, \gamma Z$ (this latter being divided by two to take into account that only one photon per annihilation is emitted). Additionally, we show the projected sensitivity for line searches towards the Galactic Centre for CTA (assuming consistently an Einasto dark matter density profile) from~\cite{Lefranc:2016fgn} (CTA $\gamma\gamma, \gamma Z$).
\begin{figure*}[t!]
    \centering
    \includegraphics[width=0.7\columnwidth]{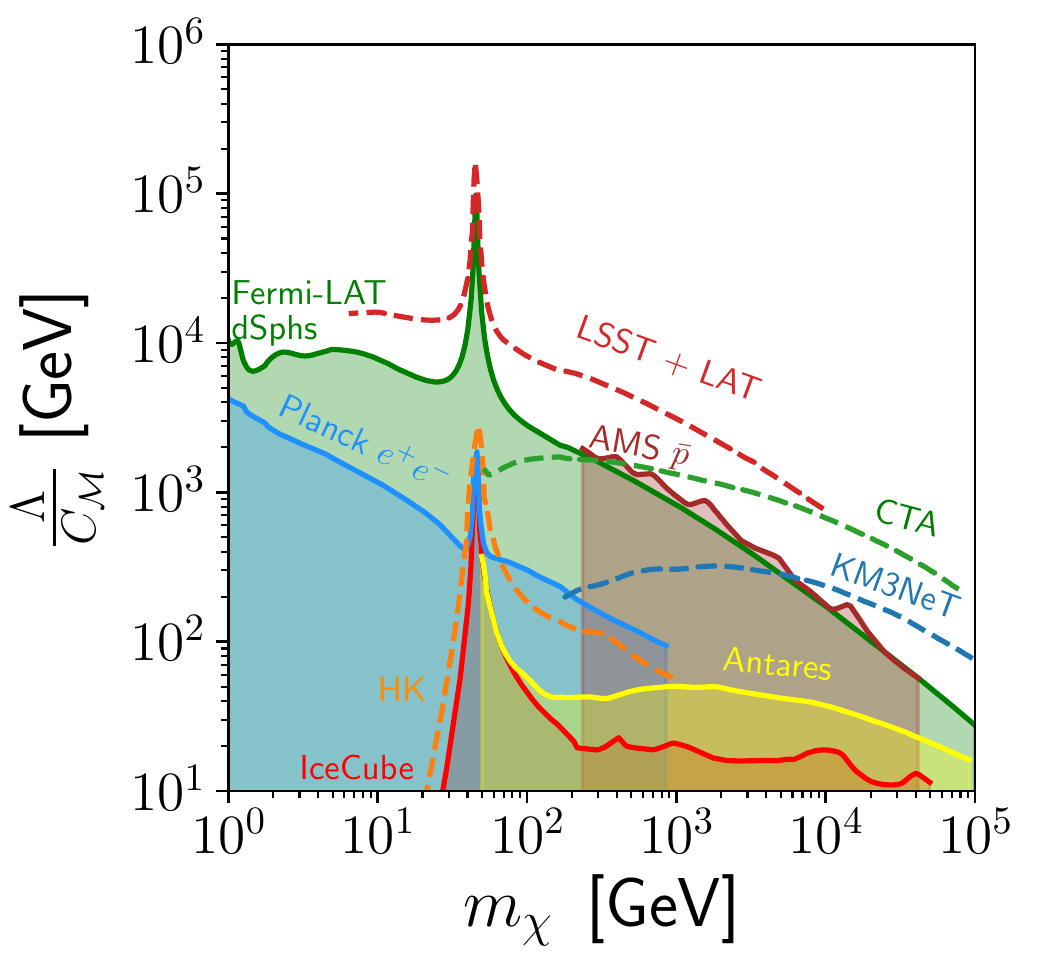}
    \includegraphics[width=0.7\columnwidth]{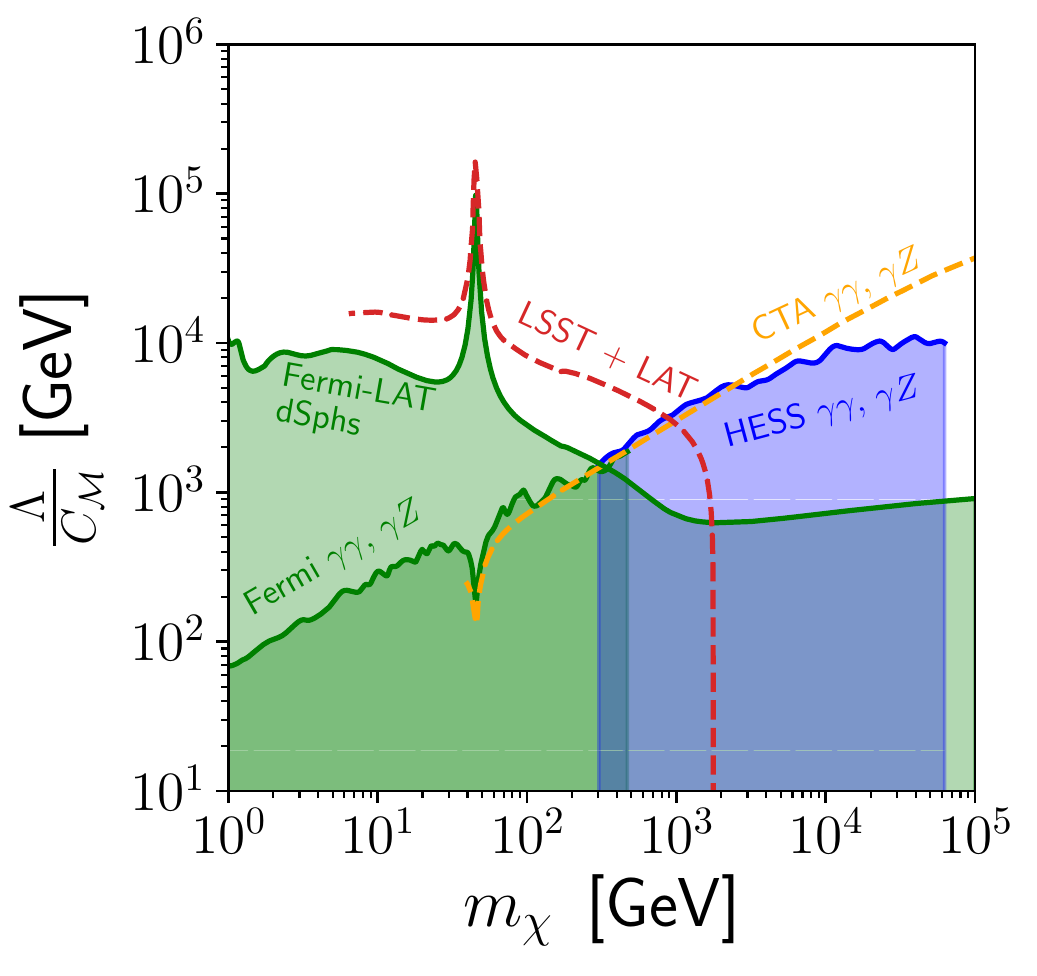}
    \includegraphics[width=0.7\columnwidth]{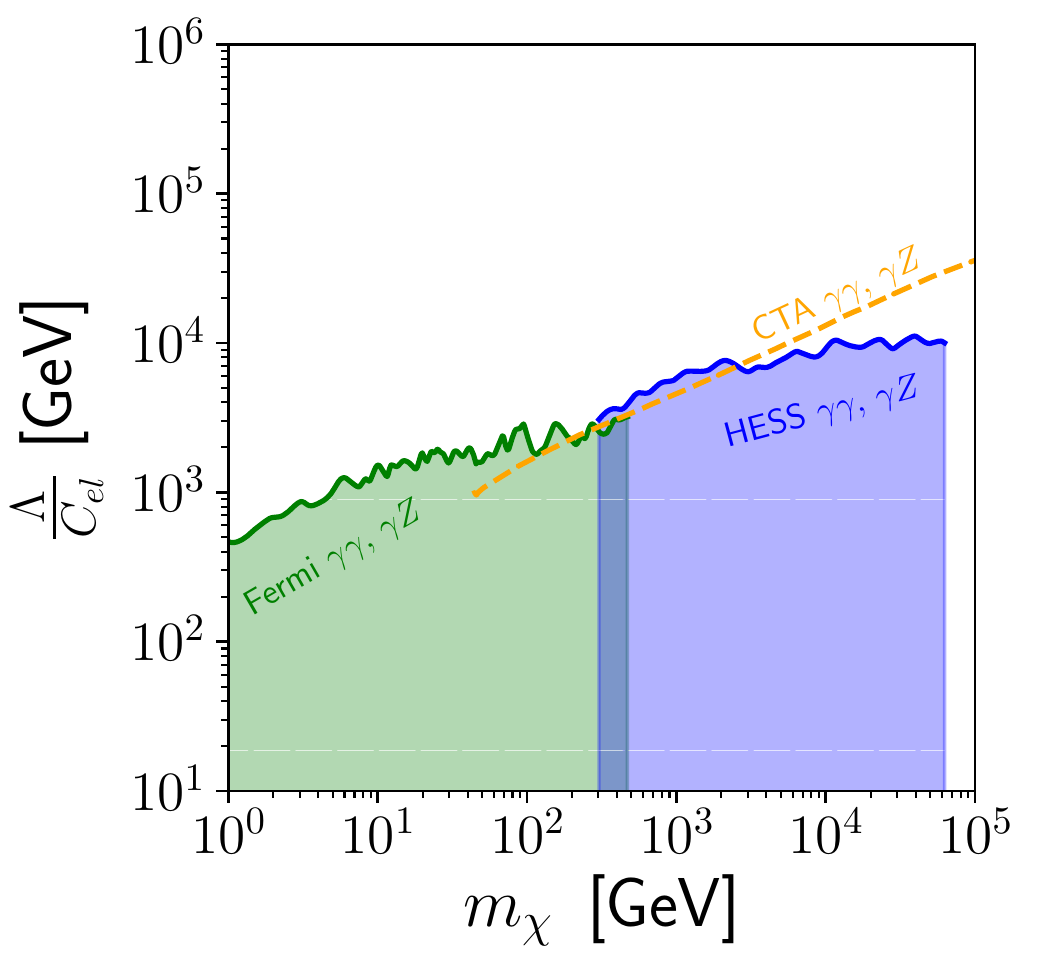}
    \includegraphics[width=0.7\columnwidth]{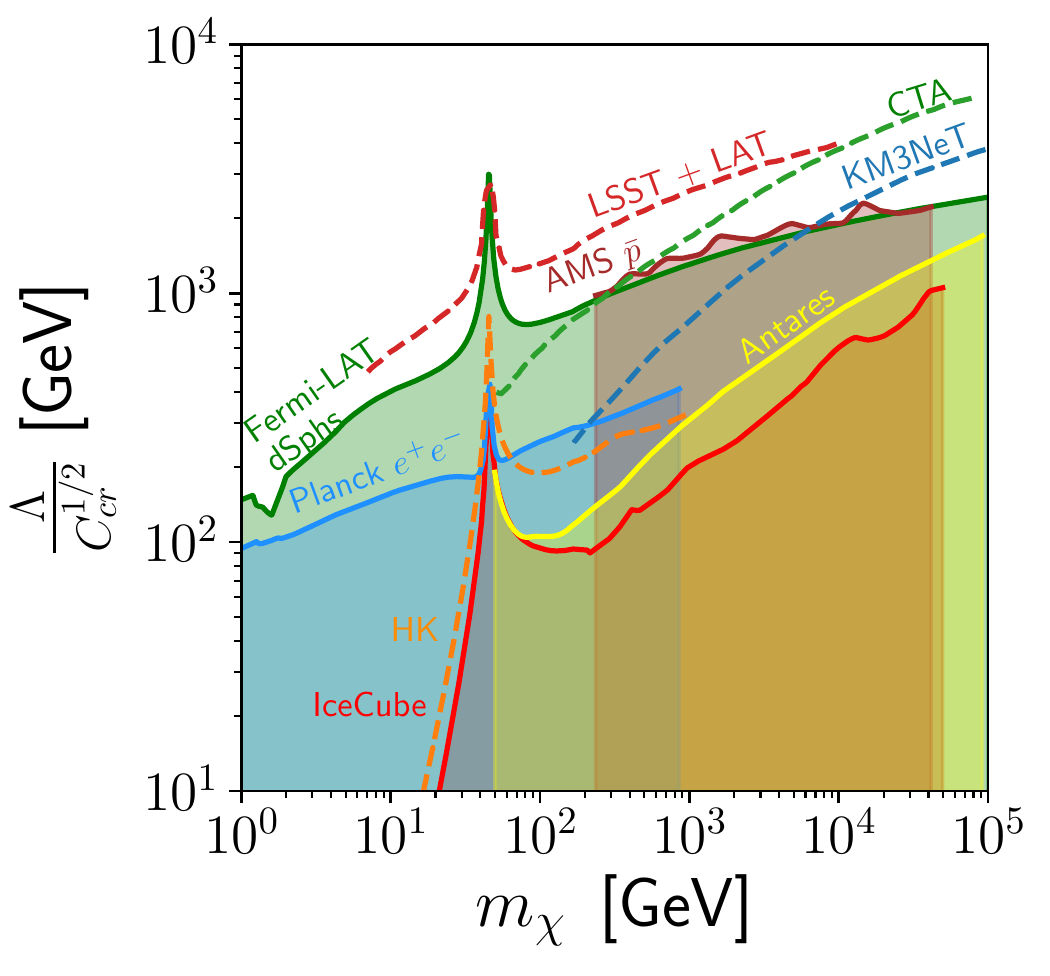}
    \caption{\textbf{Top left:} Indirect detection constraints for the magnetic hypercharge moment interaction in the plane $\Lambda/\mathcal{C_{M}}$ and dark matter mass $m_\chi$. \textbf{Top right:} Same as left including the dimension-6 operator leading to the diphoton final state.
   \textbf{Bottom left:}   Same as top right for the electric dipole operator, notice that only diphoton exclusion bounds are relevant for this operator.  \textbf{Bottom right:} Same as top left for the charge radius interaction. The experimental constraints shown by shaded region are actual bounds, while dashed lines are projected sensitivities, as in Figure~\ref{fig:prod_and_DD}.  {\it Gamma-ray continuum bounds}: Fermi-LAT dwarf spheroidal galaxies~\cite{Ambrogi:2018jqj} (green), CTA projection~\cite{Acharya:2017ttl} (green) and LSST+Fermi-LAT dwarf spheroidal galaxies sensitivity~\cite{Drlica-Wagner:2019xan} (dark red); {\it Neutrino bounds}: IceCube~\cite{ElAisati:2017ppn} (red), Antares~\cite{Arguelles:2019ouk,Adrian-Martinez:2015wey} (yellow), Hyper-Kamiokande (HK)~\cite{Abe:2018uyc} (orange) and KM3NeT~\cite{Adrian-Martinez:2016fdl} (blue). {\it Additional constraints}: Planck~\cite{Slatyer:2015jla} (light blue), AMS 02 cosmic rays~\cite{Cuoco:2017iax} (brown), gamma-ray lines from the Galactic Centre from Fermi-LAT~\cite{Ackermann:2015lka} (green), HESS~\cite{Rinchiuso:2019rrh} (dark blue) and expected CTA sensitivity~\cite{Lefranc:2016fgn}. Unlike Figure~\ref{fig:prod_and_DD} the relic density curve is not shown to avoid additional cluttering. Details on the CL of each exclusion limit are provided in the text.}
    \label{fig:ID_summary}
\end{figure*}

We report in Figure~\ref{fig:ID_summary} the constraints from indirect searches as well as the reach of future probes for the hypercharge EFT models, in a comprehensive fashion. By considering first the results for the magnetic dipole at dimension-5 only (top left panel) we see that current dSph Fermi-LAT limits are the most constraining, together with cosmic-ray antiproton bounds at high masses. We see that the current neutrino bounds are substantially weaker than that of Fermi-LAT and AMS $\bar{p}$.  It is important to note however, that future experiments such as KM3NeT will be competitive with CTA for heavy dark matter while LSST discovery of new dSphs will increase the current Fermi-LAT bounds (LSST+Fermi-LAT). Furthermore, complimentary across annihilation channels will prove hugely important in the event of positive signal. This description is qualitatively unchanged for the charge radius operator (bottom right panel).

As discussed above, the magnetic dipole phenomenology should be considered up to dimension-6, this shown in the top right panel. When the branching ratio of $\langle \sigma v \rangle_{\gamma\gamma}, \gamma Z$ start to dominate, all continuum searches weaken, as expected from Figure~\ref{fig:sigma_branching_di}. When this occurs depends on Equation~\eqref{eq:branch_ratio_double}. Interestingly, up to this certain mass value the magnetic dipole behaves like a pure dimension-5 operator, while above it the HESS limit dominates in constraining the model parameter space. The experimental sensitivity from continuum searches however doesn't completely drop to zero as it is still sourced by the $ZZ$ final state. Here, for simplicity, we only show the behaviour of the dominant Fermi-LAT dSph limit.

The projected sensitivity of CTA only slightly improves the sensitivity of gamma-ray line searches to the magnetic moment operator at very large dark matter masses. Notice that the LSST+Fermi-LAT sensitivity drops artificially to zero because it is obtained from a $b\bar{b}$ final state and could not be easily translated into a $ZZ$ ones. We do not expect our conclusion to be changed if this bound were to be properly included, as that portion of the parameter space is already excluded by HESS.
The case of the electric dipole interaction (bottom left) is peculiar as, as stated above, the diphoton  and $\gamma Z$ channels are actually the only $s$-wave interactions, so this introduces the possibility of constraining the model via indirect probes which would not be possible otherwise. Also notice that unlike the magnetic and charge radius, there is no Z-peak (or dip for the line searches), due to the p-wave suppression of Z mediating diagrams. Here we do not consider the $ZZ$ gamma-ray continuum as it would give a subdominant contribution as for the magnetic dipole operator, only showing the effect of line searches.

\section{Discussion and conclusions}
\label{sec:concl}

\begin{figure*}[t!]
    \centering
    \includegraphics[width=0.7\columnwidth]{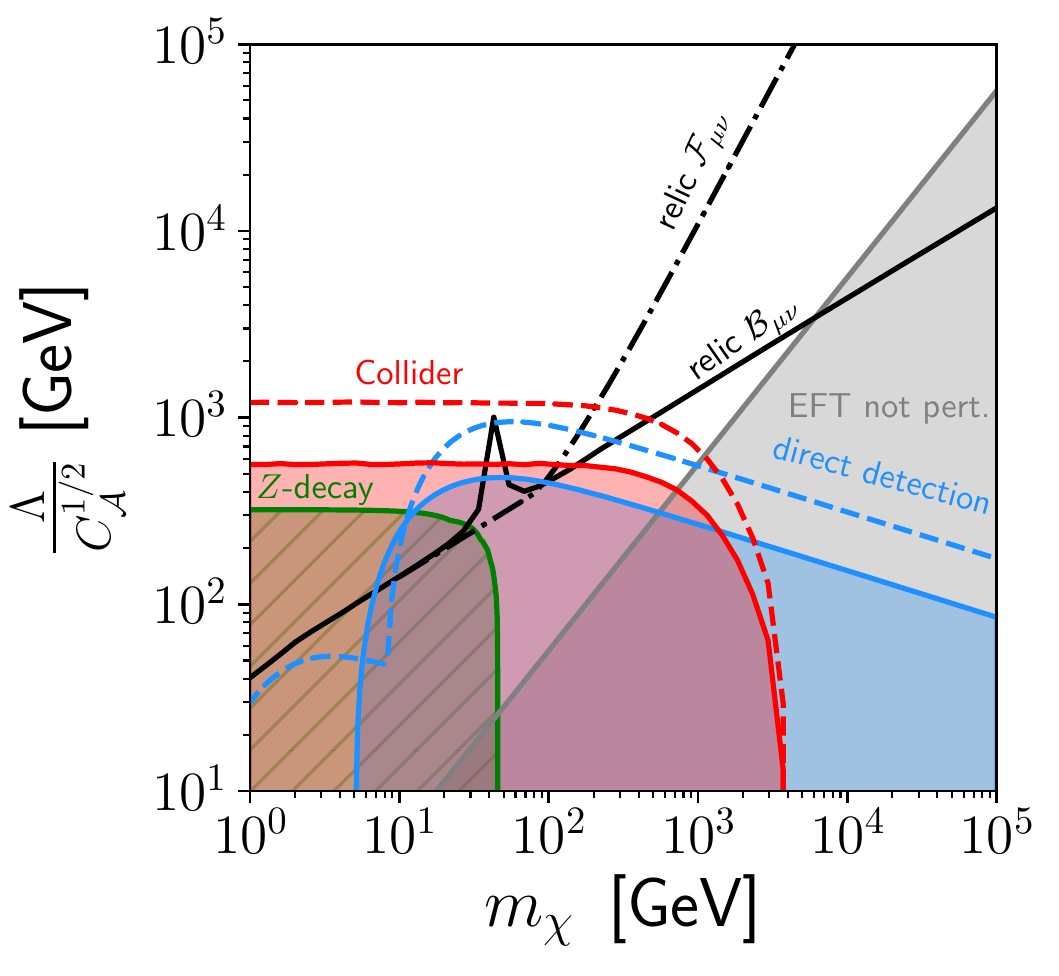}
    \includegraphics[width=0.7\columnwidth]{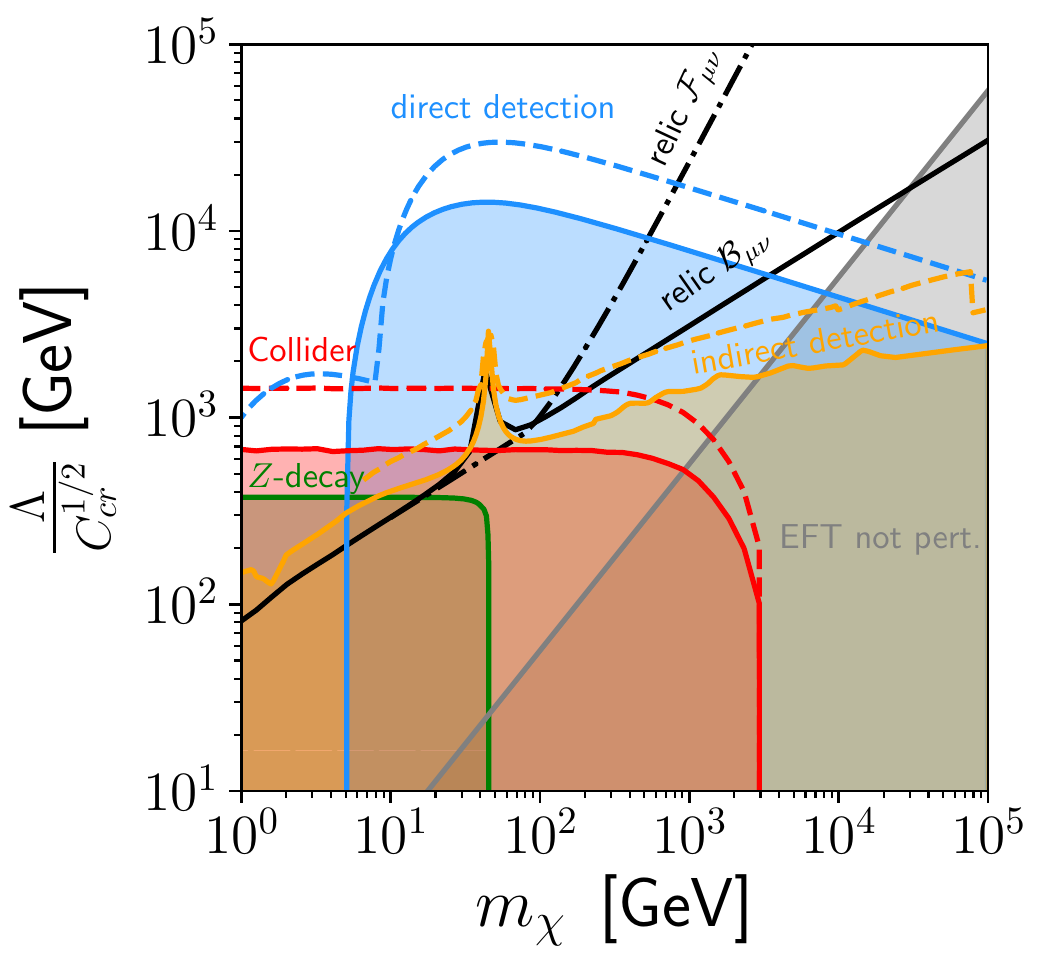}
        \includegraphics[width=0.7\columnwidth]{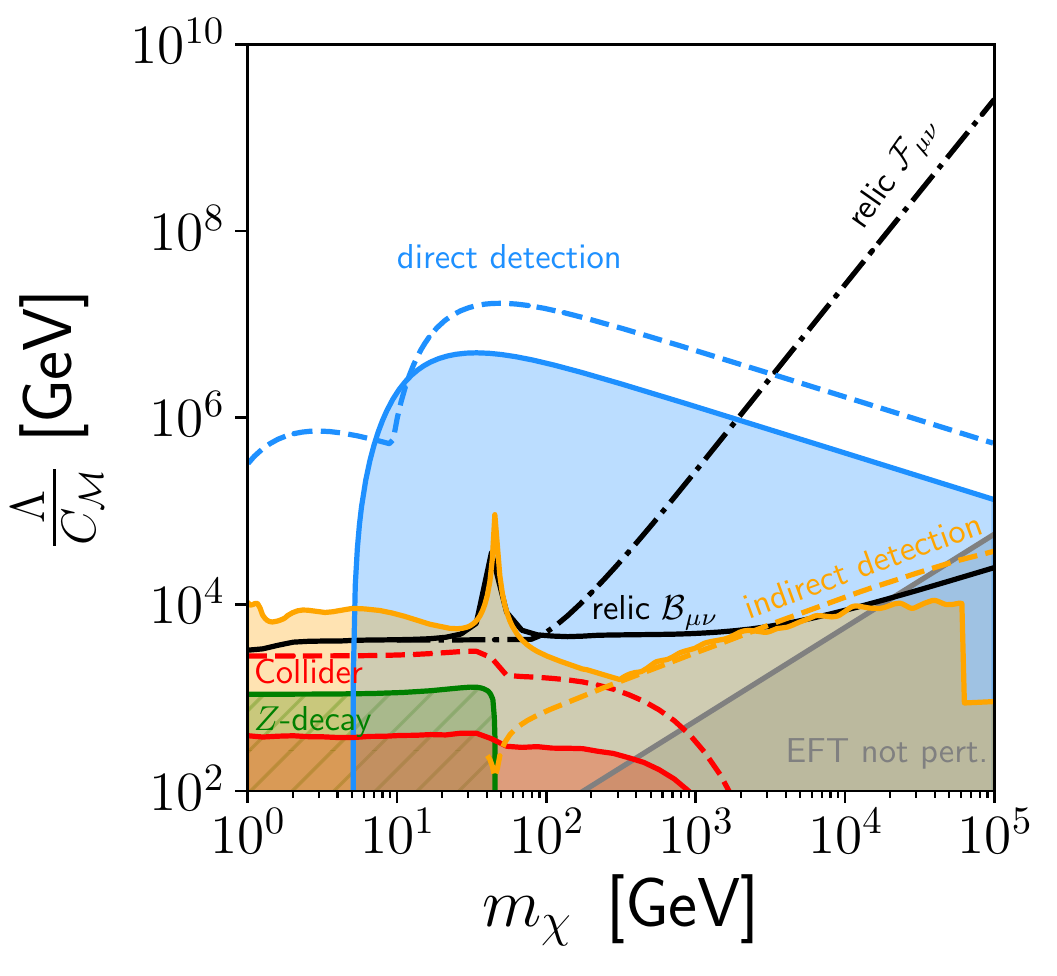}
    \includegraphics[width=0.7\columnwidth]{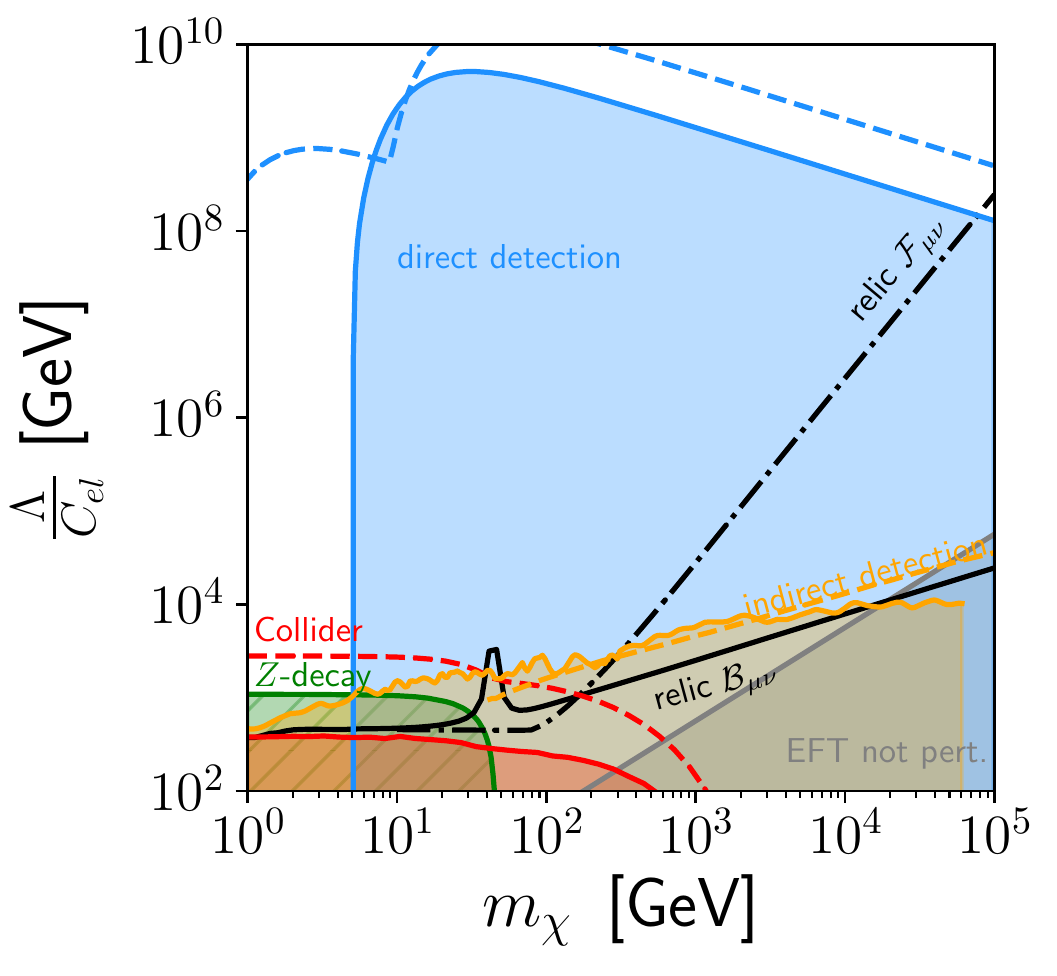}
    \caption{\textbf{Left:} Summary of the most constraining search from direct (blue), indirect  (yellow) detection and collider searches (red for LHC searches and green for LEP searches) in the plane $\Lambda/\CA^{1/2}$ versus $m_{\chi}$ for the anapole moment operator. Current experimental bounds are denoted with solid lines and shaded regions, while projected sensitivities are shown with dashed lines. The relic density with denoted by a black line for the hypercharge (solid) and electromagnetic (dot-dashed) scenarios. \textbf{Top right, bottom left and right:} Same as top left for the charge radius operator and for the magnetic and electric dipoles respectively.}
    \label{fig:all_summary}
\end{figure*}

In this paper we have considered effective interactions between fermionic dark matter $\chi$ (both Dirac and Majorana) and photons. The only effective interaction which is not zero for Marojana dark matter is the anapole moment at dimension-6 while for Dirac dark matter the magnetic and electric dipole at dimension-5 and the charge radius at dimension-6 also exist.
These so-called electromagnetic form factors might have seemed to be thoroughly studied in the literature however at certain energies the EFT treatment has not been properly addressed, leading to the wrong conclusions. Our analysis amends these issues and results in a proper mapping of the operator parameter space in the light of current and future dark matter and collider searches. The results from each section are collated and presented together in Figure~\ref{fig:all_summary}, to summarise our main findings and bring to light possible caveats.

Starting with the primary issue, a naive treatment of the electromagnetic operator, dark matter coupling to $F_{\mu\nu}$, signals gauge-violating processes at large energies or dark matter masses.
Gauge invariance is simply and correctly retrieved by coupling the dark matter to the $U(1)_Y$ gauge boson $B_{\mu\nu}$ of the Standard Model instead of the photon at energies above the $W$ boson mass. This is dictated by the proper choice of low energy symmetries of the EFT given the energy scales of the processes relevant for dark matter phenomenology. The price or indeed recompense of the consistent description is a set of interactions for dark matter with the $Z$ boson. An immediate consequence is that constraints apply from the $Z$ invisible decay width, having important consequences for the parameter space at low dark matter masses. The description also leads to a richer set of final states ($Z\gamma$ \& $ZZ$ in addition to $\gamma\gamma$) for indirect detection via the dimension-5 interactions.

This issue is represented in the plots by the relic density lines (black) from both the gauge-violating $F_{\mu\nu}$ and $B_{\mu\nu}$ and tells us that the gauge violating annihilation cross-sections leads to a completely different picture for where viable thermal dark matter candidates are in parameter space. Additionally, the $Z$-width bound from LEP (green region) closes the window of freeze-out dark matter for masses $\lesssim 45$ GeV for all but the magnetic dipole. The gauge violating process, namely $W^+W^- \to \chi \chi$, also provides large, unphysical contributions in collider experiments, which would lead to incorrect conclusions concerning the most sensitive searches (cf. VBF instead of mono-jet searches, as described in section~\ref{sec:pheno_colliders}). The same scattering process would lead to the incorrect impression that indirect detection experiments have a better sensitivity that direct searches at high dark matter masses. The hypercharge form factors correctly describe the electromagnetic effective interaction of dark matter at energies relevant for dark matter and collider searches, so in the following we only consider those.

The search that is most dominant in Figure~\ref{fig:all_summary} is direct detection. These experiments (here shown XENON1T and projected LZ and SuperCDMS sensitivities) are able to basically cover the parameter space up until the perturbative limit for the EFT for magnetic and electric dipoles. More generally, direct detection is the strongest current constraint above $m_{\chi}\sim 6 $ GeV for all but the anapole interaction. Notice that for direct detection the pure electromagnetic description is valid, as the relevant energies are much below the EW scale.

At low masses, direct detection is likely more sensitive than $Z$ invisible decay width for all interactions but the anapole moment. Above $m_{\chi}\sim100$ GeV, experimental sensitivity can be improved by analysing recoil energies up to 500 keV as shown in Ref.~\cite{Bozorgnia:2018jep}. This is of particular relevance for the anapole interaction since improvements in this region could constrain the thermal freeze-out scenario.

It is important, however, to emphasise that astrophysical assumptions are at play in these bounds. For example, there is a plausible level of uncertainty in the density of dark matter in the solar system, see {\it e.g.}~\cite{Bidin:2012vt,Pitjev:2013sfa,Read:2014qva,Benito:2016kyp,Ibarra:2018yxq}. Given that the limits for direct detection are so sensitive in the case of dimension-5 operators, it is still likely that direct detection remains the most constraining search also in the case of huge variation of the local density. The only region which might escape direct detection is possibly the low $m_{\chi}$ region for the magnetic dipole, where indirect limits (Fermi-LAT dSphs) dutifully cover the relic line and overcome as well the $Z$-decay bound.

Astrophysical uncertainty is certainly at play in indirect searches as well, but in a completely different domain. Limits on the continuum of gamma rays coming from the Fermi-LAT telescope are derived from a set of dwarf spheroidal galaxies, which are dark matter dominated objects. The argument used to weaken limits in direct detection experiments is simply not available in the indirect case. For this reason, the complimentary nature of the two searches is important, especially since large portions of the thermal relic line is covered by indirect searches in the cases of magnetic, electric and charge radius interactions.

The strongest mono-jet bounds from section~\ref{subsec:pheno_monojet} and their high luminosity projections are shown in Figure~\ref{fig:all_summary} as ``collider bounds''.
When we compare our new mono-jet bounds to other dark matter searches, we achieve competitive results apart from the impressive hierarchy between the direct detection sensitivity to dimension-5 operators and all others. The best case is for the dimension-6 operators, in fact, for the anapole interaction, the constraints currently are more sensitive for the whole valid parameter space. We also point out that these limits either already, or will eventually surpass those coming from invisible $Z$ decays in all cases. We observe an interesting complementarity between the high luminosity LHC and direct detection bounds not only for the anapole but also for the charge radius interaction, which was not naively expected given the fact that it induces spin-independent nucleon scattering.

One limitation of our results is that our study focuses only on one interaction at a time, rather than allowing all operators to vary at once. In such a global study, renormalization group running and operator mixing may well change the picture, since dark matter annihilation, $Z$-decay, collider production and nuclear scattering all take place at different scales (See Refs.~\cite{Matsumoto:2014rxa,DEramo:2014nmf,DEramo:2016gos,Cao:2020ihv} for relevant studies). Nevertheless, the results presented here can provide useful inputs to a global DM-EFT analysis, in order to properly assess how much parameter space is left for a thermally produced dark matter candidate. In such a case, the Dirac dark matter scenario will be severely impeded by the dimension-5 moments, with no good reason, a priori to suppress them. Data therefore seems to be pushing us towards a Majorana dark matter candidate.

We have also endeavoured to assess the validity of the EFT description given the sensitivity obtained by each experiment. We use naive perturbativity arguments to suggest regions in which predictions are not expected to be reliable. The validity issue is especially important for the collider bounds, since a range of energies are naturally probed by the LHC. Therein we discuss the viable range of Wilson coefficients that admit a valid EFT interpretation, concluding that couplings of order one are required.
This means that, a thermal relic produced by loops is not likely to be compatible with a viable EFT interpretation. On the other hand, tree level processes via a $U(1)^{\prime}$-mixing are unavailable to Majorana particles. Assuming a simple thermal history of dark matter may well lead to exotic model building territory, the full implications of which can only be assessed after a full global analysis.

\section*{Acknowledgments}
Thanks to B. Kavanagh and J. Heisig for communications on direct and AMS $\bar{p}$ searches respectively and to C. Zhang for useful discussions on the EFT expansion. Thanks to B. Fuks for helping setup the FeynRules wiki page. CA is supported by the Innoviris  ATTRACT 2018 104 BECAP 2 agreement. AC and KM are supported by the F.R.S.-FNRS under the Excellence of Science EOS be.h project n. 30820817.

%
\bibliographystyle{JHEP}
\bibliography{./refs}

\end{document}